\documentclass[twocolumn,epjc3]{svjour3}
\usepackage[dvipsnames]{xcolor}
\usepackage{microtype}
\usepackage{eso-pic}
\usepackage{feynmp}
\usepackage[T1]{fontenc}
\usepackage[utf8]{inputenc} % omit if using lualatex/xelatex
\usepackage[section]{placeins} % or without [section] if you prefer manual only
\usepackage{capt-of}
\usepackage{float}
\usepackage{subfigure}
\newcommand{\preprintA}{SI-HEP-2024-34}

\newcommand{\preprintfont}{\normalfont\rmfamily\footnotesize}
\newcommand{\preprintcolor}{black}
\newcommand{\lb}{\left(}
\newcommand{\rb}{\right)}
\newlength{\preprintTopMargin}    
\newlength{\preprintRightMargin}  
\makeatletter
\@ifundefined{shortstackgap}{\newlength{\shortstackgap}}{}
\makeatother

\newcommand{\PutPreprintFirstPage}{%
  \AddToShipoutPictureFG*{%
    \put(
      \LenToUnit{\dimexpr\paperwidth-\preprintRightMargin\relax},
      \LenToUnit{\dimexpr\paperheight-\preprintTopMargin\relax}
    ){%
      \makebox[0pt][r]{%
        \preprintfont \color{\preprintcolor}%
        \shortstack[r]{\preprintA\\\preprintB}%
      }%
    }%
  }%
}

\journalname{Eur. Phys. J. C}
\usepackage{arydshln, cite, graphicx, hyperref, amsmath, newtxtext, newtxmath, relsize, booktabs, tabularx, array, rotating}
\hypersetup{
  colorlinks=true,
  linkcolor=ForestGreen,
  citecolor=ForestGreen,
  filecolor=ForestGreen,
  urlcolor=ForestGreen,
  pdftitle={Triple Higgs Boson Production with Two Heavy Scalars at the LHC via a Simplified Approach},
  pdfauthor={Andreas Papaefstathiou and Gilberto Tetlalmatzi-Xolocotzi},
  pdfsubject={High Energy Physics - Phenomenology}
}
\newcolumntype{Y}{>{\raggedright\arraybackslash}X}
\begin{document}

\PutPreprintFirstPage

\title{Triple Higgs Boson Production with Two Heavy Scalars at the LHC via a Simplified Approach}

\author{Andreas Papaefstathiou\thanksref{i1,e1} \and
        Gilberto Tetlalmatzi-Xolocotzi\thanksref{i2,e2}}

\institute{\hypertarget{i1}{}Department of Physics, Kennesaw State University, 830 Polytechnic Lane, Marietta, GA 30060, USA\label{i1} \and
           \hypertarget{i2}{}Theoretische Physik 1, Center for Particle Physics Siegen (CPPS), Universit\"at Siegen, Walter-Flex-Str. 3, 57068 Siegen, Germany\label{i2}}

\thankstext{e1}{\email{apapaefs@kennesaw.edu}}
\thankstext{e2}{\email{gtx@physik.uni-siegen.de}}

\date{}
\maketitle
\begin{abstract}
      We investigate triple Higgs boson production at the CERN LHC, in models containing two new heavy neutral scalar particles. We apply a simplified factorized approach in the narrow-width approximation to double-resonant triple Higgs boson production, and demonstrate that relevant constraints can be derived on models with extended scalar sectors, during the high-luminosity phase of the LHC. We also find, within an explicit two-real-singlet model and for the cut-based 6 $b$-jet analysis considered here, that the double-resonant contribution captures the dominant features of the signal, while non-resonant components have a limited impact on the expected sensitivity. The method therefore provides a fast approximate framework for models in which triple Higgs boson production is dominated by a narrow double-resonant topology.
\end{abstract}

\section{Introduction}\label{sec:intro}
Multi-Higgs boson production processes at colliders, such as the CERN Large Hadron Collider (LHC), can provide further insight into the electroweak and scalar sectors of the standard model (SM), going beyond the information harnessed by studying single Higgs production alone ~\cite{ATLAS:2012yve,CMS:2012qbp,Higgs:1964pj,Englert:1964et,Guralnik:1964eu}. The two primary multi-Higgs boson production processes, pair production and triple production, can be used within the SM to yield a consistency check of the triple and quartic self-interactions, respectively, verifying the ``standard'' shape of the Higgs boson's ($h_1$) potential,
\begin{equation}
\mathcal{V}(h_1) = \frac{1}{2} m_{1}^2 h_1^2 + \frac{ m_{1}^2 } { 2 v } h_1^3 + \frac{ m_{1}^2 } { 8 v^2 }  h_1^4\;,
\end{equation}
where $m_{1}\approx 125$~GeV is the Higgs boson mass, and $v\approx 246$~GeV is the Higgs vacuum expectation value. In the realm of such self-coupling measurements, within and beyond the SM, Higgs boson pair production has received considerable attention through experimental (e.g.~\cite{CMS:2017yfv,CMS:2017hea,CMS:2017rpp,CMS:2018tla,CMS:2018vjd,CMS:2018sxu,CMS:2018ipl,CMS:2020tkr,CMS:2022cpr,CMS:2022hgz,CMS:2022kdx,CMS:2022omp,ATLAS:2014pjm,ATLAS:2015zug,ATLAS:2015sxd,ATLAS:2016paq,ATLAS:2018rnh,ATLAS:2018dpp,ATLAS:2018hqk,ATLAS:2018uni,ATLAS:2018fpd,ATLAS:2018ili,ATLAS:2019qdc,ATLAS:2019vwv,ATLAS:2020jgy,ATLAS:2021ifb,ATLAS:2022xzm,ATLAS:2022jtk,ATLAS:2023qzf,ATLAS:2023gzn}) and theoretical (e.g.~\cite{Baur:2002rb,Baur:2002qd,Baur:2003gp,Dolan:2012ac,Papaefstathiou:2012qe,Cao:2013si,Goertz:2013kp,Arbey:2013jla,deFlorian:2013uza,Gupta:2013zza,Ellwanger:2013ova,Barr:2013tda,Maierhofer:2013sha,deFlorian:2013jea,Dolan:2013rja,Goertz:2013eka,Goertz:2014qta,Azatov:2015oxa,Frederix:2014hta,Baglio:2014nea,FerreiradeLima:2014qkf,deFlorian:2014rta,Hespel:2014sla,Barger:2014taa,Godunov:2014waa,Liu:2014rba,Maltoni:2014eza,Chen:2014ask,Barr:2014sga,Kumar:2015kca,Kumar:2015tua,MartinLozano:2015vtq,Papaefstathiou:2015iba,Dawson:2015oha,Kotwal:2015rba,Lu:2015jza,Carvalho:2015ttv,Cao:2015oxx,Batell:2015koa,Dawson:2015haa,Cao:2015oaa,Kanemura:2016tan,Contino:2016spe,Cao:2016zob,Banerjee:2016nzb,Huang:2017jws,Nakamura:2017irk,Lewis:2017dme,DiLuzio:2017tfn,Grober:2017gut,Zurita:2017sfg,Arganda:2017wjh,Adhikary:2017jtu,Bauer:2017cov,Maltoni:2018ttu,Borowka:2018pxx,Goncalves:2018qas,Chang:2018uwu,Basler:2018dac,Adhikary:2018ise,DiMicco:2019ngk,Li:2019uyy,Cheung:2020xij,Iguro:2022fel,Davies:2023npk,Bhattiprolu:2024tsq}) studies. Primarily due to its tiny SM cross section at hadron colliders~\cite{Maltoni:2014eza}, the triple Higgs boson process has received much less attention. To the best of our knowledge, it was first investigated in~\cite{Plehn:2005nk}, where it was demonstrated that the prospects for the measurement of the quartic self-coupling will be very challenging, both at the LHC and at a future ``VLHC'', with a centre-of-mass energy of 200 TeV. Subsequent studies of triple Higgs boson production at future colliders~\cite{Papaefstathiou:2015paa,Chen:2015gva,Fuks:2015hna,Papaefstathiou:2017hsb,Fuks:2017zkg,Liu:2018peg,Papaefstathiou:2019ofh,deFlorian:2019app,Chiesa:2020awd,Abdughani:2020xfo,Fuks:2025gjv}, armed with the knowledge of the value of the Higgs boson mass, and with the prospects for Higgs boson pair production measurements at hand, further quantified the difficulty of this measurement. Nevertheless, these studies demonstrated that \textit{some} useful information may be obtained via the triple Higgs boson final state.
{We also note that, in extended scalar sectors, the scalar self-couplings entering such interpretations may receive sizeable electroweak radiative corrections. Recent work has developed precision one- and two-loop predictions for trilinear Higgs and more general scalar self-interactions, and has studied their impact on Higgs-sector observables~\cite{Bahl:2023eau,Aiko:2023nqj,Bahl:2025wzj,Braathen:2025qxf,Bahl:2026aou}.}

\begin{figure}[htp]
    \centering
    \setlength{\unitlength}{0.8cm}

\subfigure{
\begin{fmffile}{triangle2i2hhh}
    \begin{fmfgraph*}(7,4)
   \fmfstraight
    \fmfleft{i1,i2}
    \fmfright{o1,m,o2,o3}
    \fmf{gluon,tension=2}{i1,t1}
    \fmf{gluon,tension=2}{t2,i2}
    \fmf{fermion,tension=1}{t1,t2,t3,t1}
    \fmf{phantom,tension=1}{t3,m}
    \fmffreeze
    \fmf{dashes,label=$h_3$,tension=2.5}{t3,h}
    \fmf{dashes,tension=1.2}{h,o1}
    \fmf{dashes,label=$h_2$,l.s=left,tension=3}{h,v1}
    \fmf{dashes,tension=2.2}{v1,o2}
    \fmf{dashes,tension=1.0}{v1,o3}
    \fmfv{decor.shape=circle,decor.filled=full,decor.size=4,f=(.0,,.13,,.98),
          l=$\color{blue}\lambda_{112}$,l.d=6,l.a=-45}{v1}
    \fmfv{decor.shape=circle,decor.filled=full,decor.size=4,f=(.0,,.13,,.98),
          l=$\color{blue}\lambda_{123}$,l.d=8,l.l=2,l.a=-90}{h}
    \fmfv{decor.shape=circle,decor.filled=full,decor.size=4,f=(.0,,.13,,.98),
          l=$\color{blue}\kappa_{3}$,l.d=8,l.a=-90}{t3}
    \fmfv{l.a=-20,l=$h_1$}{o1}
    \fmfv{l.a=-20,l=$h_1$}{o2}
    \fmfv{l.a=-20,l=$h_1$}{o3}
    \fmfv{l.a=160,l=$g$}{i1}
    \fmfv{l.a=160,l=$g$}{i2}
    \end{fmfgraph*}
\end{fmffile}
}
\caption{Double-resonant triple SM-like Higgs boson ($h_1$) production in a model with two heavy scalars $h_3$ and $h_2$, with $m_3 > m_2 + m_1$ and $m_2 > 2 m_1$.}
    \label{fig:hhhdoubleres}
\end{figure}

At LHC energies, the situation for triple Higgs boson production is much more dire, owing to the tiny cross section, about $\sigma^{hhh}_{\rm SM}\approx 0.1~\rm{fb}$ in the SM~\cite{Maltoni:2014eza,deFlorian:2019app}. Substantial enhancement of the triple-Higgs-boson production rate at the LHC could arise due to resonant contributions from new particles in specific new physics models~\cite{Robens:2019kga,Papaefstathiou:2020lyp,Biermann:2024oyy,Abouabid:2024gms,Aaboud:2017gsl}, or due to large anomalous contributions in the context of effective field theories, including modifications of the trilinear and quartic Higgs self-couplings~\cite{Stylianou:2023xit,Papaefstathiou:2023uum}, as well as of the Higgs-heavy-quark or Higgs-gluon interactions~\cite{Papaefstathiou:2023uum}.
 More specifically, in the context of explicit models of new physics, it has been shown, within a two-real singlet field extension of the SM, that triple Higgs boson production may be enhanced to a level sufficient for observation, and exploration, at the LHC~\cite{Robens:2019kga,Papaefstathiou:2020lyp,Karkout:2024ojx}. This is possible through the opening of the double-resonant process $gg \rightarrow h_3 \rightarrow (h_2 \rightarrow h_1 h_1) h_1$ when $m_3 > m_2 + m_1$ and $m_2 > 2 m_1$. In this case, $h_1$ is the SM-like Higgs boson, and $h_2$ and $h_3$ are two new heavy scalar particles (see Fig.~\ref{fig:hhhdoubleres}). Examination of this channel was first proposed in~\cite{Robens:2019kga} and investigated further in~\cite{Papaefstathiou:2020lyp,Karkout:2024ojx}. Scalar-to-scalar cascade topologies have also been discussed in~\cite{Low:2020iua,Chen:2022vac}, while alignment-driven multi-Higgs scenarios, in which higher-multiplicity Higgs boson final states can become leading LHC search channels, were recently studied in~\cite{Roy:2026amhp}. This channel has also been considered in recent experimental studies, including the first-ever results on triple Higgs boson searches by the ATLAS and CMS collaborations~\cite{ATLAS:2024xcs,CMS:2025gos,CMS:2025jkb} in the 6 $b$-jet and $4b+2\tau$ final states. Here, we focus on the 6 $b$-jet final state~\cite{Papaefstathiou:2019ofh,Fuks:2025gjv}, providing one of the largest branching ratios. We note that other processes might also yield important information, such as the $4b+2\gamma$~\cite{Papaefstathiou:2015paa,Chen:2015gva,Fuks:2015hna}, $4b+2\tau$~\cite{Fuks:2015hna,Fuks:2017zkg}, $2b+4\tau$~\cite{Dong:2025lkm} or final states with $b$-jets and light leptons~\cite{Abdughani:2020xfo}. These processes have the additional advantage of reduced combinatoric ambiguities that arise in the 6 $b$-jet final state.

\begin{figure}[htp]
\centering
\includegraphics[width=\columnwidth]{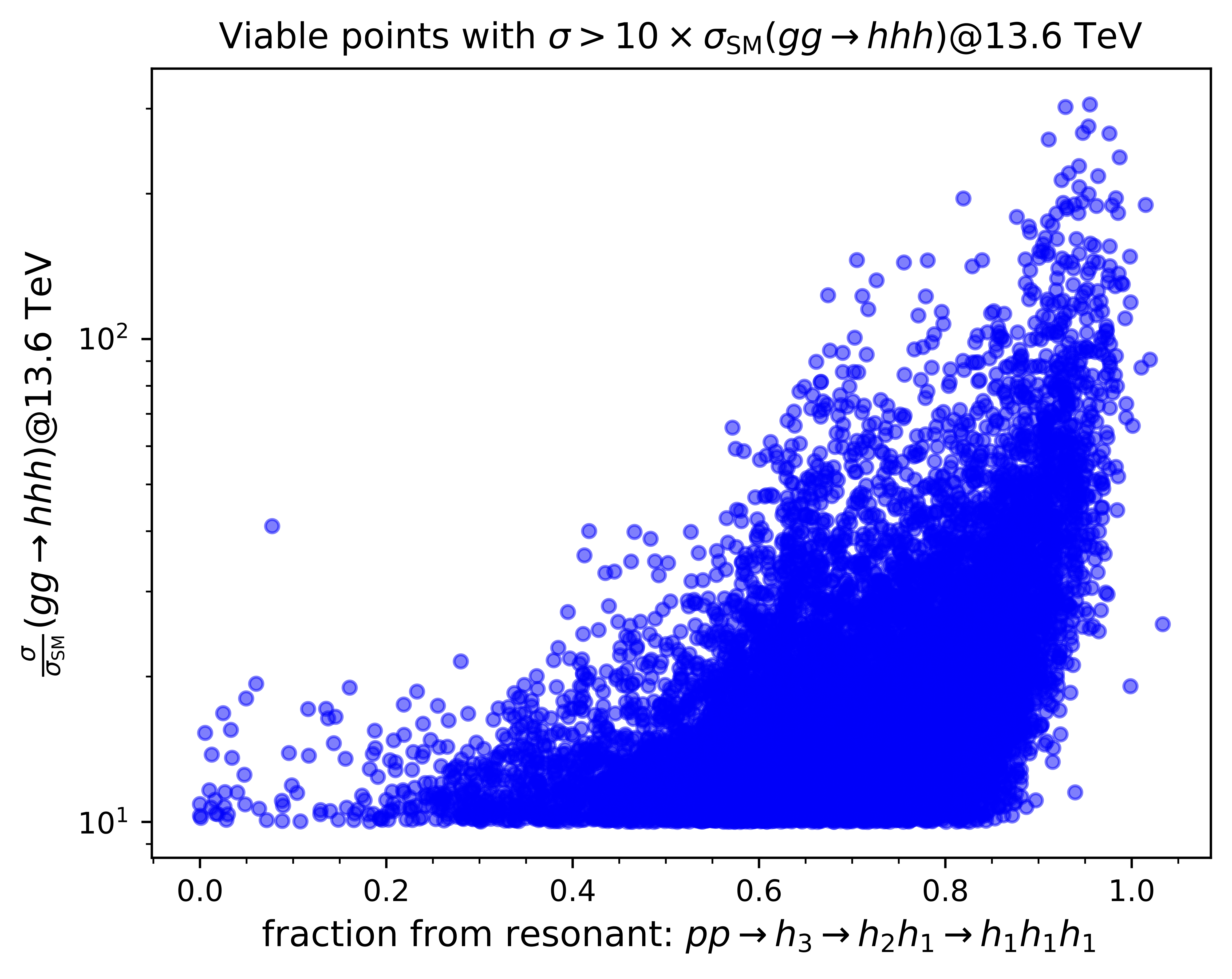}
\caption{Enhancement of the triple Higgs boson production cross section $\sigma(p p \rightarrow h_1 h_1 h_1)$ at 13.6~TeV, given in terms of multiples of the SM value, and the resonant fraction contribution from $ p p \rightarrow h_3 \rightarrow h_2 h_1 \rightarrow h_1 h_1 h_1$ within the TRSM (discussed in Section~\ref{sec:trsm}). The plot was adapted from Ref.~\cite{Karkout:2024ojx}. Only points with a factor 10 enhancement or greater over the SM triple Higgs boson cross section are shown. The distribution demonstrates a clear positive correlation between the enhancement of the cross section over the SM and the contribution of the double-resonant process. {The points correspond to benchmark scenarios from our parameter scan and are used only to illustrate the behaviour of the cross section; their density in the figure should not be interpreted as a probability distribution or prior over the BSM parameter space.}}
\label{fig:EnhancementvsresonantFraction}
\end{figure}

At the LHC, and within a model with two new scalar resonances, such as the TRSM discussed in the present article, it has also been shown in Ref.~\cite{Karkout:2024ojx} that there is a strong positive correlation between the enhancement of the total cross section and the contribution of the double-resonant process. Therefore, without the double-resonant enhancement, the cross section will not be sufficiently large for observation at the LHC, within models akin to the TRSM. This is exemplified by examining Fig.~\ref{fig:EnhancementvsresonantFraction}, taken from Ref.~\cite{Karkout:2024ojx}, which shows a scatter plot of the cross section enhancement over the SM value as a function of the (approximate) fraction of the double-resonant process, found during a parameter-space scan. It is clear that enhancements of $\mathcal{O}(>20\times)$ over the SM typically require a contribution of $\mathcal{O}(>50\%)$ from the double-resonant process. This motivates our present focus on the phenomenology of the double-resonant process.

{For the TRSM benchmark points considered in this work, after imposing the theoretical and experimental constraints used in the scan, the widths of the new scalars are small compared with their masses, corresponding to narrow resonances. This motivates the use of the narrow-width approximation for these points. We stress, however, that this is not a model-independent statement: in other extended scalar sectors, or in different regions of parameter space, the condition $\Gamma_i/m_i \ll 1$ should be checked explicitly, or imposed as an additional requirement.} Therefore, the double-resonant process can be factorized at each step to a good approximation, leading to substantial simplifications to its phenomenological treatment. This factorization, made possible due to the narrow-width approximation, is what we call the ``simplified approach''. In this paper, we demonstrate these simplifications, providing a simple phenomenological analysis, and deriving approximate constraints that can be reinterpreted in models realizing the same narrow double-resonant topology, subject to the validity of the NWA and to possible finite-width, off-shell, interference, and acceptance effects. Within this framework we examine the prospect for excluding example benchmark parameter-space points within the TRSM, generated following the treatment of Ref.~\cite{Karkout:2024ojx}. With the developed analysis, we examine, and contrast, the effect of the non-resonant part of the process in explicit case studies within the TRSM.  We do not attempt a parameter extraction in this work; our goal is instead to provide a fast, reusable framework to translate experimental sensitivity into constraints.

The paper is organized as follows: in Section~\ref{sec:trsm} we discuss extended scalar sectors, outlining some of the properties of the case study model, the TRSM. In Section~\ref{sec:simplified} we discuss the simplified approach to the phenomenology of this process, and in Section~\ref{sec:pheno} we perform an analysis for the LHC over the allowed range of masses for the processes, as well as a case study examining the effects of the non-resonant parts. We present our conclusions in Section~\ref{sec:conclusions}. \ref{app:ME_NWA_gg_h3_h2_3h1} demonstrates the factorization that we employ at the differential level, \ref{app:Broad_Peak_Limits} derives the edges of the di-Higgs boson invariant mass distribution discussed in the main text, and \ref{app:benchmarks} presents a set of high cross-section benchmark points.

\section{Extending the Standard Model by Real Singlet Scalar Fields}\label{sec:trsm}

The scalar potential of the SM can be extended by an additional sector of
scalar fields. These can transform as singlets, doublets, triplets, and so on. For the sake of simplicity, we focus on the case that the new fields $\phi_i$ transform  as singlets under the SM gauge group, leading to an extension of the SM scalar potential, $\mathcal{V}_\mathrm{SM}$, by $\mathcal{V}_{\rm singlets}(\Phi, \phi_i)$ as follows:
\begin{eqnarray}
\mathcal{V}(\Phi, \phi_i)&=& \mathcal{V}_{\rm SM}(\Phi) +\mathcal{V}_{\rm singlets}(\Phi, \phi_i)\;,
\end{eqnarray}
where $\Phi$ is the Higgs complex doublet field, with the most general renormalizable expression for $\mathcal{V}_{\rm singlets}$ given by
\begin{eqnarray}
\mathcal{V}_{\rm singlets}(\Phi, \phi_i) &=& a_i \phi_i + m_{ij}\phi_i \phi_j \nonumber \\
&&+ \tilde{\lambda}_{ijk}\phi_i\phi_j\phi_k + \tilde{\lambda}_{ijkl}\phi_i\phi_j\phi_k\phi_l  \nonumber\\
&&+ \tilde{\kappa}_{i H H}\phi_i(\Phi^{\dagger}\Phi)
+ \tilde{\kappa}_{ij H H}\phi_i \phi_j(\Phi^{\dagger}\Phi)\;,
\end{eqnarray}
where $a_i$, $m_{ij}$, $\tilde{\lambda}_{ijk}$, $\tilde{\lambda}_{ijkl}$, $\tilde{\kappa}_{iHH}$ and $\tilde{\kappa}_{ijHH}$ are new couplings.

The goal of the present paper is to study enhanced triple Higgs boson production at the LHC in models that contain at least two additional physical scalar particles, beyond the SM-like Higgs boson.  As a particular example of such a model, we will use the TRSM~\cite{Robens:2019kga,Papaefstathiou:2020lyp,Karkout:2024ojx}, where two extra real scalar singlet fields $S$ and $X$ are introduced. To reduce the number of free parameters, the following discrete $\mathbb{Z}_2$ symmetries are imposed:\footnote{As was discussed in~\cite{Lopez-Val:2014jva} within the context of a singlet extension of the SM, the breaking of such discrete symmetries during the electroweak phase transition in the early universe may in principle lead to problematic weak-scale cosmic domain walls, see, e.g.~\cite{Kobzarev:1974cp,Kibble:1976sj,Kibble:1980mv}. Nevertheless, analyses of the stability, and evolution, of such topological defects in multi-scalar extensions of the SM, e.g.\ discussed in Refs.~\cite{Preskill:1991kd,Abel:1995wk,Panagiotakopoulos:1998yw}, identify mechanisms that may overcome these issues, see also~\cite{Barger:2008jx}.  Here, we follow~\cite{Lopez-Val:2014jva}, in interpreting the TRSM as a low-energy effective scalar sector of a more fundamental UV-completion, whose specific details are not relevant for the purposes of our study.}
\begin{eqnarray}\label{eq:symm}
\mathbb{Z}^S_2 &:& S\rightarrow -S\;,\quad X \rightarrow X\;, \nonumber\\
\mathbb{Z}^X_2 &:& X\rightarrow -X\;,\quad S \rightarrow S\;,
\end{eqnarray}
with all SM particles transforming evenly under both symmetries. The methods developed herein could also be applied to non-singlet scalar extensions of the SM scalar sector.
The application of the discrete symmetries of Eq.~\eqref{eq:symm} reduces the scalar potential for two real singlet fields to:
\begin{eqnarray}\label{eq:reduced}
\mathcal{V}(\Phi, X, S) &=&\mu^2_{\Phi}\Phi^{\dagger}\Phi +\lambda_{\Phi}\Bigl(\Phi^{\dagger}\Phi\Bigl)^2 \nonumber \\
&& +\mu^2_S S^2 +\lambda_S S^4 + \mu^2_X X^2  + \lambda_X X^4  \nonumber\\
&&  +\lambda_{\Phi S}\Phi^{\dagger}\Phi S^2 +
\lambda_{\Phi X} \Phi^{\dagger}\Phi X^2 + \lambda_{S X} S^2 X^2\;,
\end{eqnarray}
which is characterized by nine real free couplings: $\mu_{\Phi}$, $\lambda_{\Phi}$ representing the couplings already present in the SM scalar potential,
$\mu_S$, $\lambda_S$, $\mu_X$, $\lambda_X$, representing equivalent couplings for the $S$ and $X$ singlet fields, respectively, and $\lambda_{\Phi S}$,  $\lambda_{\Phi X}$,  $\lambda_{S X}$ representing portal couplings, i.e. renormalizable quartic interactions that link the SM Higgs bilinear $\Phi^\dagger\Phi$ to singlet bilinears ($S^2$, $X^2$) and thereby provide the (Higgs) ``portal'' through which otherwise SM-neutral scalars communicate with the SM.\footnote{The ``Higgs portal'' terminology is widely used; an early discussion emphasizing the Higgs sector as a portal to singlet hidden sectors is~\cite{Patt:2006fw}. Early singlet-scalar models featuring the renormalizable operator $(\Phi^\dagger\Phi)S^2$ include~\cite{Silveira:1985rk}.} All fields are assumed to acquire vacuum expectation values (vevs), represented by $v$ for the Higgs field (in the appropriate direction) and $v_{S,X}$ for the singlet fields $S$, $X$, respectively. The physical gauge-eigenstates $\phi_{h,S,X}$ then follow by expanding around these vevs according to:
\begin{equation}
    \Phi = \begin{pmatrix} 0\\\frac{\phi_h + v}{\sqrt{2}}\end{pmatrix}, \quad
    S = \frac{\phi_S + v_S}{\sqrt{2}}, \quad
    X = \frac{\phi_X + v_X}{\sqrt{2}}\;.
    \label{eq:fields}
\end{equation}
We will work in the broken phase in which, generically, $v_S, v_X\neq 0$
and $v = v_{\rm SM} \simeq 246$~GeV.
Then, the discrete symmetries $\mathbb{Z}^S_2$ and $\mathbb{Z}^X_2$ are spontaneously
broken, and the scalars $\phi_h$, $\phi_S$, $\phi_X$ mix into the physical
states $h_1$, $h_2$ and $h_3$ according to the rotation:
\begin{equation}
    \begin{pmatrix}
        h_1 \\h_2\\h_3
    \end{pmatrix} = R \begin{pmatrix}
        \phi_h \\\phi_S\\\phi_X
    \end{pmatrix}\;,
\end{equation}

with the rotation matrix $R$ given by

\begin{equation}\label{eq:Rmat}
    R = \begin{pmatrix}
        c_1 c_2             & -s_1 c_2             & -s_2     \\
        s_1 c_3-c_1 s_2 s_3 & c_1 c_3+ s_1 s_2 s_3 & -c_2 s_3 \\
        c_1 s_2 c_3+s_1 s_3 & c_1 s_3-s_1 s_2 c_3  & c_2 c_3
    \end{pmatrix}\;,
\end{equation}
where we have used as shorthand:
\begin{eqnarray}
s_1\equiv\sin\theta_{12}\;,&\quad& s_2\equiv\sin\theta_{13}\;,\quad s_3\equiv\sin\theta_{23}\;,
\nonumber\\
c_1\equiv\cos\theta_{12}\;,&\quad& c_2\equiv\cos\theta_{13}\;,\quad c_3\equiv\cos\theta_{23}\;,
\end{eqnarray}
with the mixing angles $\theta_{12}$, $\theta_{13}$ and $\theta_{23}$, within the range:
\begin{eqnarray}
-\frac{\pi}{2} < \theta_{12}\;, \theta_{13}\;, \theta_{23} < \frac{\pi}{2}\;.
\end{eqnarray}
Using the same notation as in~\cite{Robens:2019kga,Papaefstathiou:2020lyp}, the entries of the first row in
the matrix $R$ are denoted as $\kappa_i\equiv R_{i 1}$ for $i=1,2,3$. The $h_1$ state can be identified with the SM-like Higgs boson, and $h_2$ and  $h_3$ are two new physical \textit{heavier} scalars obeying the mass hierarchy
\begin{eqnarray}
m_1 \leq m_2\leq m_3\;.
\end{eqnarray}
There are \textit{nine} real parameters that characterize the TRSM. However, the identification of $h_1$ as the SM Higgs boson fixes
\begin{eqnarray}
m_1&\backsimeq& 125\,\rm{GeV},\nonumber\\
v&\backsimeq&246\,\rm{GeV}.
\end{eqnarray}
This leaves us with seven independent parameters, which can be chosen to be:
\begin{eqnarray}
m_2\;, m_3\;, \theta_{12}\;, \theta_{13}\;, \theta_{23}\;, v_S, v_X\;,
\label{eq:freeparam}
\end{eqnarray}
where, respectively, these correspond to: the masses of the physical scalars $h_2$ and $h_3$, the three mixing angles, and the two new vevs for the singlet fields $S$ and $X$.
In this model all couplings for the mass eigenstates $h_i$ to SM particles are inherited from the SM-like Higgs doublet through the rotation from the gauge to the mass eigenstates, such that their couplings are rescaled with respect to the SM Higgs boson couplings by a mixing factor $\kappa_i$:
\begin{equation}\label{eq:mixfac}
    g_i\,\equiv\,\kappa_i\,g^\text{SM}\;.
\end{equation}
For example, in a factorized approach, this leads to predictions for production cross sections of the form
\begin{equation}
\sigma\lb p p \rightarrow h_i \rb\,=\,\kappa_i^2\,\sigma^\text{SM}\lb p p \rightarrow h_\text{SM} \rb\,\lb m_i\rb,
\end{equation}
 where $\sigma^\text{SM}\lb m_i \rb$ denotes the production cross section of an SM-like Higgs boson of mass $m_i$ for $i=1,2,3$. Furthermore, the total width of the $h_i$ scalars ($i=1,2,3$) is given by:
\begin{equation}\label{eq:totwidth}
\Gamma_{h_i} = \kappa_{i}^2 ~ \Gamma^{\mathrm{SM}} (m_i) + \sum_{j,k,(l) \neq i} \Gamma_{h_i \rightarrow h_j h_k (h_l)},
\end{equation}
where $\Gamma^{\mathrm{SM}} (m_i)$ corresponds to the width of a scalar boson of mass $m_i$ possessing the same decay modes as a SM Higgs boson of mass $m_i$. The branching ratios corresponding to $h_i \rightarrow xx$ (where $x$ excludes other scalars) are then given by:
\begin{equation}
\mathrm{BR}(h_i \rightarrow x x) = \kappa_{i}^2 \frac{\Gamma_{xx}^{\mathrm{SM}} (m_i)}{\Gamma_{h_i}} \;,
\end{equation}
where $\Gamma_{xx}^{\mathrm{SM}} (m_i)$ corresponds to the SM-like partial decay width of a scalar boson of mass $m_i$ for the final state $xx$.
The scalar-to-scalar branching ratios are equivalently obtained via
\begin{equation}
\mathrm{BR}(h_i \rightarrow h_j h_k (h_l)) =  \frac{\Gamma_{h_i\,\rightarrow\,h_j\,h_k (h_l)}}{\Gamma_{h_i}}\;,
\end{equation}
where $h_i\,\rightarrow\,h_j\,h_k (h_l)$ indicates the decay of $h_i$ into $h_j h_k$ (or  $h_i$ into $h_j h_k h_l$) for $i,j,k (l) = 1,2,3$. For the rest of the article, we will use the shorthand notation $\Gamma_{h_i} \equiv \Gamma_i$ for simplicity.

The triple couplings between scalars $ijk$ have been derived in~\cite{Robens:2019kga}, and the quartic couplings between scalars $ijkl$ have been derived in~\cite{Papaefstathiou:2020lyp}, both in terms of the parameters of Eq.~\eqref{eq:freeparam}.

\section{A Simplified Approach to Double-Resonant Triple Higgs Boson Production}\label{sec:simplified}

In the present study, we focus on the largest enhancement in triple Higgs boson production via gluon fusion, i.e.\ $pp\rightarrow h_1 h_1 h_1$, coming through the double-resonant production $gg \rightarrow h_3 \rightarrow (h_2 \rightarrow h_1 h_1) h_1$, in a model where the masses of the three scalars satisfy $m_3 > m_2 + m_1$ and $m_2 > 2 m_1$, such that all particles are produced on-shell, see Fig.~\ref{fig:hhhdoubleres}. In this case, the cross section corresponding to this process can be written as:
\begin{equation}
    \sigma(m_2, m_3) = \sigma_u(m_2,m_3) \times \kappa_3^2 \lambda_{123}^2 \lambda_{112}^2\;,
\end{equation}
where $\sigma_u(m_2,m_3)$ is the cross section for the process when the $\kappa_3$ parameter is set to unity, and the couplings $\lambda_{123}$ and $\lambda_{112}$ are set to 1~GeV. Furthermore, if we assume that $h_2$ and $h_3$ have narrow widths, such that $\Gamma_i \ll m_i$, then they are both produced  near on-shell, and we can replace the related Breit-Wigner factors in the cross section by $\delta$-functions via the standard narrow-width approximation (NWA) substitution:
\begin{equation}\label{eq:BWdelta}
\frac{ \mathrm{d} q_i^2 } { (q_i^2 - m_i^2)^2 + m_i^2 \Gamma_i^2} \;\xrightarrow{\rm NWA}\;\frac{ \pi }{m_i \Gamma_i} \delta(q_i^2 - m_i^2) \mathrm{d} q_i^2 \;.
\end{equation}
With the substitution of Eq.~\eqref{eq:BWdelta}, we can then write the cross section for the double-resonant production as follows:
\begin{eqnarray}\label{eq:rescaling}
    \sigma(m_2, m_3, \Gamma_2, \Gamma_3, \kappa_3, \lambda_{123}, \lambda_{112})
    &=& \hat{\sigma}_u(m_2,m_3) \times \rho^2\;,
\end{eqnarray}
where now $\hat{\sigma}_u(m_2,m_3)$ is the cross section for the process for $\kappa_3=1$, $\lambda_{123}=\lambda_{112}=1$~GeV, \textit{and} $\Gamma_2=\Gamma_3=1$~GeV, and the second line defines the rescaling factor
\begin{equation}\label{eq:rhosqdef}
\rho^2 \equiv \kappa_3^2 \lambda_{123}^2 \lambda_{112}^2 / (\Gamma_2 \Gamma_3)\;.
\end{equation}
We call $\hat{\sigma}_u(m_2,m_3)$ the ``unity'' cross section for brevity. We have opted for this factorization in order to absorb all mass dependence into $\hat{\sigma}_u(m_2,m_3)$, which can then be generated once-and-for-all within the NWA. Note that a sufficiently large cross section for a parameter-space point in any model realizing this topology requires both the unity cross section to be large, implying substantial double-resonant effects, as well as the scalar couplings. We would like to emphasize that the arguments that lead to the rescaling of the total cross section also work at the differential level, see~\ref{app:ME_NWA_gg_h3_h2_3h1} for a detailed derivation.
\begin{figure*}[htp]
\begin{center}
  \includegraphics[width=0.6\textwidth]{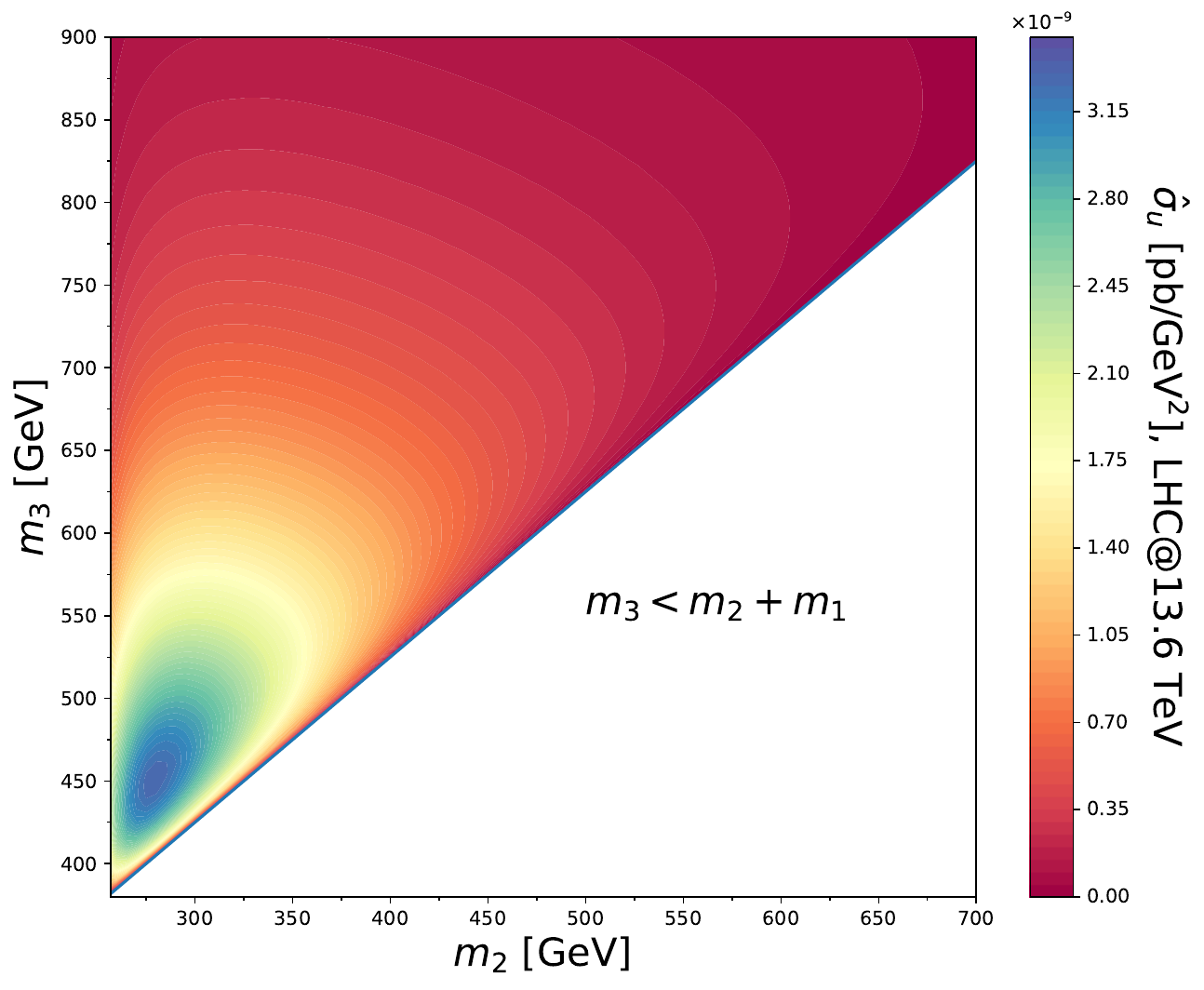}
\caption{\label{fig:sigmau} The unity cross section $\hat{\sigma}_u$, defined in Eq.~\eqref{eq:rescaling}, for the double-resonant triple Higgs boson production process, $gg \rightarrow h_3 \rightarrow h_2 h_1 \rightarrow h_1 h_1 h_1$ at a 13.6 TeV LHC, for $\kappa_3=1$, $\lambda_{123}=\lambda_{112}=1$~GeV, and $\Gamma_2=\Gamma_3=1$~GeV, interpolated over the $(m_2, m_3)$-plane. This quantity represents the factor of the cross section that depends explicitly on the masses of the scalars $h_{2,3}$, $m_{2,3}$, respectively. Points just above the resonant region $m_2 \gtrsim 2m_1$ and $m_3 \gtrsim 3m_1$ exhibit a larger unity cross section, as expected. To find the cross section for a specific parameter-space point, one has to multiply by the appropriate rescaling factor, $\rho^2$, defined by Eq.~\eqref{eq:rhosqdef}.}
\end{center}
\end{figure*}

For the selected subset of benchmark points derived in Ref.~\cite{Karkout:2024ojx}, we give the rescaling factor $\rho^2$ together with the defining parameters reproduced in Table~\ref{tab:benchmarks_model}, which appears in~\ref{app:benchmarks} for clarity of presentation: the masses $m_2$ and $m_3$, the vacuum expectation values $v_S$ and $v_X$ (in GeV), and the three mixing angles $\theta_{12}$, $\theta_{13}$ and $\theta_{23}$. The selected benchmark points satisfy state-of-the-art experimental and theoretical constraints, which we discuss briefly below in Section~\ref{sec:pheno}; see Ref.~\cite{Karkout:2024ojx} for further details.
We note that the given value of $\rho^2$ represents a leading-order calculation; higher-order corrections can be applied via a $k$-factor, by calculating them for the case of on-shell $h_3$ production, i.e.\ $(\rho^2)^\mathrm{HO} = k_\mathrm{fac} \times (\rho^2)^\mathrm{LO}$, where $(\rho^2)^\mathrm{HO}$ is the higher-order rescaling factor. In the present article, we opt for the simple choice of $k_\mathrm{fac}=2$ for all signal processes. The enhancement factors over the expected, non-resonant SM triple Higgs boson production at a 13.6~TeV LHC, and the resonant fraction (R.F.), corresponding to the approximate contribution of the double-resonant process to the total cross section are given in Table~\ref{tab:benchmarks_pheno}.\footnote{For orientation, among the 2012 scan points satisfying $m_3\ge m_2$, $\sigma/\sigma_{\rm SM}\ge 20$, and R.F.\ $\ge 0.2$, the median value of $\rho^2$ is $1.07$ in units of $10^6~\mathrm{GeV}^2$; the central 68\% and 90\% ranges are $0.642$--$2.05$ and $0.500$--$3.36$, respectively, and the full span is $0.268$--$9.06$.} One can observe the high degree of correlation between points with large cross section enhancements and the presence of the double-resonant process (i.e.\ large ``R.F.''), first mentioned in reference to Fig.~\ref{fig:EnhancementvsresonantFraction}. Table~\ref{tab:benchmarks_pheno} shows the corresponding parameters relevant to double-resonant triple Higgs boson production: $\Gamma_2$, $\Gamma_3$, $\kappa_3$, $\lambda_{123}$ and $\lambda_{112}$, as well as the rescaling factor, $\rho^2$, as defined by Eq.~\eqref{eq:rescaling}, and the unity cross section for the given combination of masses, $\hat{\sigma}_u(m_2,m_3)$. This table can be used, e.g., to obtain the cross section for a given parameter-space point, through the product of the unity cross section with the rescaling factor $\rho^2$. We show a contour plot of the unity cross section, $\hat{\sigma}_u$, in Fig.~\ref{fig:sigmau} for proton collisions at a 13.6~TeV LHC, representing the factor of the cross section that depends only on the masses of the scalars $h_{2,3}$, $m_{2,3}$, respectively. Points just above the resonant region $m_2 \gtrsim 2m_1$ and $m_3 \gtrsim 3m_1$ exhibit a larger unity cross section, as expected. The region $m_3 < m_2 + m_1$ is excluded since the double-resonant process is zero by definition there, as it is kinematically disallowed.

\begin{figure*}[htp]
\begin{center}
  \includegraphics[width=0.45\textwidth]{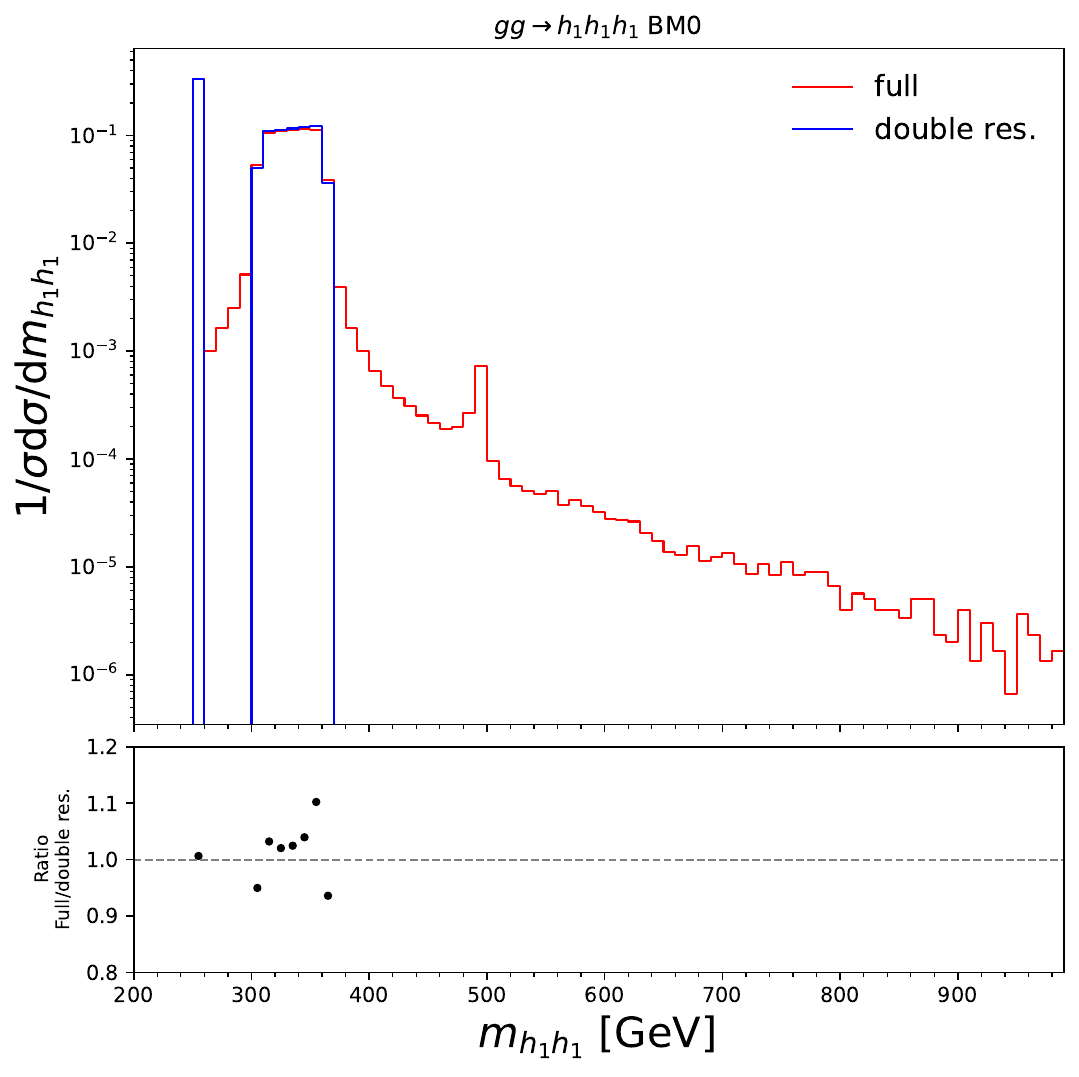}
  \includegraphics[width=0.45\textwidth]{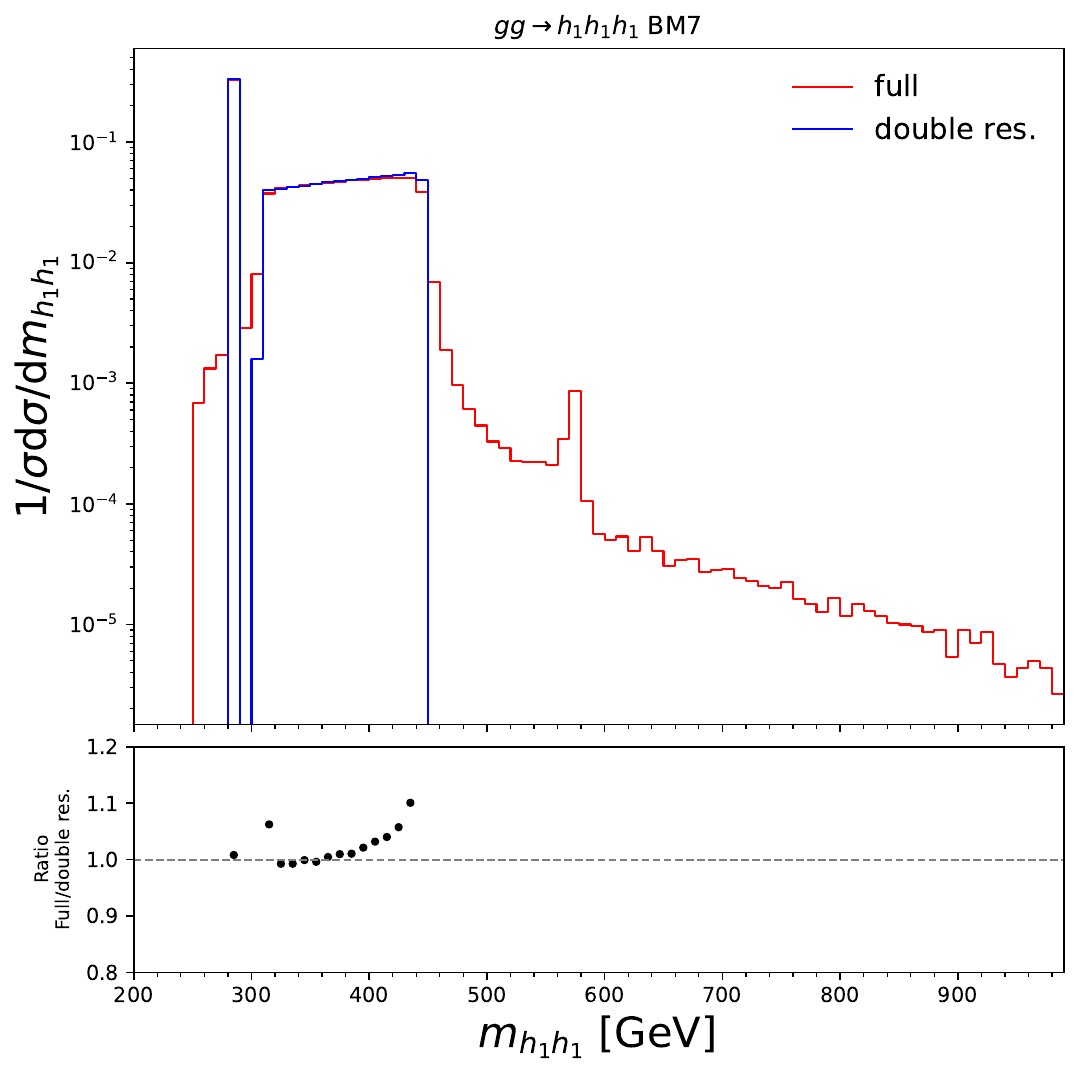}
\caption{\label{fig:restest} A comparison of the normalized $m_{h_1 h_1}$ distributions, between the full $gg \rightarrow h_1 h_1 h_1$ process and the double-resonant process $gg \rightarrow h_3 \rightarrow h_2 h_1 \rightarrow h_1 h_1 h_1$ for the benchmark points BM0 (left) and BM7 (right) given in Tables~\ref{tab:benchmarks_model} and~\ref{tab:benchmarks_pheno}. The lower part of the plot shows the ratio of the (normalized) full to double-resonant process, demonstrating only minor differences in shape. The sharp peak is due to the two Higgs bosons reconstructing the $h_2 \rightarrow h_1 h_1$ decay, whereas the broad blue structure in the double-resonant case comes from the two ``wrong'' Higgs bosons being combined; see~\ref{app:Broad_Peak_Limits} for a derivation of its endpoints. For the TRSM benchmark points shown, the double-resonant contribution represents the bulk of the cross section and gives similar normalized shapes for this stable-Higgs-level observable.}
\end{center}
\end{figure*}

\begin{figure*}[htp]
\begin{center}
  \includegraphics[width=0.45\textwidth]{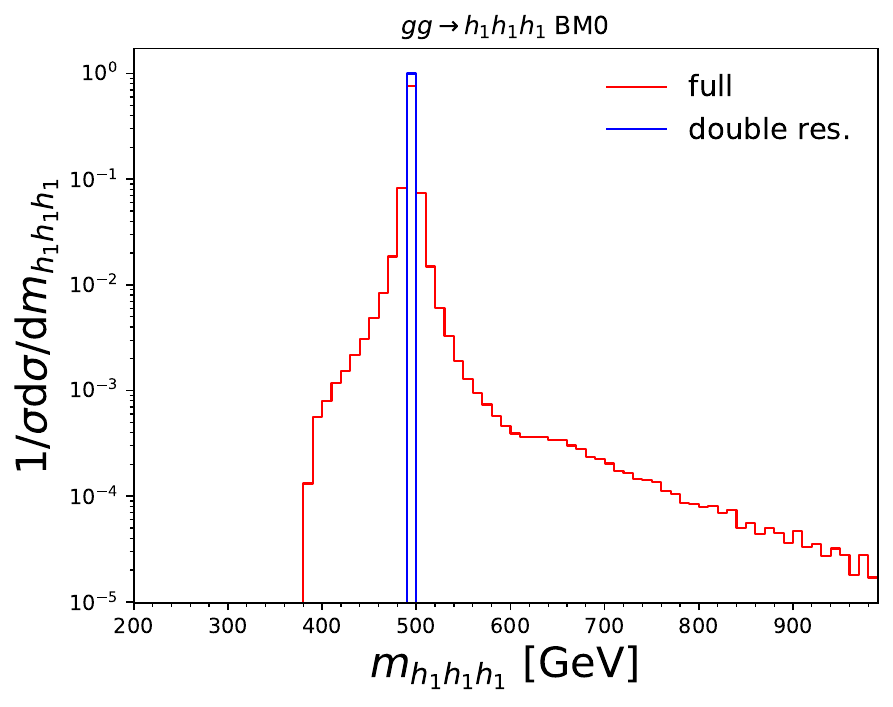}
  \includegraphics[width=0.45\textwidth]{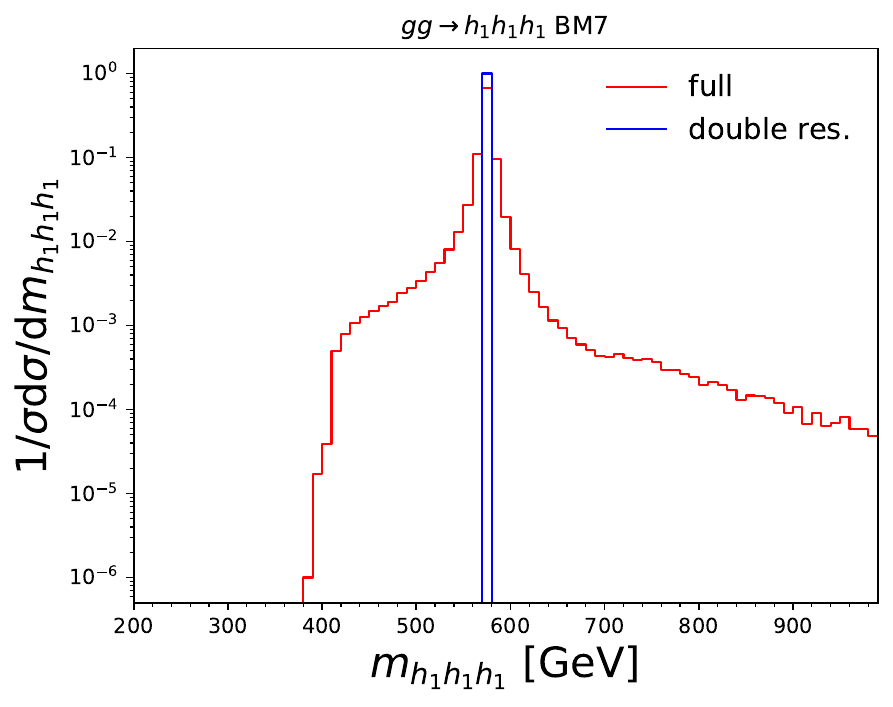}
\caption{\label{fig:restest_mhhh}{A comparison of the normalized triple-Higgs invariant-mass distributions, $m_{h_1h_1h_1}$, between the full $gg \rightarrow h_1 h_1 h_1$ process and the double-resonant process $gg \rightarrow h_3 \rightarrow h_2 h_1 \rightarrow h_1 h_1 h_1$ for the benchmark points BM0 (left) and BM7 (right). These stable-Higgs-boson-level comparisons provide an additional check of the extent to which the double-resonant contribution reproduces the kinematics of the full leading-order process before analysis cuts. Due to the $\delta$ function-like behaviour of the double-resonant process, we omit the ratio panel.}}
\end{center}
\end{figure*}

\begin{figure*}[htp]
\begin{center}
  \includegraphics[width=0.45\textwidth]{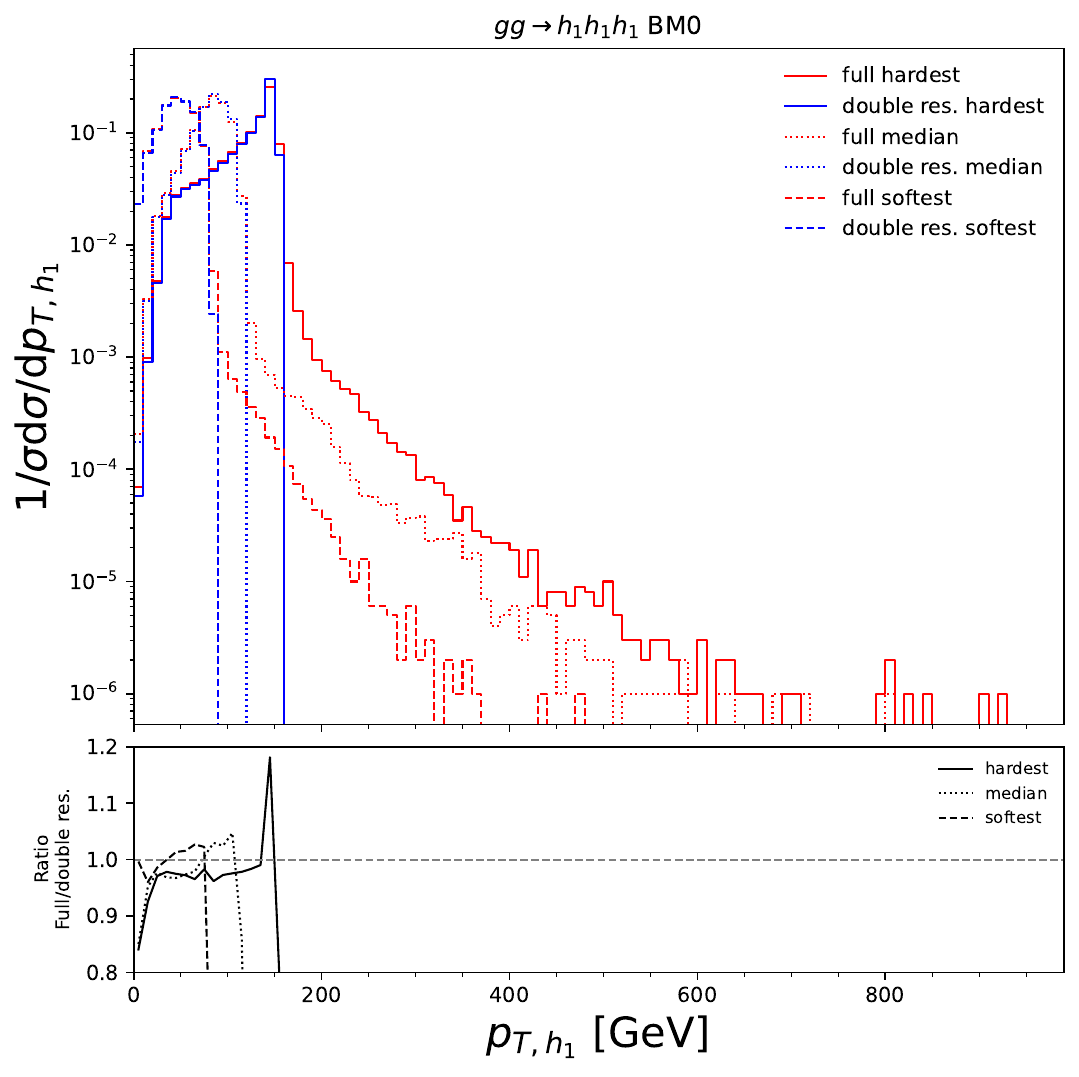}
  \includegraphics[width=0.45\textwidth]{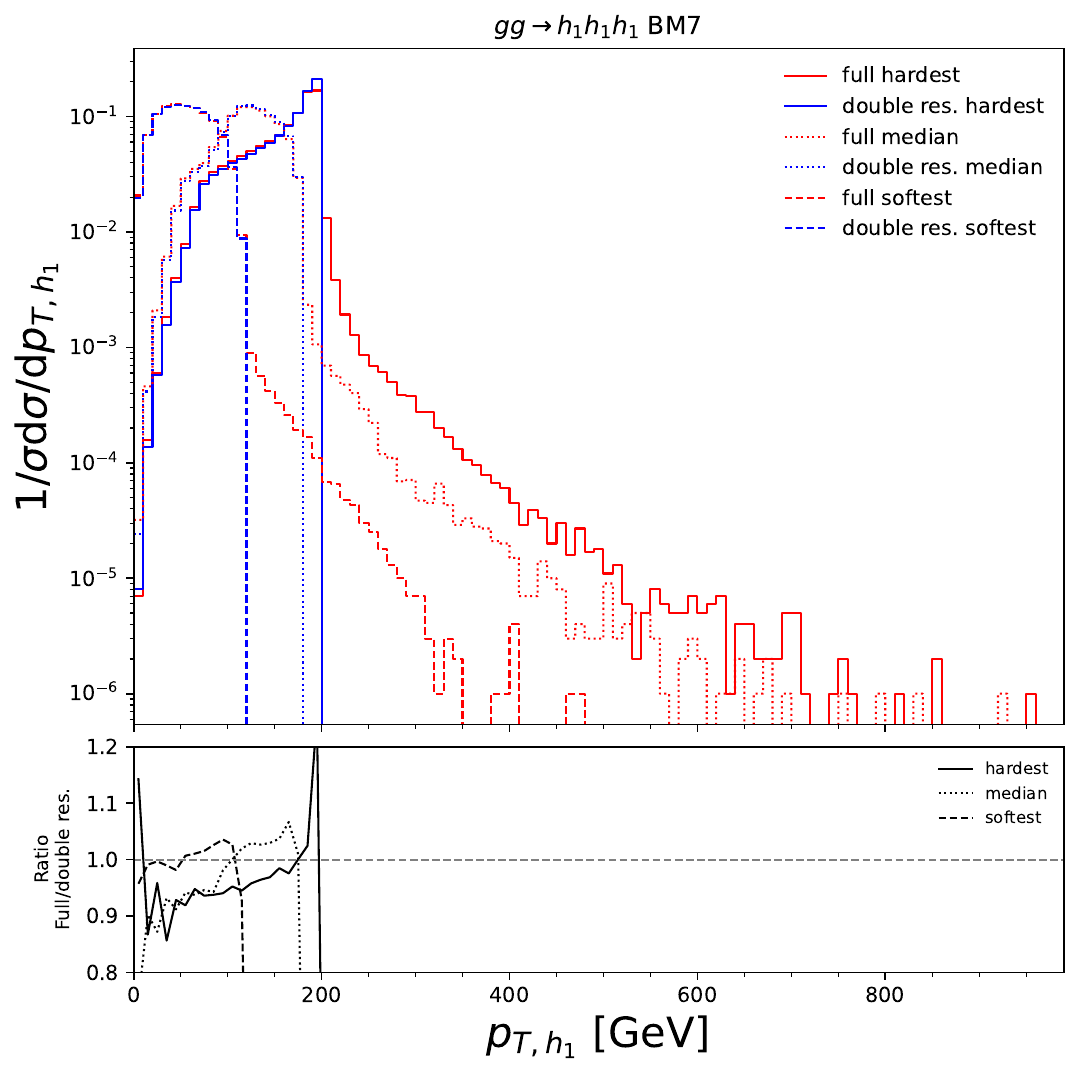}
  \includegraphics[width=0.45\textwidth]{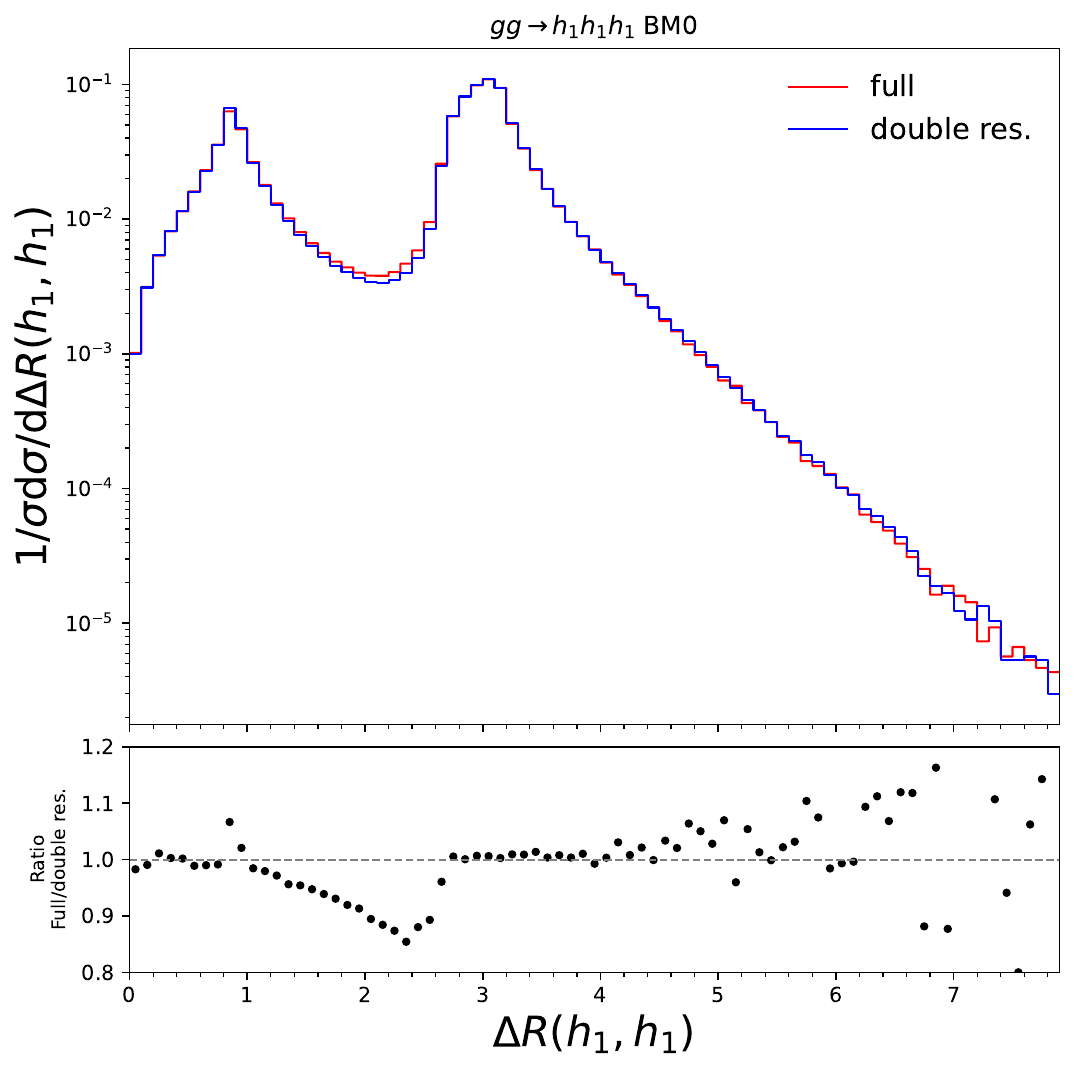}
  \includegraphics[width=0.45\textwidth]{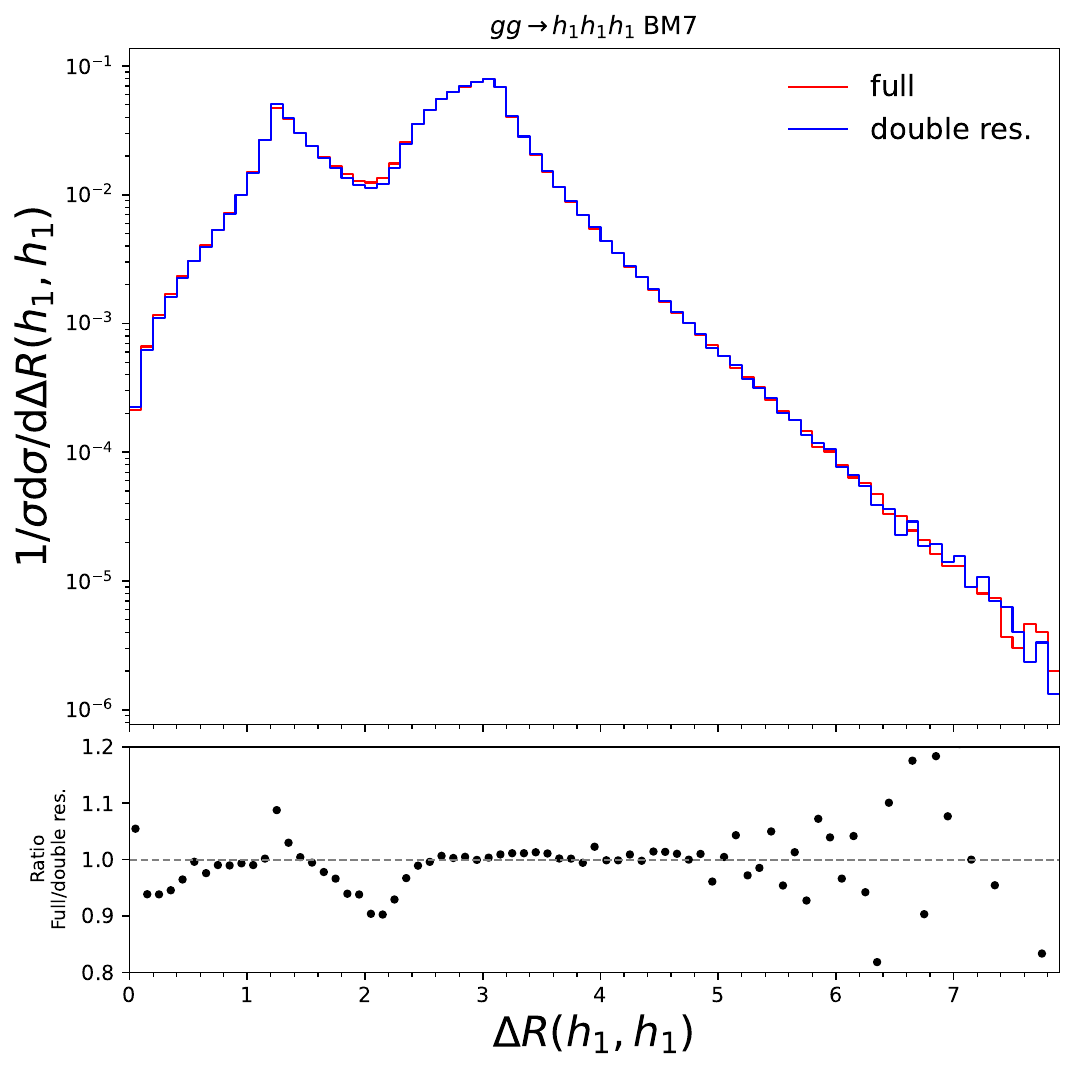}
\caption{\label{fig:restest_kinematics}{A comparison of additional normalized stable-Higgs boson-level kinematic distributions between the full $gg \rightarrow h_1 h_1 h_1$ process and the double-resonant process $gg \rightarrow h_3 \rightarrow h_2 h_1 \rightarrow h_1 h_1 h_1$ for the benchmark points BM0 (left column) and BM7 (right column). The upper row shows the ordered transverse momenta of the three Higgs bosons, while the lower row shows the angular separations $\Delta R(h_1,h_1)$ for all Higgs boson pairings. The lower panels show the ratio of the normalized full and double-resonant distributions.}}
\end{center}
\end{figure*}

Fig.~\ref{fig:restest} shows a normalized comparison of the $m_{h_1h_1}$ distributions, constructed for undecayed Higgs bosons (i.e.\ at parton level), for \textit{any} two Higgs bosons $h_1$ (i.e.\ there are three entries per event), for the benchmark points BM0 and BM7 of Tables~\ref{tab:benchmarks_model} and~\ref{tab:benchmarks_pheno}. One can observe that the ``full'' process distributions exhibit non-resonant components, represented by the small ``continuum'' part of the distributions. There also exist smaller contributions from $h_3^* \rightarrow h_3 h_1 \rightarrow h_1 h_1 h_1$, where the first $h_3$ is off shell, which appear as peaks on top of the continuum around $m_{h_1 h_1} \sim m_3 = 495/575$~GeV for BM0/BM7, respectively. The bulk of the cross section lies in the lower-$m_{h_1h_1}$ peak, that represents the combination of two Higgs bosons that reconstruct $h_2$ in the double-resonant process, and a broader ``peak'', where the two Higgs bosons do not reconstruct a resonance (i.e.\ one $h_1$ comes from the $h_2$ decay, and the other does not). The edges of this broader peak can be found by considering the limiting case where the $h_2 \rightarrow h_1 h_1$ decays in such a way that the $h_1$s are travelling collinearly to the directly-produced $h_1$, in the centre-of-mass frame of $h_3$. See~\ref{app:Broad_Peak_Limits} for a brief derivation of the edge values. {We show additional stable-Higgs boson-level comparisons in Figs.~\ref{fig:restest_mhhh} and~\ref{fig:restest_kinematics}, including the triple Higgs boson invariant mass, the ordered Higgs boson transverse momenta, and the angular separations between Higgs boson pairs, before analysis cuts.} These comparisons indicate that, for the TRSM benchmark points considered here, the double-resonant contribution captures the dominant stable-Higgs boson-level features of the full leading-order process. This motivates the use of double-resonant samples as the baseline for our phenomenological study. In particular, for each mass combination $(m_2,m_3)$ we generate a reference sample with $\kappa_3=1$, $\lambda_{123}=\lambda_{112}=1$~GeV, and $\Gamma_2=\Gamma_3 = 1$~GeV, and reinterpret it through the rescaling of Eqs.~\eqref{eq:rescaling} and~\eqref{eq:rhosqdef}. We stress, however, that the comparison with the full process is empirical for the benchmarks and observables considered, whereas the formal rescaling of the double-resonant contribution follows from the narrow-width approximation in Eq.~\eqref{eq:BWdelta} and from the corresponding differential factorization derived in~\ref{app:ME_NWA_gg_h3_h2_3h1}. In models or parameter regions where finite-width effects, off-shell contributions, interference, or non-resonant diagrams become larger, a dedicated validation with the full process would be required. This point is also illustrated by a recent N2HDM study, in which the kinematic structure and interference effects in resonant triple Higgs boson production were investigated in detail~\cite{Naskar:2026n2hdm}. We investigate the impact of the non-resonant pieces on our phenomenological analysis in Subsection~\ref{sec:nonres}.

In practice, one should check the narrow-width requirement on $h_3$ and $h_2$ in any explicit model within which the double-resonant process can be generated. Since the scalars are experimentally constrained to possess small mixing parameters with the SM-like Higgs boson, one way larger widths can occur, while remaining consistent with experimental bounds, is through scalar-to-scalar decays via scalar triple couplings, i.e.\ through $h_i \rightarrow h_j h_k$ for $i,j,k=1,2,3$. When the decay chain $h_3\rightarrow h_2 h_1 \rightarrow h_1 h_1 h_1$ becomes dominant in triple Higgs boson production (i.e., much larger than the contributions from fermion or gauge boson decays: $h_3 \rightarrow f\bar{f}$ or $h_3 \rightarrow ZZ/WW$), $\rho^2$ tends to a constant value for a given mass combination $(m_2,m_3)$. {This can be seen by writing, in this limit, $\Gamma_2 \simeq C_2(m_2)\lambda_{112}^2$ and $\Gamma_3 \simeq C_3(m_2,m_3)\lambda_{123}^2$, where $C_2$ and $C_3$ contain the corresponding phase-space and mass factors. Eq.~\eqref{eq:rhosqdef} then gives $\rho^2\rightarrow \kappa_3^2/(C_2 C_3)\equiv\rho_S^2$. Thus, once the scalar decays dominate the total widths, increasing the trilinear couplings further increases the widths rather than $\rho^2$.}
We will refer to this limit as the ``scalar-dominated region''. {For $\kappa_3=1$ and the mass range considered here, the limiting value is typically $\rho_S^2=\mathcal{O}(10^8-10^9)~\mathrm{GeV}^2$; for other values of $\kappa_3$ it scales as $\kappa_3^2$. Therefore, if the experimental limits imposed on $\rho^2$ are sufficiently below the corresponding $\rho_S^2$ for the masses and mixing under consideration, the limit is reached before entering the scalar-dominated large-width region.} This is the case for the expected limits that we derive here through our phenomenological analysis. Nevertheless, we emphasize that this statement is somewhat model-dependent and therefore should be investigated in models beyond the one under consideration in the present article.\footnote{For further discussion on cases where the narrow-width approximation breaks down, see Refs.~\cite{Berdine:2007uv,Fuchs:2014ola}.}
\begin{figure}[htp]
\begin{center}
  \includegraphics[width=0.95\columnwidth]{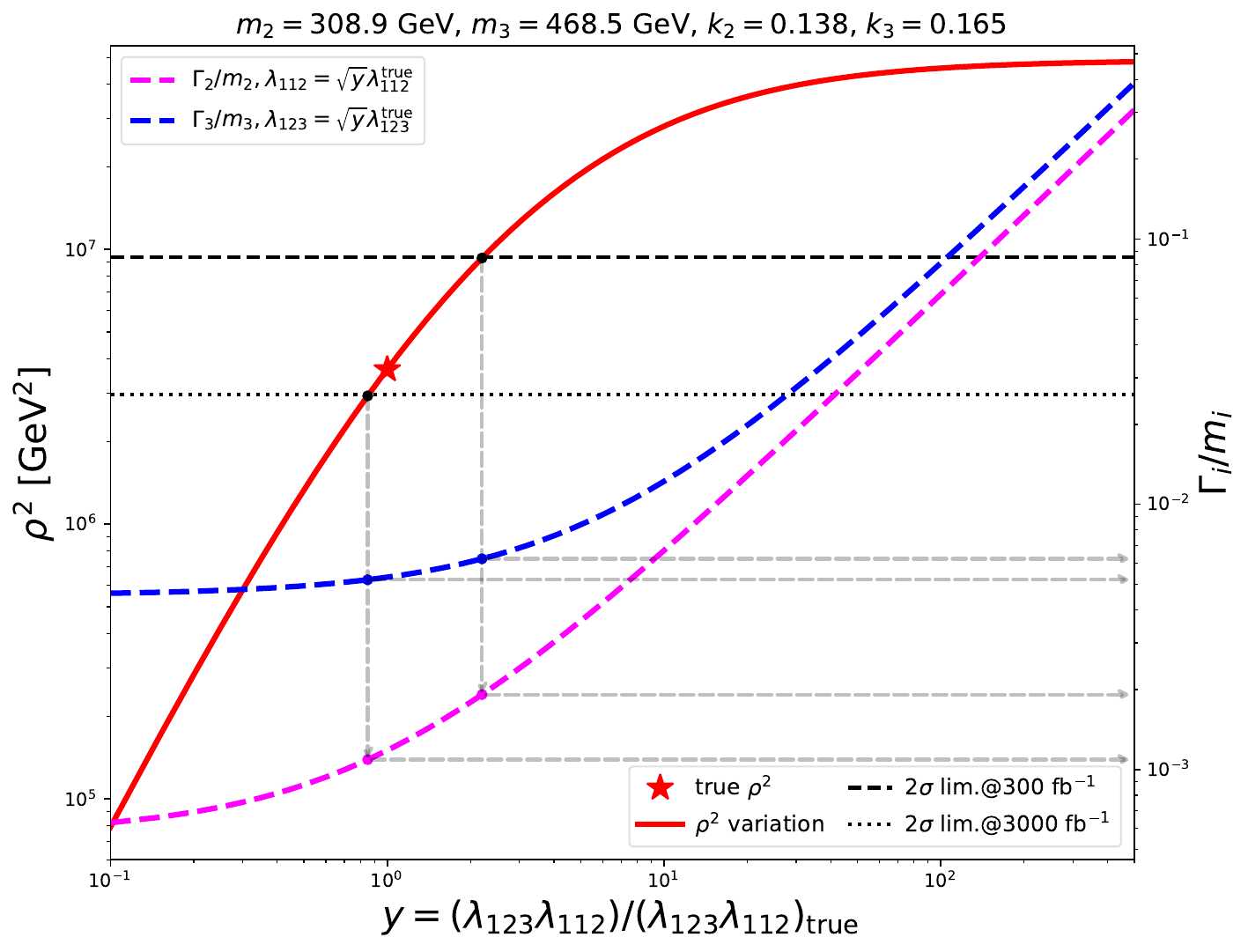}
\caption{\label{fig:rhosqvsy}
Case study of the coupling deformation $\lambda_{xyz}=\sqrt{y}\,\lambda^{\mathrm{true}}_{xyz}$ around a TRSM benchmark point. The red curve shows $\rho^2$ as a function of $y=(\lambda_{123}\lambda_{112})/(\lambda_{123}\lambda_{112})_{\mathrm{true}}$, while the red star marks the benchmark value. The black horizontal lines show the $2\sigma$ sensitivity thresholds at $300~\mathrm{fb}^{-1}$ and $3000~\mathrm{fb}^{-1}$. The magenta- and blue-dashed curves give $\Gamma_2/m_2$ and $\Gamma_3/m_3$ on the right axis; the light-grey guides indicate the inferred fractional widths at the threshold crossings.}
\end{center}
\end{figure}

As an explicit case study of the above discussion, we pick a viable parameter-space point with $m_2 = 308.9$~GeV, $m_3=468.5$~GeV, $\kappa_2=0.138$ and $\kappa_3=0.165$, which through our phenomenological analysis (described in the next section), is expected to be excluded at the high-luminosity run of the LHC. This parameter-space point holds a true value of the rescaling parameter $\rho^2 \simeq 3.66\times 10^6$~GeV$^{2}$ within the TRSM. Fixing the values of the mixing factors $\kappa_{2,3}$, defined in Eq.~\eqref{eq:mixfac}, we then consider the consequences of ``deforming'' the scalar couplings $\lambda_{123}$ and $\lambda_{112}$ \textit{away} from their predicted values in the TRSM, via $\lambda_{xyz} = \sqrt{y} \lambda_{xyz}^\mathrm{true}$, where $y$ is a deformation parameter that we vary. Note that this is a particular choice of a deformation that we have made as part of this illustrative case study.\footnote{Note that we do not interpret this curve as a scan over perturbative TRSM parameter points: the original point at $y=1$ satisfies the constraints used in our scan, while the perturbativity of other values of $y$ would have to be checked in a complete scalar potential through the corresponding dimensionless couplings or scattering eigenvalues. As a rough orientation only, at $y=1$, $|\lambda_{112}|/v\simeq 50.0/246\simeq 0.20$ and $|\lambda_{123}|/v\simeq 156.1/246\simeq 0.63$. Along the illustrative deformation these ratios scale as $\sqrt{y}$, so the larger ratio becomes $\mathcal{O}(1)$ for $y\simeq 2.5$, while a loose $4\pi$-type estimate would correspond to $y\simeq 390$. This estimate is normalization-dependent and is not a substitute for the perturbativity test of the full scalar potential.} This allows us to control the widths of the $h_2$ and $h_3$, and calculate the corresponding $\rho^2$ for each $y$. A departure from the TRSM predictions could be achieved by, e.g., removing the imposed  $\mathbb{Z}^S_2$ and $\mathbb{Z}^X_2$ discrete symmetries. We study the variation of $\rho^2$ with $y$ (red curve in Fig.~\ref{fig:rhosqvsy}, corresponding to the left vertical axis), as well as with respect to the widths of $h_2$ and $h_3$, as fractions of the respective masses of the particles (pink-dashed and blue-dashed curves in Fig.~\ref{fig:rhosqvsy}, corresponding to the right vertical axis). It can be seen that the imposed limits at this parameter space point, for the selected luminosities of 300~fb$^{-1}$ and 3000~fb$^{-1}$ (horizontal black-dashed and black-dotted lines respectively), restrict the ``deformed'' TRSM point to lie within the narrow-width regime, where both $h_2$ and $h_3$ would have $\Gamma_{2,3}/m_{2,3} < 10^{-2}$. To find the widths to which the limits correspond, we trace vertically downwards from the intersection of the $\rho^2$ curve and the horizontal limit lines, then find the points of intersection with the fractional width lines $\Gamma_{2,3}/m_{2,3}$. This is illustrated by the light-grey guide lines in Fig.~\ref{fig:rhosqvsy}. The explicit corresponding values on the right axis are, for $300~\mathrm{fb}^{-1}$ and $3000~\mathrm{fb}^{-1}$ respectively, $\Gamma_2/m_2 \simeq  1.9\times 10^{-3}, ~1.1\times 10^{-3}$ and $\Gamma_3/m_3 \simeq 6.2\times 10^{-3},~ 5.2\times 10^{-3}$.

\section{Phenomenological Analysis at the LHC} \label{sec:pheno}

\subsection{Limits on the $(m_2,m_3)$-plane through the 6 $b$-jet final state}
To extract limits at the LHC by employing the simplified model defined in Section~\ref{sec:simplified}, we perform a phenomenological analysis of the $pp \rightarrow 6$ $b$-jet final state, over the $(m_2, m_3)$-plane, using Monte Carlo signal event samples generated with $\kappa_3=1$, $\lambda_{123}=\lambda_{112}=1$~GeV, and $\Gamma_2=\Gamma_3=1$~GeV. The parton-level samples were generated using \texttt{MadGraph5\_aMC@NLO} (version 2.9.22)~\cite{Alwall:2011uj}, and the parton shower and non-perturbative effects were simulated through the \texttt{HERWIG 7} event generator (version 7.3.0)~\cite{Bahr:2008pv,Bellm:2017bvx,Gieseke:2011na,Arnold:2012fq,Bellm:2013hwb,Bellm:2019zci,Bewick:2023tfi}. We consider the dominant QCD-initiated 6 $b$-jet background, $pp \rightarrow 6b$ as well as the sub-dominant $pp \rightarrow Zb\bar{b} b\bar{b}$, $pp\rightarrow ZZb\bar{b}$, $pp\rightarrow h_1Zb\bar{b}$, $pp \rightarrow h_1 h_1 b\bar{b}$, $pp \rightarrow h_1h_1 Z$ and $pp \rightarrow h_1 ZZ$, with all the bosons decaying into $b\bar{b}$. As with the case of the signal, we applied a flat $k$-factor of 2 on all backgrounds. We note here that we have assumed that all backgrounds are SM-like, i.e.\ we did not consider modifications on the backgrounds due to new physics effects or explicit new physics backgrounds, e.g.\ arising through ``single'' production $pp\rightarrow h_2 \rightarrow h_1 h_1$. Since the dominant background by far is the QCD-initiated 6 $b$-jet background, we do not expect this to impact our results. We also note that we did not apply any detector effects beyond the detector acceptance.

\begin{table}[t]
\small
\centering
\setlength{\tabcolsep}{3pt}
\renewcommand{\arraystretch}{1.05}
\begin{tabularx}{\columnwidth}{@{}Y r r@{}}
\toprule
\textbf{Process}, $pp\rightarrow$ & $\sigma_\mathrm{GEN}$ [pb] & $\sigma_\mathrm{GEN}\times \mathcal{P}(\mathrm{6b\text{-}tags})$ [pb] \\
\midrule
$ 6b\text{-jets}$ & 2.97 & 1.12 \\
$4\,b\text{-jets}+2\,\mathrm{light\ jets}$ & 876.8 & 0.0458 \\
$4\,b\text{-jets}+2\,c\text{-jets}$ & 4.39 & 0.0229 \\
$4\,b\text{-jets}+1\,c\text{-jet}+1\,\mathrm{light\ jet}$ & 13.3 & 0.0694 \\
$5\,b\text{-jets}+1\,c\text{-jet}$ (electroweak) & $\mathcal{O}(10^{-5})$ & $\mathcal{O}(10^{-6})$ \\
\bottomrule
\end{tabularx}
\caption{We show the \texttt{MadGraph5\_aMC@NLO} generator-level, leading-order cross section at the LHC at 13.6~TeV, $\sigma_\mathrm{GEN}$, for multi-jet backgrounds with at least four true $b$-jets, and the resulting generator-level cross section after the application of $b$-tagging efficiencies, taken to be flat in transverse momentum and pseudorapidity. For a single jet, these were taken to be: $\mathcal{P}(b\rightarrow b) = 0.85$ for true $b$-jet identification, $\mathcal{P}(c\rightarrow b) = 0.1$ for charm-to-$b$-jet mistagging, and $\mathcal{P}(j\rightarrow b) = 0.01$ for light-to-$b$-jet mistagging. It is clear that the 4 $b$-jet + two non-$b$-jet contribution to the total six $b$-jet rate will be $\mathcal{O}(10\%)$ in any phenomenological analysis, assuming similar kinematic behaviour of light and charm-jets as for $b$-jets. Note that the 5 $b$-jet process that includes a charm-jet is electroweak and therefore suppressed.}
\label{tb:6jxsecs}
\end{table}

All jets were reconstructed using the anti-$k_T$ algorithm, with parameter $R=0.4$ using the \texttt{FastJet} library~\cite{Cacciari:2008gp}. The $b$-tagging efficiency, $\mathcal{P}(b\rightarrow b)$, was taken to be 85\% for true $b$-jets.\footnote{See, e.g.\ Table 1 in Ref.~\cite{ATLAS:2025nyf}.} We apply this efficiency flat in transverse momentum and pseudorapidity, by considering $b$-jets from Monte Carlo truth. Assuming that the light-to-$b$-jet mis-identification efficiency is $\sim$1\% and for charm-to-$b$-jets $\sim 10\%$, we obtain the cross sections shown in Table~\ref{tb:6jxsecs}, starting from initial cross sections with identical generation-level cuts on the $b$, charm or light quarks: $p_T(q) > 20$~GeV, $|\eta(q)| < 5.0$, and the minimum distance between any two quarks $\Delta R(q,q) > 0.2$, where we collectively refer to quarks as $q$. Assuming that the kinematic behaviour of the light and charm-jets would be similar as that of the $b$-jets, similar efficiencies as for the $6$ b-jet sample would be obtained through any analysis. Note that this was indeed verified in the ATLAS data-driven approach of~\cite{ATLAS:2024xcs}. By examining the third column in Table~\ref{tb:6jxsecs}, we can deduce that the light and charm-jet samples will contribute $\mathcal{O}(10\%)$ to the total QCD 6 $b$-jet background. Therefore, we do not consider Monte Carlo samples with six jets that contain light or charm-jets in our analysis. We note that even for the efficiencies considered in the analysis of~\cite{ATLAS:2024xcs}, i.e.: $\mathcal{P}(b\rightarrow b)=0.77$, $\mathcal{P}(c\rightarrow b) = 0.155$, and $\mathcal{P}(j\rightarrow b) = 0.0038$, this conclusion would still be valid, with $\mathcal{O}(10\%)$ contributions from 4 $b$-jet+2 jet processes to the total 6 $b$-jet rate.

Given that, at leading order in the (double) narrow-width approximation, the fully differential rate for the double-resonant topology factorizes into a kinematic function times an overall coupling factor $\rho^2$, see~\ref{app:ME_NWA_gg_h3_h2_3h1}, one may construct the event selection and discriminating observables using Monte Carlo signal samples generated at a convenient reference point (e.g.\ $\kappa_3=1$, $\lambda_{123}=\lambda_{112}=1~\mathrm{GeV}$, $\Gamma_2=\Gamma_3=1~\mathrm{GeV}$). The resulting analysis can then be reinterpreted by a rescaling of the signal normalization for other parameter choices, provided the resonances are sufficiently narrow for the NWA to hold and the mass hierarchy satisfies $m_3>m_2+m_1$ and $m_2>2m_1$; residual deviations may arise from finite-width, off-shell, interference, and acceptance effects. We emphasize that our study is intended as a fast, reusable baseline for reinterpretation/scouting across models that realize this double-resonant topology; a dedicated experimental optimization (including detector effects and multivariate strategies) could further improve the attainable reach. The analysis can be performed point-by-point on the $(m_2,m_3)$-plane to obtain the optimal signal-versus-background discrimination. Instead, for the sake of clarity and simplicity of presentation, we pursue the definition of a \textit{single} set of cuts, apart from explicitly mass-dependent invariant-mass windows, that can be applied everywhere in the interesting region of the $(m_2,m_3)$-plane. This is in contrast to what was done in Ref.~\cite{Papaefstathiou:2020lyp}, where the analysis was optimized for points over the so-called ``BP3'' plane. {The product of significances introduced below is used only as an optimization criterion for choosing this representative set of cuts; it is not intended as a statistical combination of different mass hypotheses. Maximizing the product, or equivalently the sum of $\log\Sigma_i$, rewards selections that perform reasonably well over the full grid and penalizes choices that give a high significance at only a small number of mass points. Other global criteria are possible, and a dedicated experimental analysis could instead optimize point by point.} To define the quantity to be maximized over the $(m_2,m_3)$-plane, consider a pre-selected set of parameter-space points $i=1,...,N$ that span the $(m_2,m_3)$-plane, e.g.\ on a grid, while satisfying the necessary mass hierarchies. Then, we construct the \textit{product} of significances of each point, $\Sigma_\Pi$ as:
\begin{equation}
\Sigma_\Pi \equiv \prod_{i=1}^N \Sigma_i = \prod_{i=1}^N \frac{S_i}{\sqrt{B_i}}\;,
\end{equation}
where $S_i$ represent the expected number of signal events for the point $i$ on the $(m_2,m_3)$-plane, and $B_i$ the expected number of background events, at an integrated luminosity $\mathcal{L}$. If the analysis efficiency is $\varepsilon_i$ for signal point $i$, with cross section $\sigma_i$ before cuts, and $\varepsilon_{B,i}$ is the efficiency for the background with the same cuts, with cross section $\sigma_B$ before cuts, then we can rewrite $\Sigma_\Pi$ as:
\begin{equation}
\Sigma_\Pi =  \left(\prod_{j=1}^N \frac{\sqrt{\mathcal{L}} \sigma_j}{\sqrt{\sigma_B}}\right)\mathcal{E}_\Pi\;.
\end{equation}
where we have defined
\begin{equation}
\mathcal{E}_\Pi \equiv \prod_{i=1}^N \frac{\varepsilon_i}{{\sqrt{\varepsilon_{B,i}}}}\;,
\end{equation}
which we maximize to obtain the optimal set of cuts over the pre-defined set of points on the $(m_2,m_3)$-plane. For optimized values, we found that the condition $\frac{\varepsilon_i}{{\sqrt{\varepsilon_{B,i}}}} > 1$ can easily be satisfied. Therefore, to avoid large numbers in  $\mathcal{E}_\Pi$, we optimize $\log \mathcal{E}_\Pi$. We would like to reiterate at this point that this optimization will depend on the set of chosen combinations $(m_2,m_3)$ on the grid, and a full analysis should be performed over the whole plane defined by the masses $(m_2,m_3)$, ideally optimizing at each point.

The optimization performed in the present study is based on a subset of 280 combinations for $(m_2,m_3)$ on a regular grid, i.e.\ 280 signal samples are considered, within the mass ranges $260 \leq m_2 \leq 622.3$~GeV and $425 \leq m_3 \leq 1150$~GeV. The mass ranges are arbitrary, and chosen to span the plane in the region where triple Higgs boson production is of reasonable size within the TRSM. We select events based on the same procedure followed in Ref.~\cite{Papaefstathiou:2020lyp}, and for a detailed description of the observables see Section 4.3 therein. Here, we limit ourselves to summarizing the main features of the selection criteria that distinguish the present study from the analysis of Ref.~\cite{Papaefstathiou:2020lyp}. Just as in Ref.~\cite{Papaefstathiou:2020lyp}, we select events with at least six $b$-tagged jets and consider those with the highest transverse momentum as our six $b$-jet candidates. For the purposes of control and manipulation each $b$-tagged jet gets an index.
Our optimization imposes a lower cut $p_{T\mathrm{min},b}$ on the transverse momentum of a $b$-jet, a window on the invariant mass of the 6 $b$-jets, $m^{inv}_{6b}$, and upper limits on the observables $\chi^{2,(6)}$, $\chi^{2,(4)}$, defined by:
\begin{eqnarray}
\chi^{2, (6)}&=&\sum_{q r\in J}\Bigl(M_{q r} - m_1 \Bigl)^2\;,\label{eq:chi26}\\
\chi^{2, (4)}&=&\sum_{q r \in I}\Bigl(M_{q r} - m_1 \Bigl)^2\label{eq:chi24}\;.
\end{eqnarray}

Here $I=\{i_1 i_2, i_3 i_4\}$ and $J=\{j_1 j_2, j_3 j_4, j_5 j_6\}$ denote pairings of 4 and 6 $b$-jets respectively. Thus, $i_k, j_k$ refer to the indices assigned to the corresponding jets. Moreover, $M_{q r}$ is the invariant mass of the respective jet pairing, $q r$, and $m_1$ is the Higgs boson mass. It should be understood that the index corresponding to each $b$-jet can appear only in a single arrangement inside the pairings $I$ and $J$. The number of possible $n$ pairings given the 6 $b$-jets with the highest $p_T$ is given by $\frac{1}{n !}\,\binom{6}{2}\,\binom{4}{2}$. This translates into 45 different combinations for $I$ and 15 combinations for $J$, respectively.
We then select the combinations of $b$-tagged jets entering in $I$ and $J$ based on the minimisation of the sum
 \begin{eqnarray}
 \chi^{2, (6)} + \chi^{2, (4)}\;.
 \end{eqnarray}
Subsequently, we ``identify'' candidates for the scalars $h_2$ and $h_3$ with the pairing configurations $I_{\rm min}$ and $J_{\rm min}$, which minimize $\chi^{2,(4)}$ and $\chi^{2, (6)}$ respectively, as described above. Since each pairing inside $J_{\rm min}$ ``defines'' a Higgs boson candidate $h^{i}_1$, we determine the absolute differences between the invariant mass of each pairing and $m_1$, i.e.~the mass of the SM Higgs boson. Each one of these differences is sorted from minimum to maximum, $(\Delta m_{\rm min}, \Delta m_{\rm med}, \Delta m_{\rm max})$. These deviations allow us to prioritize the best-reconstructed individual SM Higgs bosons. We also obtain the transverse momentum $p_T(h^i_1)$ of the $h^{i}_1$ candidate, constructed from the pairings inside $J_{\rm min}$. These transverse momenta are then ordered from hardest to softest, and used as variables for signal and background discrimination. Notice that the sortings on $p_T$ and on $\Delta m$ are independent of each other. We expect that signal has a more hierarchical structure on these variables in comparison with background which is completely random. Similarly, we make use of the angular distance $\Delta R(h^i_1, h^j_1)$ between the $h_1$ candidates  $h^{i}_1$ and $h^j_1$, and additional angular cuts $\Delta R_{bb}(h^i_1) $ are enforced between the $b$-jet pairs that define each of the $h^{i}_1$ candidates.

One difference with the procedure followed in~\cite{Papaefstathiou:2020lyp} is that we explore the parameter space of cuts on the observables: the minimum transverse momentum of all $b$-jets, $p_{T\mathrm{min},b}$, and the maximum values of the observables $\chi^{2,(6)}$, $\chi^{2,(4)}$, by using a random sampling procedure, instead of selecting values sequentially on a fixed grid, which should result in a more efficient exploration of the cut space for the maximum significance. Thus, we consider random points drawn from the following intervals $p_{T\mathrm{min},b}\in [25, 40]~\rm{GeV}$, $\chi^{2,(6)}\in [5, 60]~\rm{GeV}^2$, $\chi^{2,(4)}\in [5, 40]~\rm{GeV}^2$. After this first stage, we improve our optimization by imposing a window on $m^{\mathrm{inv}}_{6b}$. We also tested the effect of imposing constraints on $m^{inv}_{4b}$, but concluded that, in our ``universal'' cuts approach, this observable does not improve the discrimination power. In particular, when considering the linear rule $m^{inv}_{4b}\leq m_2 + (m_3-m_2) \times a$, for $ 0\leq a\leq 1$, we found that $\log \mathcal{E}_\Pi$ is maximized when $a=1$, rendering a cut on the  $m^\mathrm{inv}_{4b}$ observable irrelevant. {This endpoint corresponds to the loosest allowed version of this requirement, so the global optimization does not favour an additional restrictive cut on $m^\mathrm{inv}_{4b}$ once the $m^\mathrm{inv}_{6b}$ window and the Higgs boson reconstruction variables are included.} As in Ref.~\cite{Papaefstathiou:2020lyp}, we consider additional observables, which, however, are not optimized. A summary of the values considered is presented in Table~\ref{tb:cuts}, along with a short description for convenience.

\begin{table}[t]
\small
\centering
\setlength{\tabcolsep}{3.5pt}
\renewcommand{\arraystretch}{1.08}
\begin{tabularx}{\columnwidth}{@{}l Y r@{}}
\toprule
\textbf{Quantity} & \textbf{Description} & \textbf{Constraint} \\
\midrule
$p_{T\mathrm{min},b}$ & $b$-jet transverse momentum minimum & $=37.0$ \\
$|\eta_{b,\mathrm{max}}|$ & $b$-jet absolute pseudorapidity maximum & $=2.5$ \\
$\chi^{2,(6)}$ & Defined in Eq.~\eqref{eq:chi26} & $<12.0$ \\
$\chi^{2,(4)}$ & Defined in Eq.~\eqref{eq:chi24} & $<34.0$ \\
$m_{6b}^\mathrm{inv}$ & Invariant mass of the 6 $b$-jet combination & $\in[m_3-50.0,\,m_3+38.0]$ \\
$p_T(h_1^i)$ & Transverse momentum of the $i$-th Higgs candidate (ordered max/med/min) & $\geq[50,\,50,\,0]$ \\
$\Delta m_{\rm min,med,max}$ & Ordered deviation from the Higgs mass & $\leq[14,\,15,\,20]$ \\
$\Delta R(h_1^i,h_1^j)$ & Distance between Higgs candidates $i$ and $j$ & $\leq 3.5$ \\
$\Delta R_{bb}(h_1^i)$ & Distance between the $b$-jets in the $i$-th Higgs candidate & $\leq 3.5$ \\
\bottomrule
\end{tabularx}
\caption{The optimized universal selection cuts used in our phenomenological analysis. As in Ref.~\cite{Papaefstathiou:2020lyp}, the indices $i,j$ run over the values $1,2,3$, and refer to the three reconstructed SM-like Higgs bosons, $h_1$. The quantities with units of energy are given in GeV. The observables $\chi^{2,(6)}$ and $\chi^{2,(4)}$ are presented in $\mathrm{GeV}^2$. The optimization procedure considered $p_{T\mathrm{min},b}$, $\chi^{2,(6)}$, $\chi^{2,(4)}$, and $m_{6b}^\mathrm{inv}$, as described in the main text.}
\label{tb:cuts}
\end{table}

\begin{figure*}[htp]
\begin{center}
  \includegraphics[width=0.45\textwidth]{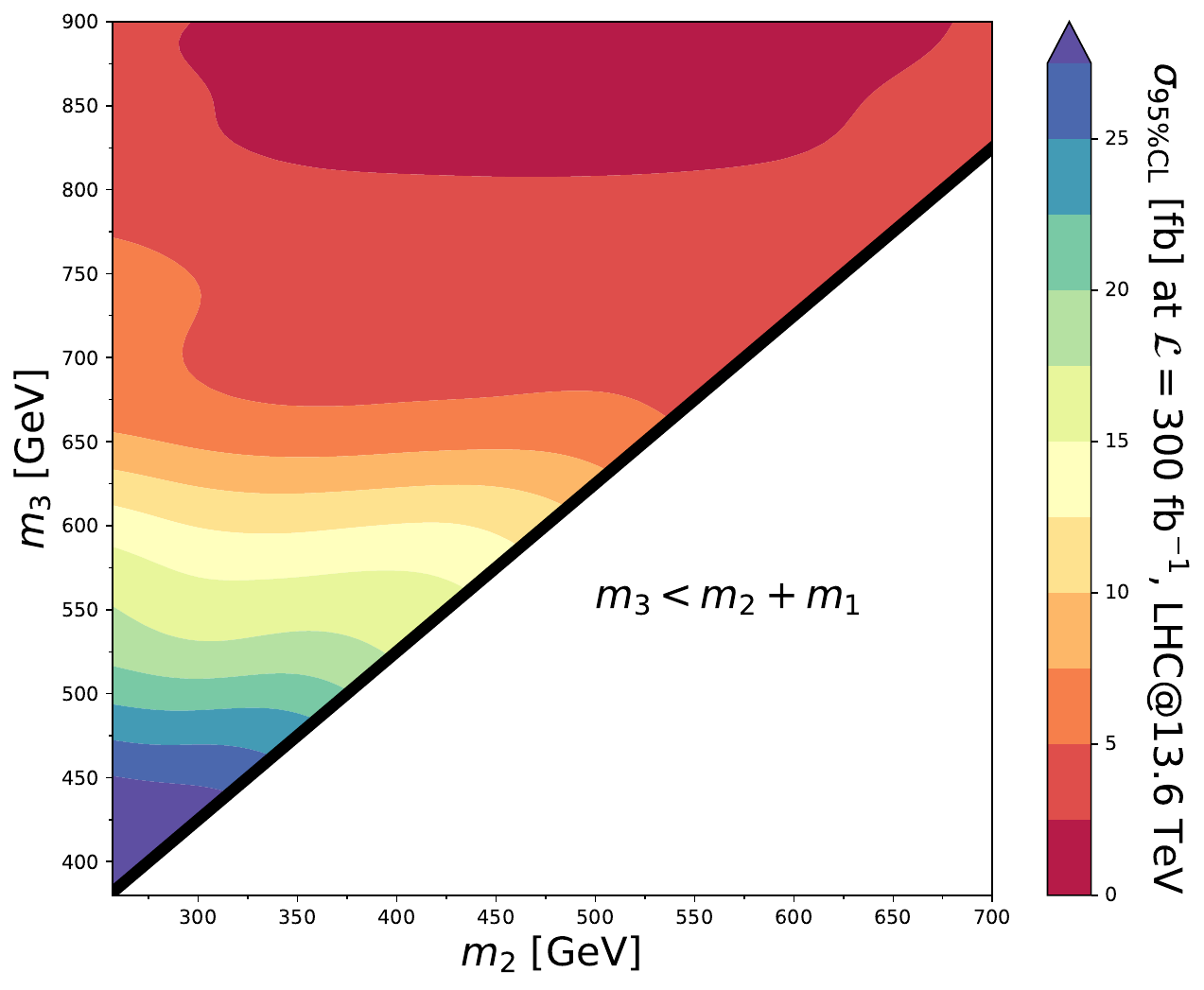}
  \includegraphics[width=0.45\textwidth]{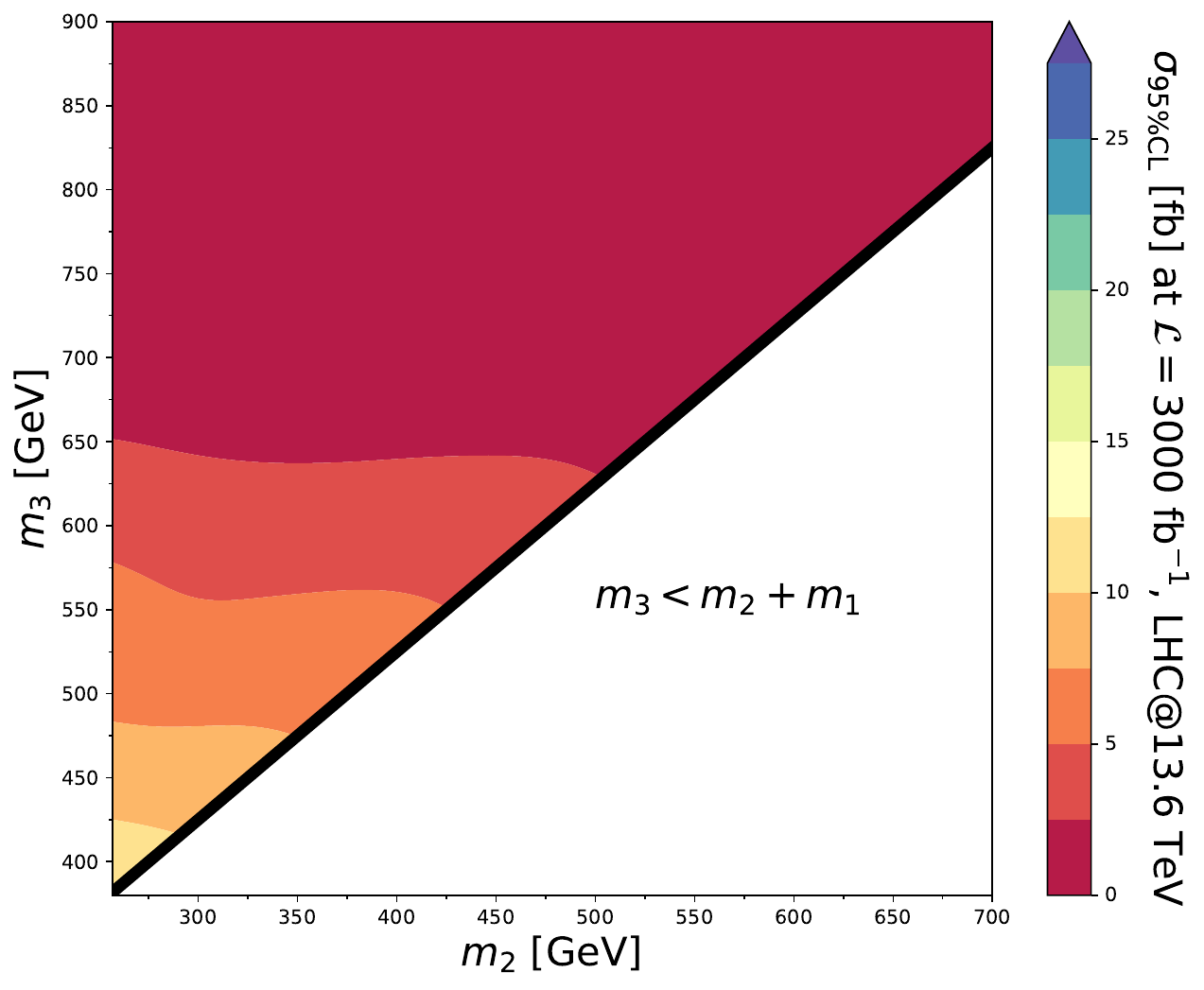}
\caption{\label{fig:sigmalim} The 95\% C.L. on the cross section for double-resonant triple Higgs boson production process, $gg \rightarrow h_3 \rightarrow h_2 h_1 \rightarrow h_1 h_1 h_1$ at a 13.6 TeV LHC, at integrated luminosities of $\mathcal{L}=300$~fb$^{-1}$ (left) and $3000$~fb$^{-1}$ (right). The scale of the left panel is broadly compatible with the luminosity-rescaled expected ATLAS direct $hhh\to 6b$ narrow-width heavy-resonance limits in the overlapping mass region~\cite{ATLAS:2024xcs}; see the text for details.}
\end{center}
\end{figure*}

 Given the described analysis and the optimized set of cuts, we can derive an expected limit on the $\rho^2$ parameter over the $(m_2,m_3)$-plane. To do so, we follow a simple approach here, first calculating the 95\% confidence limit (C.L.) on the cross section, $\sigma_{95\%\mathrm{CL}}$, by requiring that the significance,
 \begin{equation}
    \Sigma = \frac{\varepsilon_S  \sigma_S\mathcal{L}}{\sqrt{\varepsilon_B \sigma_B \mathcal{L}}}\;,
\end{equation}
 equals 2 everywhere on the $(m_2,m_3)$-plane. We remind the reader that the area under a Gaussian distribution covering two standard deviations away from its mean on either side, corresponds to approximately 95\% probability, see, e.g.~\cite{Lista:2016chp}. We then obtain $\sigma_S = \sigma_{95\%\mathrm{CL}}$ as follows:
\begin{equation}
    \sigma_{95\%\mathrm{CL}} = 2 \frac{ \sqrt{\varepsilon_B \sigma_B}}{\varepsilon_S\sqrt{\mathcal{L}}}\;,
\end{equation}
where $\sigma_B$ is the total expected background cross section. The limit on the signal cross section to the particular final state is easily translated into a limit on the total cross section for double-resonant triple Higgs boson production, by taking into account the branching ratios and the $b$-tagging efficiencies of the Higgs boson into the $b\bar{b}$ final state. All results presented here represent limits on the total cross section. This limit is shown in Fig.~\ref{fig:sigmalim}, for integrated luminosities of $\mathcal{L}=300$~fb$^{-1}$ and $3000$~fb$^{-1}$, following our analysis. To remove interpolation artefacts we apply a Gaussian smoothing filter to the results. The resulting dependence on the cross section limit on $m_2$ (horizontal direction) is mild compared to the dependence on $m_3$ (vertical direction). This is due to the fact that the initial production of $h_3$ is the determining factor for the kinematics of the process, e.g.\ how much energy is available to the subsequent decay products. As a rough validation of the scale of the projected limits, we have compared the left panel of Fig.~\ref{fig:sigmalim} with the expected 95\% C.L. limits for the narrow-width heavy-resonance interpretation shown in Fig.~8a of the ATLAS search in the 6 $b$-jet final state~\cite{ATLAS:2024xcs} at an integrated luminosity of $\mathcal{L}_1 = 126~\mathrm{fb}^{-1}$. In particular, rescaling the ATLAS results by $\sigma_{95}(\mathcal{L}_2)\simeq \sigma_{95}(\mathcal{L}_1)\sqrt{\mathcal{L}_1/\mathcal{L}_2}$, where $\mathcal{L}_2 = 300~\mathrm{fb}^{-1}$ gives a factor $\sqrt{126/300}\simeq0.65$. Accounting for the different centre-of-mass energy, and for the fact that our result is a phenomenological projection without a full detector-level treatment, the two estimates are broadly compatible in the overlapping mass region.

We can then proceed to extract the limit on the rescaling parameter, $\rho^2$, by dividing out the ``unity'' cross section (defined by Eq.~\eqref{eq:rescaling}):
\begin{equation}
    \rho^2_{95\%\mathrm{CL}} = \frac{\sigma_{95\%\mathrm{CL}}}{\hat{\sigma_u}}\;,
\end{equation}
where this operation is understood to be taken over the $(m_2,m_3)$-plane.
\begin{figure*}[htp]
\begin{center}
  \includegraphics[width=0.49\textwidth]{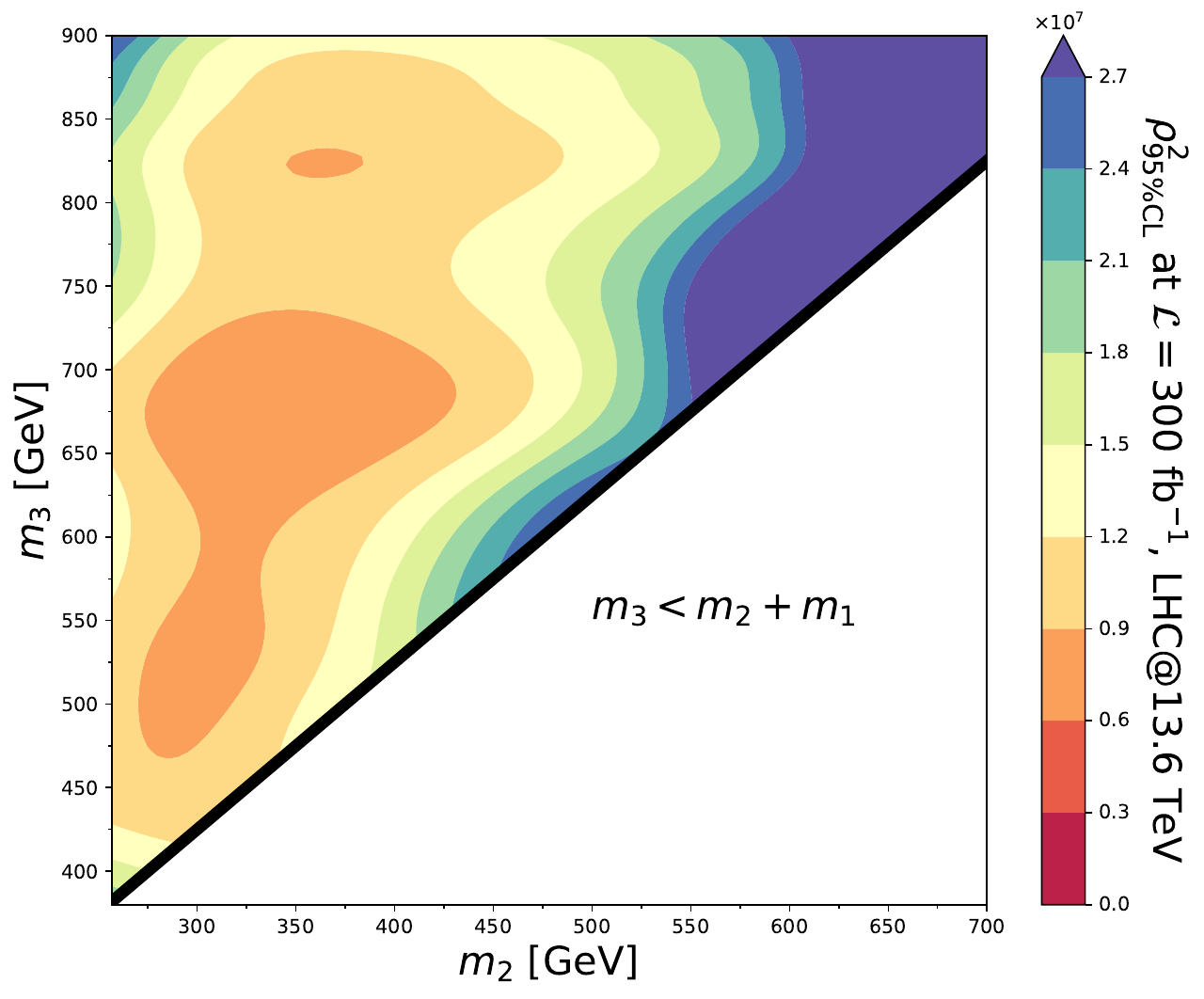}
  \includegraphics[width=0.49\textwidth]{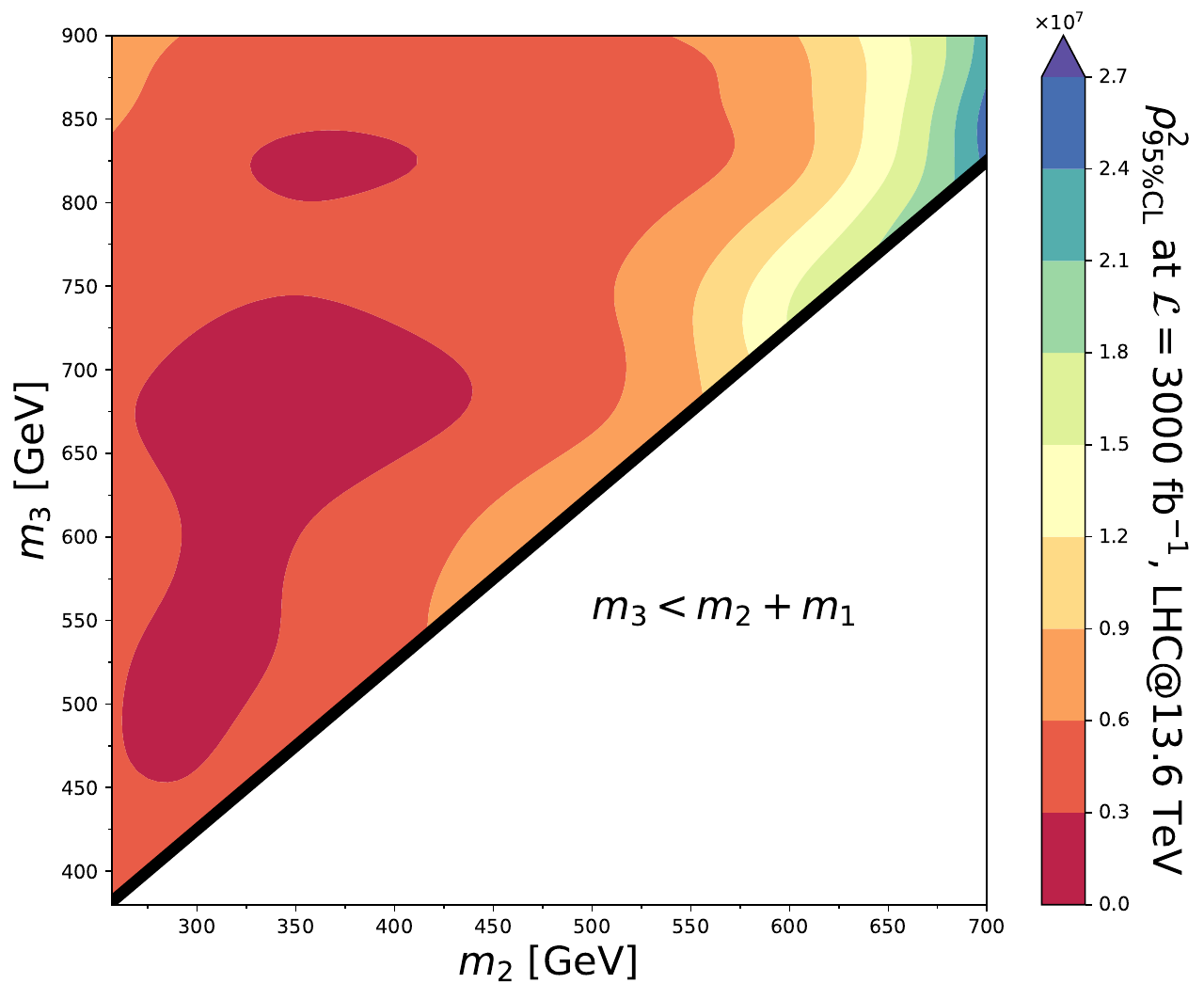}
\caption{\label{fig:rholim} The 95\% C.L. limit on the rescaling parameter, $\rho^2$ for double-resonant triple Higgs boson production process, $gg \rightarrow h_3 \rightarrow h_2 h_1 \rightarrow h_1 h_1 h_1$ at a 13.6 TeV LHC, at integrated luminosities of $\mathcal{L}=300$~fb$^{-1}$ (left) and $3000$~fb$^{-1}$ (right).}
\end{center}
\end{figure*}
The results are shown in Fig.~\ref{fig:rholim}, at a 13.6 TeV LHC, and for integrated luminosities of $\mathcal{L}=300$~fb$^{-1}$ and $3000$~fb$^{-1}$. {The limits in Figs.~\ref{fig:sigmalim} and~\ref{fig:rholim} should be interpreted as projected sensitivities in the simplified double-resonant setup, and are not by themselves filtered by direct-search constraints on a particular scalar model. When we apply these limits to the TRSM parameter scan below, the scan points are required to satisfy the existing direct scalar-resonance constraints summarized in Table~\ref{tab:expanalyses}; the resulting comparison is shown in Fig.~\ref{fig:resscan}.}
\begin{table}[t]
\small
\centering
\setlength{\tabcolsep}{5pt}
\renewcommand{\arraystretch}{1.05}
\begin{tabular}{@{}l r r@{}}
\toprule
\textbf{Parameter} & \textbf{Min} & \textbf{Max} \\
\midrule
$m_2$ & 255~GeV & 775~GeV \\
$m_3$ & 350~GeV & 900~GeV \\
$v_S$ & 0~GeV & 1000~GeV \\
$v_X$ & 0~GeV & 1000~GeV \\
$\kappa_1$ & 0.95 & 1.00 \\
$\kappa_2$ & 0.00 & 0.25 \\
$\kappa_3$ & 0.00 & 0.25 \\
\bottomrule
\end{tabular}
\caption{The range of parameters scanned over, following the theoretical and experimental constraints outlined in Ref.~\cite{Karkout:2024ojx}. The different parameters appearing in the table have been introduced in Eqs.~\eqref{eq:freeparam} and~\eqref{eq:mixfac}.}
\label{tb:scanparams}
\end{table}

To demonstrate the region of the $(m_2,m_3)$-plane where the double-resonant process can provide information on the TRSM, we have performed a new parameter-space scan that includes the theoretical and experimental constraints as they were imposed in Ref.~\cite{Karkout:2024ojx}, as follows:
\begin{table*}[t]
\centering
\setlength{\tabcolsep}{3.5pt}
\renewcommand{\arraystretch}{1.05}
\begin{tabularx}{0.6\textwidth}{@{}Y l l l@{}}
\toprule
\textbf{Process} & \textbf{Exp.} & $\boldsymbol{\mathcal{L}}$ & \textbf{Ref.} \\
\midrule
$gg\rightarrow S\rightarrow W^+W^-,\,ZZ$ & ATLAS & 139~fb$^{-1}$ & 2004.14636 \cite{ATLAS:2020fry} \\
$gg\rightarrow S\rightarrow ZZ$ & ATLAS & 139~fb$^{-1}$ & 2009.14791 \cite{ATLAS:2020tlo} \\
$gg\rightarrow S\rightarrow h_1 h_1 \rightarrow (b\bar b)(\tau^+\tau^-)$ & CMS & 137~fb$^{-1}$ & 2106.10361 \cite{CMS:2021yci} \\
$(b\bar b,\tau^+\tau^-,W^+W^-,ZZ,\gamma\gamma)(b\bar b)$ & CMS & 35.9~fb$^{-1}$ & 1811.09689 \cite{CMS:2018ipl} \\
$gg\rightarrow S\rightarrow h_1 h_1$ (combined channels) & ATLAS & 36.1~fb$^{-1}$ & 1906.02025 \cite{ATLAS:2019qdc} \\
$gg\rightarrow S\rightarrow h_1 h_1 \rightarrow (b\bar b)(\gamma\gamma)$ & ATLAS & 36.1~fb$^{-1}$ & 1807.04873 \cite{ATLAS:2018dpp} \\
$gg\rightarrow S\rightarrow W^+W^-,\,ZZ$ & ATLAS & 36.1~fb$^{-1}$ & 1808.02380 \cite{ATLAS:2018sbw} \\
$pp\rightarrow S\rightarrow ZZ$ (incl.\ VBF) & CMS & 35.9~fb$^{-1}$ & 1804.01939 \cite{CMS:2018amk} \\
$gg\rightarrow S\rightarrow h_1 h_1 \rightarrow (b\bar b)(b\bar b)$ & CMS & 35.9~fb$^{-1}$ & 1806.03548 \cite{CMS:2018qmt} \\
$gg\rightarrow S\rightarrow h_1 h_1 \rightarrow (b\bar b)(b\bar b)$ & ATLAS & 36.1~fb$^{-1}$ & 1804.06174 \cite{ATLAS:2018rnh} \\
\bottomrule
\end{tabularx}
\caption{The most constraining experimental analyses on new scalar particles in the \texttt{HiggsBounds} library, found during our scan for viable points over the parameter space of the TRSM. In the process description (first column), the particle $S$ denotes either of the $h_2$ or $h_3$ scalars, $S=\{h_2,h_3\}$.}
\label{tab:expanalyses}
\end{table*}
\begin{itemize}
\item We have used the Python-based scripts which can be found in Ref.~\cite{gitlabrepoTwoSingletScan}, allowing us to calculate all the relevant couplings, picking points randomly in the range of parameters outlined in Table~\ref{tb:scanparams}.
\item Using the scalar couplings we test the boundedness from below and the perturbativity of the theory for energy scales ranging from $\mu_0=M_Z\approx 91\,\rm{GeV}$ up to $\mu=900\,\rm{GeV}$. See Ref.~\cite{Karkout:2024ojx} for further details on how these were obtained.
\item Then, the experimental tests listed in Table~\ref{tab:expanalyses} are evaluated using \texttt{HiggsTools}~\cite{Bechtle:2008jh,Bechtle:2011sb,Bechtle:2012lvg,Bechtle:2013wla,Bechtle:2013xfa,Stal:2013hwa,Bechtle:2014ewa,Bechtle:2015pma,Bahl:2021yhk,Bechtle:2020pkv,Bechtle:2020uwn,Bahl:2022igd}.
\end{itemize}
The scan was performed ``flat'' within the regions of parameters defined by Table~\ref{tb:scanparams}. We note, however, that the density of parameter-space points found here is not meant to be representative of the true statistical distribution of the probability of a point occurring in the theory space of the TRSM, and should not be interpreted as such. Indeed, if different, but equivalent free parameters were scanned over, the density of parameter-space points would appear different.

\begin{figure*}[!t]
\begin{center}
  \includegraphics[width=0.45\textwidth]{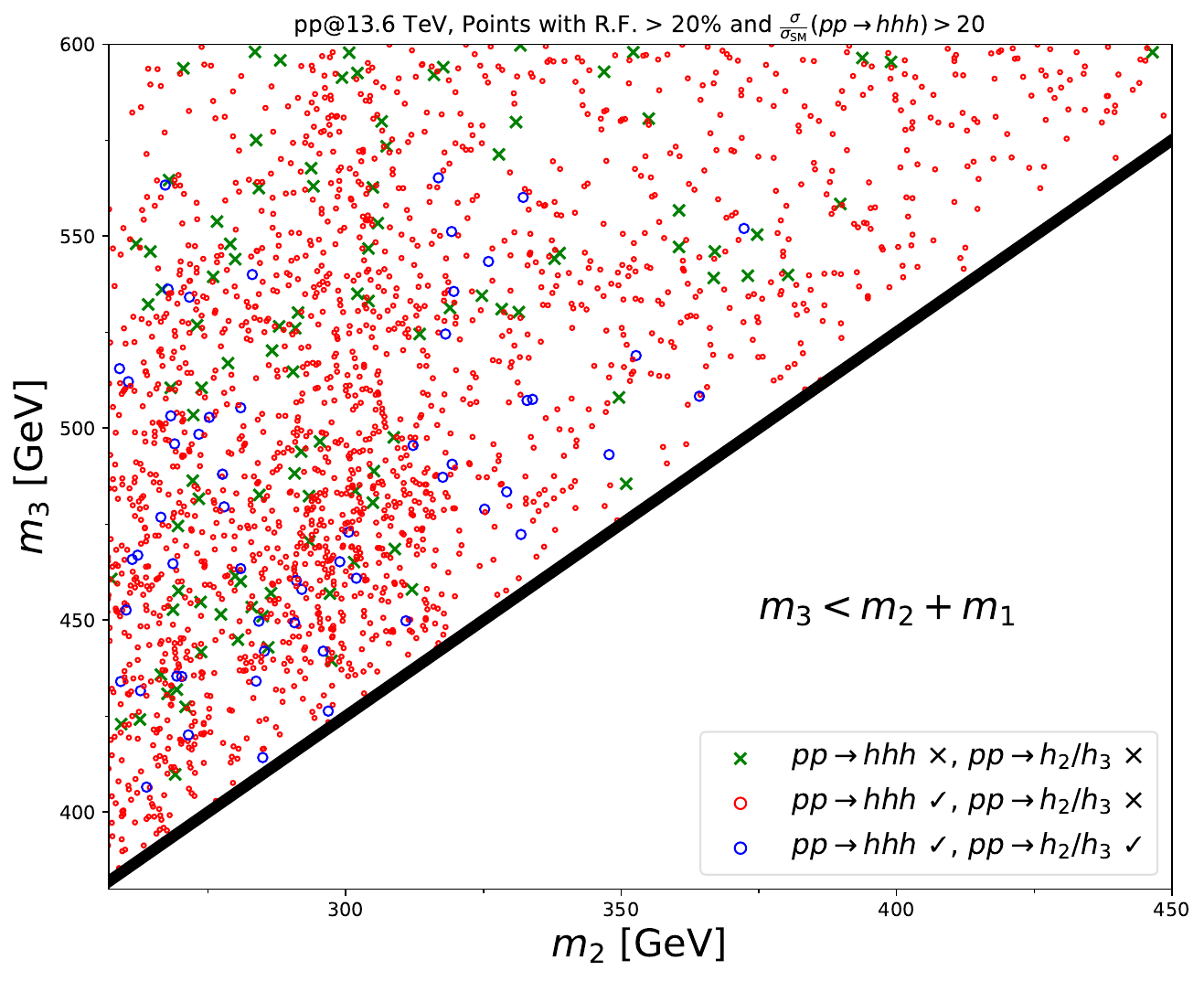}
  \includegraphics[width=0.45\textwidth]{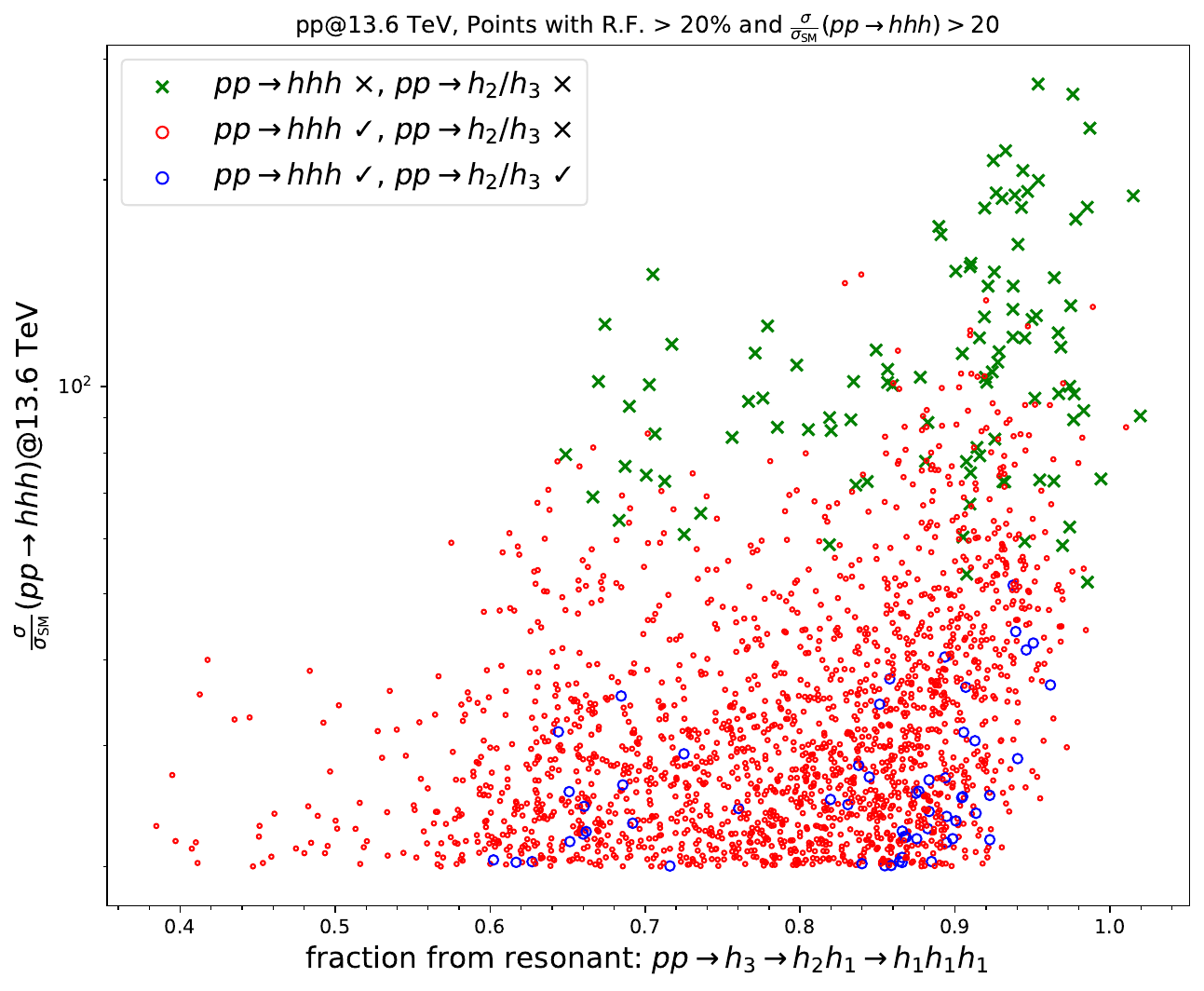}
\caption{\label{fig:resscan} Exclusion results obtained through our phenomenological analysis of the 6 $b$-jet final state for the high-luminosity LHC at 13.6~TeV, with $L=3000$~fb$^{-1}$. {The exclusion is obtained after applying the optimized selection cuts summarized in Table~\ref{tb:cuts}.} Only points with at least a factor of 20 enhancement over the SM triple Higgs boson production cross section are shown. The TRSM parameter-space points considered have been selected to satisfy the theoretical and experimental constraints following Ref.~\cite{Karkout:2024ojx}, and in addition are required to have an approximate double-resonant process contribution to the total cross section of $\mathrm{R.F.}\sim 20\%$. This choice is consistent with the correlation found in Fig.~\ref{fig:EnhancementvsresonantFraction}. We also consider constraints coming from single production of the new scalar resonances $h_{2/3}$, where green crosses indicate points that will be excluded by both double-resonant triple Higgs boson production and single scalar production, red circles indicate points that will be excluded only by single scalar production, and blue circles indicate points that will be excluded by neither. The plot on the left shows the points on the $(m_2, m_3)$-plane, whereas the plot on the right shows their total cross section enhancement plotted against the (approximate) contribution to the total cross section coming from the double-resonant process, akin to Fig.~\ref{fig:EnhancementvsresonantFraction}. The right-hand side plot indicates that in practice, the requirement of a 20$\times$ enhancement over the SM supersedes the $\mathrm{R.F.}\sim 20\%$ requirement, allowing for points with only $\mathrm{R.F.} \gtrsim 40\%$.}
\end{center}
\end{figure*}

A comparison was then made of the $\rho^2$ value corresponding to each generated point, to the $\rho^2$ limit at 95\% C.L. derived through our analysis. The results are shown in Fig.~\ref{fig:resscan} for an integrated luminosity of $\mathcal{L}=3000$~fb$^{-1}$, on the $(m_2, m_3)$-plane in the left plot, and the corresponding total cross section enhancement plotted against the (approximate) contribution to the total cross section coming from the double-resonant process, akin to Fig.~\ref{fig:EnhancementvsresonantFraction}, is shown in the right panel. We have also imposed the restrictions that the parameter-space points have a cross section at least 20 times larger than that of SM triple Higgs boson production, and they should ``carry'' at least $20\%$ of their cross section in the double-resonant process (i.e.\ their ``resonant fraction'', R.F., should be $>20\%$). These restrictions are motivated by the positive correlation between the enhancement of the cross section over the SM and the resonant fraction demonstrated in Fig.~\ref{fig:EnhancementvsresonantFraction}. We also examine the impact of additional ATLAS and CMS analyses not included in \texttt{HiggsTools}, obtained through extrapolating the luminosity following the procedure outlined in Appendix~D of Ref.~\cite{Papaefstathiou:2020iag}. These originate from the analyses considering the ``single'' production processes $pp \rightarrow h_i \rightarrow h_1 h_1$ \cite{Sirunyan:2018two,Aad:2019uzh}, $pp \rightarrow h_i \rightarrow ZZ$ \cite{Sirunyan:2018qlb,Cepeda:2019klc} and $pp \rightarrow h_i \rightarrow W^+W^-$ \cite{Aaboud:2017gsl,ATL-PHYS-PUB-2018-022}, for $i=2,3$.\footnote{We note that the double-resonant process will constitute a background to the $pp \rightarrow h_i \rightarrow h_1 h_1$ searches, but we expect its effect to be subleading.} It is interesting to note that, for the considered parameter-space points, the branching ratio $\mathrm{BR}(h_3 \rightarrow h_2 h_1 \rightarrow h_1 h_1 h_1)$ ranges from $\sim 1\%$ to $\sim 35\%$, $\mathrm{BR}(h_3 \rightarrow h_1 h_1)$ from $\sim 0.1\%$ to $\sim 45\%$, $BR(h_3 \rightarrow WW~\mathrm{or}~ZZ)$ from $4\%$ to $65\%$. A reasonable subset, $\sim 100$ out of $\sim 2000$, of viable parameter-space points can be excluded through double-resonant triple Higgs boson production (green crosses), as well as by single scalar ($h_2$ or $h_3$) production through the extrapolated analyses. We note, however, that a large number of parameter-space points will likely not be excluded through triple Higgs boson production, even at the end of the lifetime of the LHC, allowing only an upper limit to be imposed on $\rho^2$, and hence the parameters by which it is constructed. Nevertheless, a large fraction of the parameter-space points will likely be excluded through single $h_2$ or $h_3$ production (red circles), with a smaller number potentially evading these searches as well (blue circles), at least within the context of the analyses considered here. We anticipate that for the parameter-space points that would evade all searches, a future collider would be necessary, see e.g.~\cite{Papaefstathiou:2020iag}.

It is interesting to observe from the right-hand side plot of Fig.~\ref{fig:resscan}, that triple Higgs boson production can \textit{only} be excluded if the fraction of the double-resonant process is greater than $\sim 65\%$ and the corresponding cross section enhancement greater than $\sim 50\times$ the standard model value. This supports our use of the double-resonant topology as the primary target of the present cut-based study, while a full experimental analysis may exploit additional information from non-resonant contributions or other final states. An additional interesting observation is that there exist parameter-space points with large triple Higgs boson production cross sections, e.g.~$\mathcal{O}(100)\times$ the SM value, that cannot be excluded through that process. This is a consequence of differences in kinematics originating from the different masses of the scalars.

Next, we also note that we have found \textit{no} points that would not be excluded in single scalar production, while being excluded by double-resonant triple Higgs boson production. This suggests that, in view of the results obtained in the present analysis, triple Higgs boson production may not play the role of an exclusion channel for the two new scalar resonances $h_2$ and $h_3$ in the TRSM. Definitive proof of this statement would require concrete experimental analyses, which we are looking forward to.

Finally, it is also interesting to note that during our parameter-space scan for viable points within the TRSM, according to theoretical and experimental constraints, we did not obtain any with $m_3 \gtrsim 650$~GeV or $m_2 \gtrsim 450$~GeV that also satisfy the $\times 20$ enhancement over the SM triple Higgs boson production. This hints towards a rapid decrease of the resonant triple Higgs boson cross section in those regions of parameter space. {One important reason for this behaviour is that the production of the $s$-channel scalar approximately follows the gluon-fusion production rate of an SM-like scalar with the same mass, multiplied by the appropriate mixing factor, and this rate decreases rapidly as the scalar mass increases. The final size of the enhancement is also affected by the viable branching ratios and by the theoretical and experimental constraints imposed on the TRSM scan.} We would like to point out a similar finding in Ref.~\cite{Lane:2024vur}, where the complex singlet model was investigated, see Appendix C of the aforementioned reference.

\newpage
\subsection{Case Studies: Resonant Versus Full Non-Resonant Production at the LHC}\label{sec:nonres}

\begin{figure*}[!t]
\centering
\includegraphics[width=0.43\textwidth]{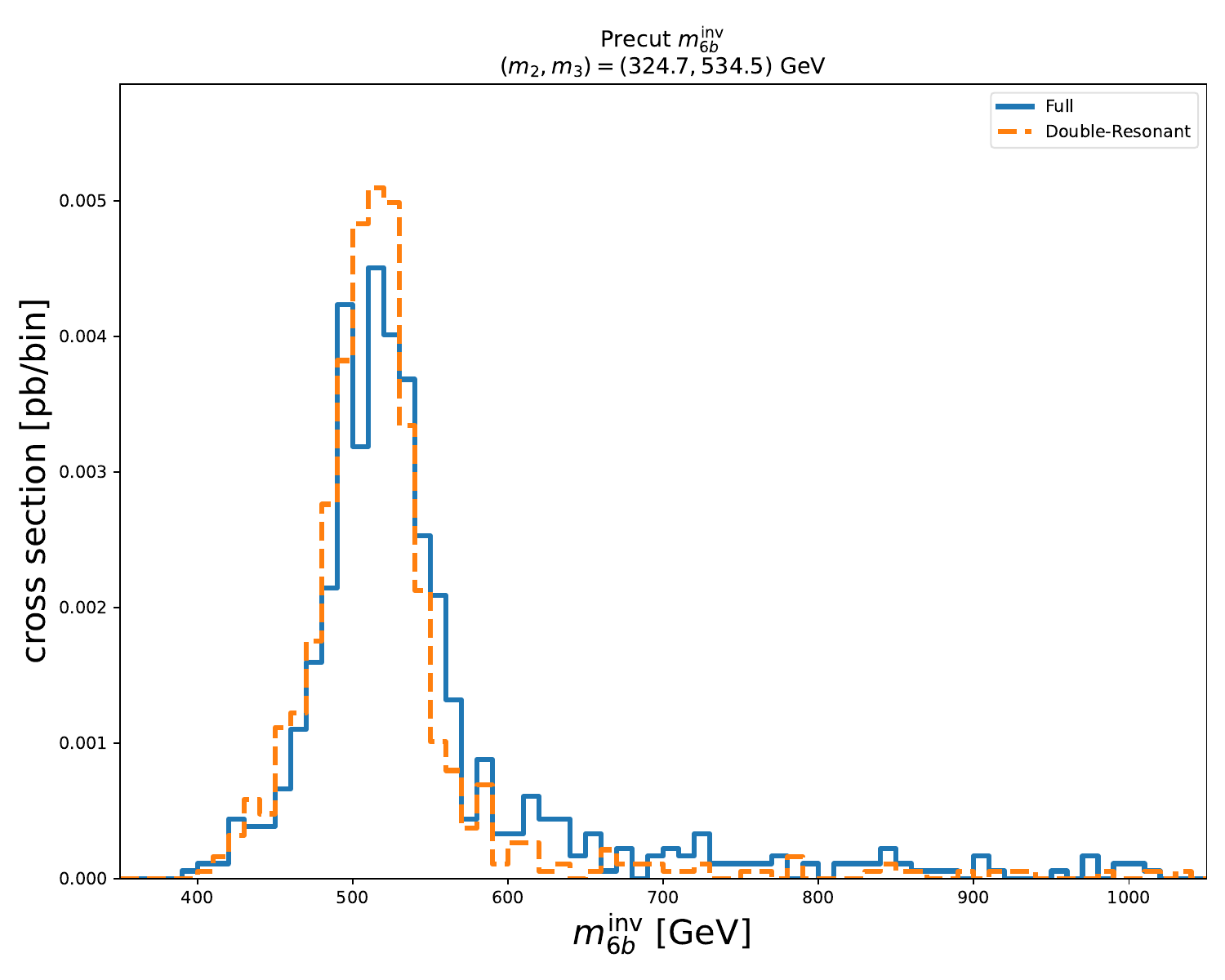}\hfill
\includegraphics[width=0.43\textwidth]{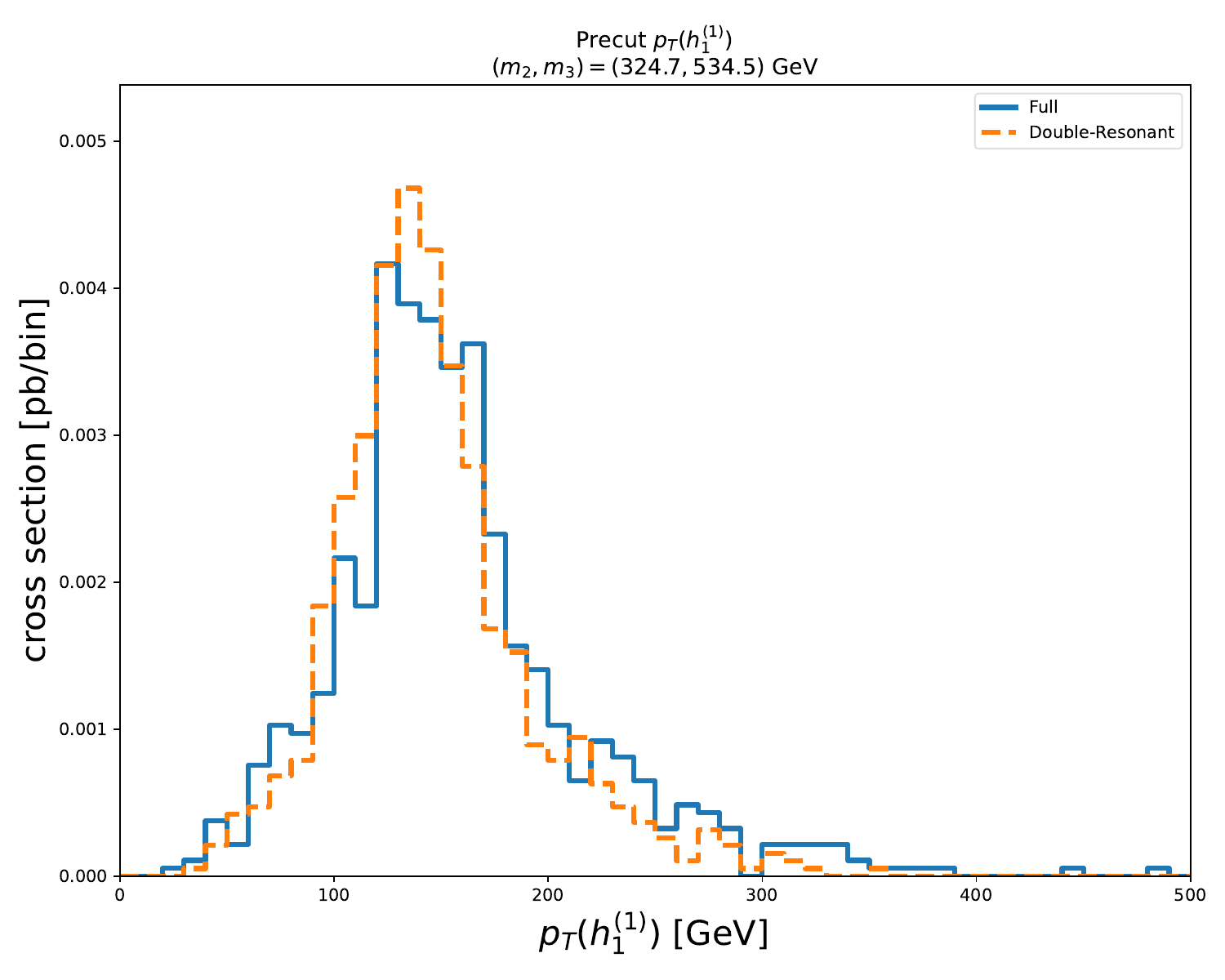}\\
\includegraphics[width=0.43\textwidth]{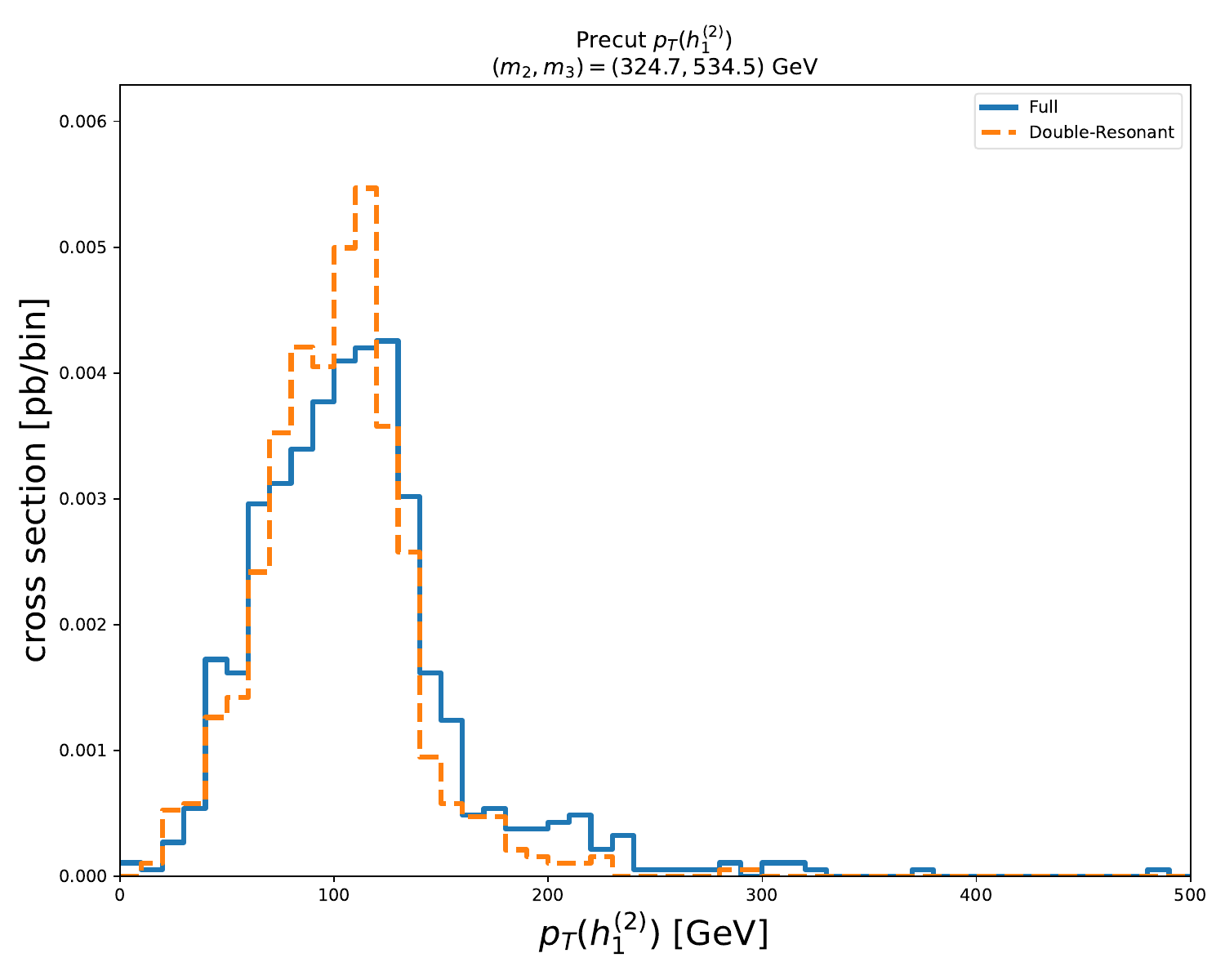}\hfill
\includegraphics[width=0.43\textwidth]{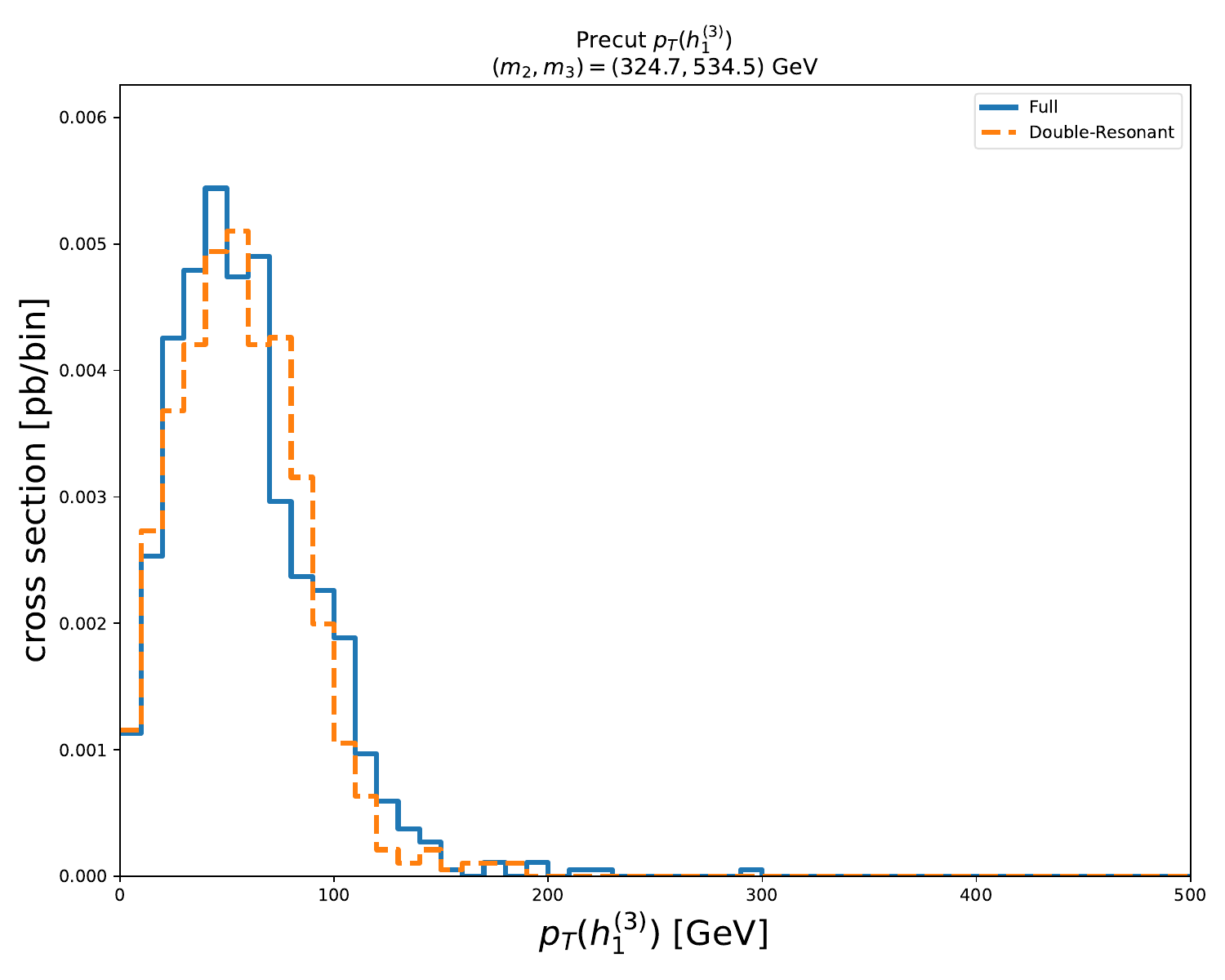}\\
\includegraphics[width=0.43\textwidth]{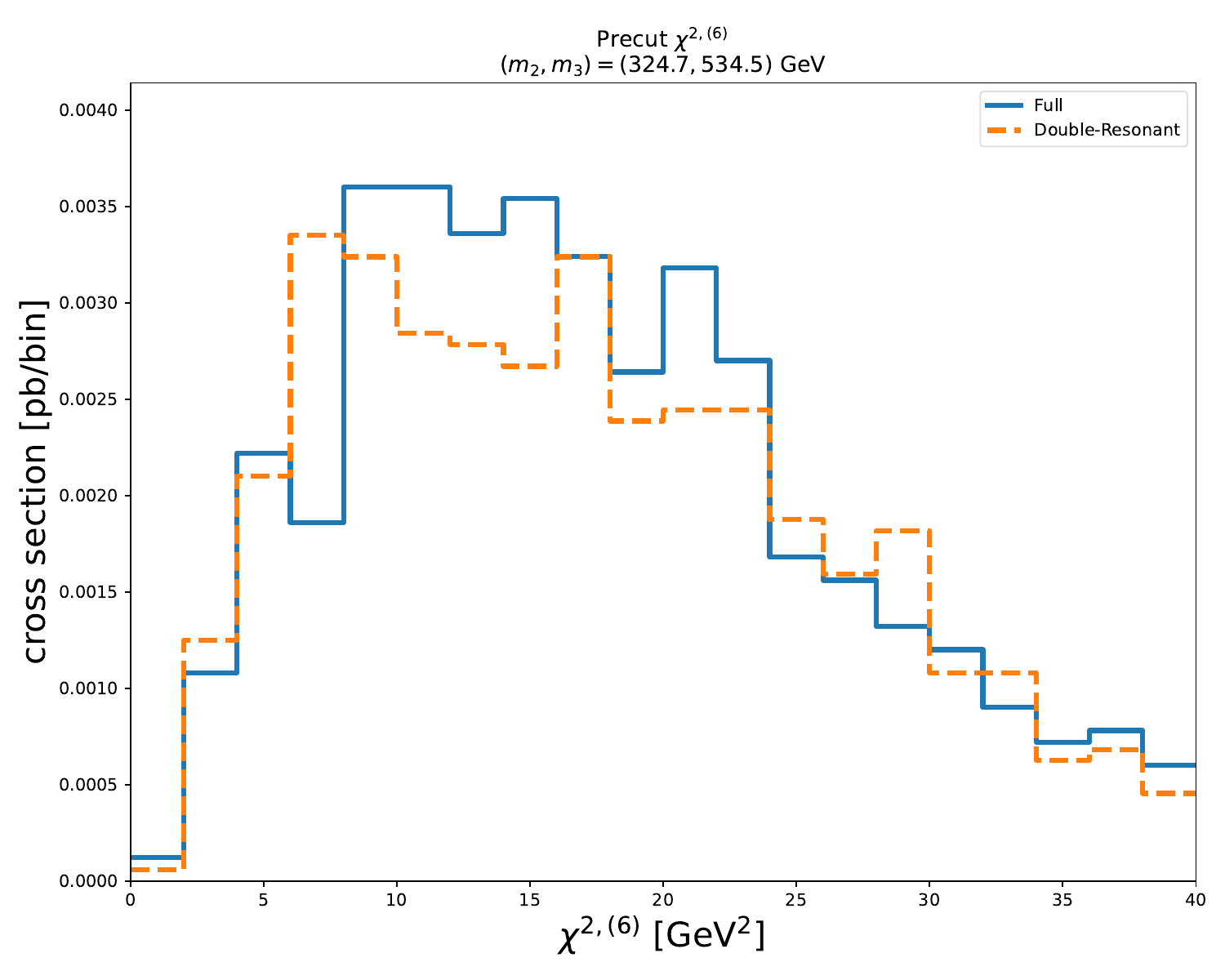}\hfill
\includegraphics[width=0.43\textwidth]{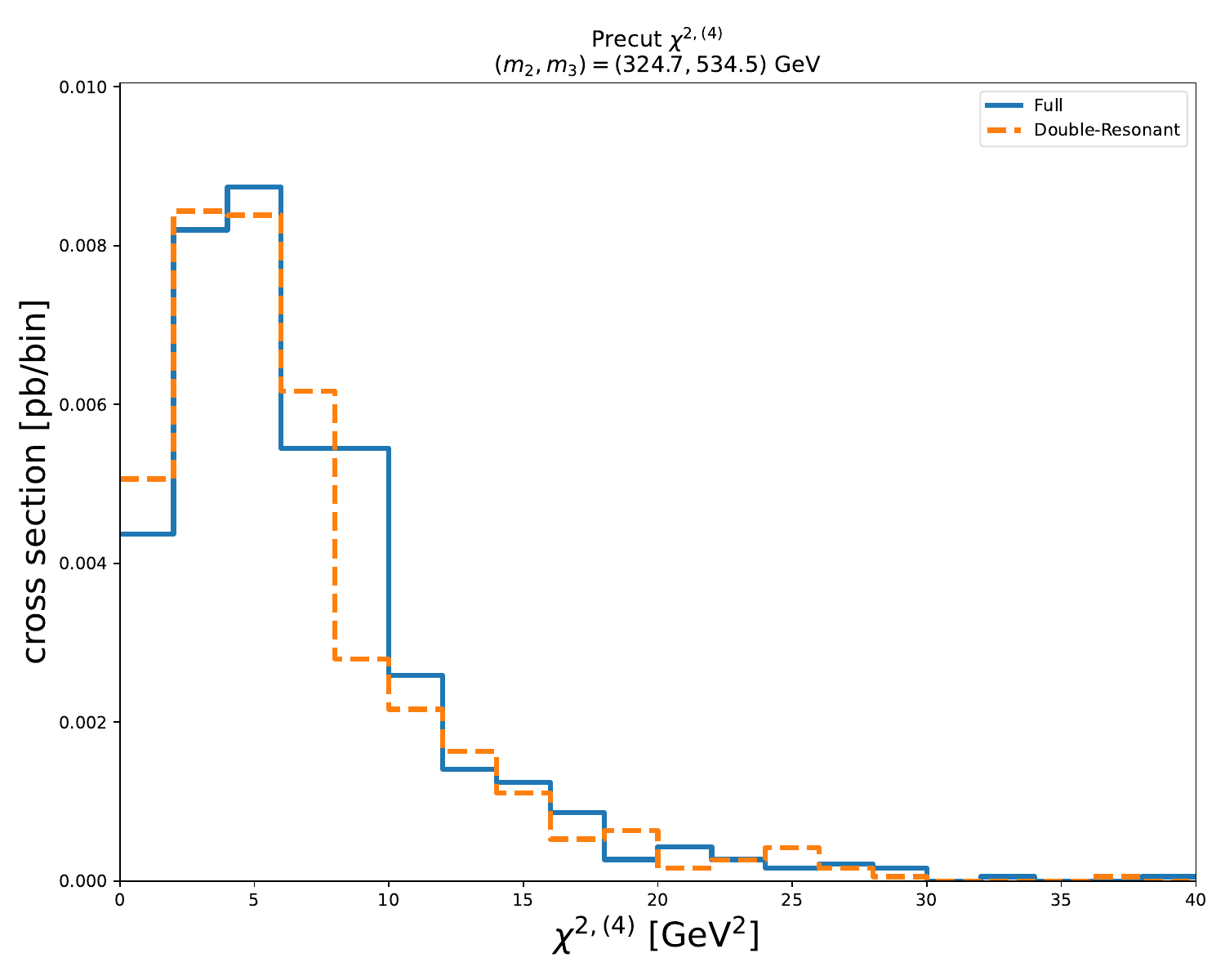}
\caption{\label{fig:nonres_precut_324p7_534p5}{Representative reconstructed-level distributions comparing the full leading-order and double-resonant processes for the TRSM point with $(m_2,m_3)=(324.7,534.5)$~GeV, before the optimized selection cuts. The panels show $m_{6b}^{\rm inv}$, the ordered transverse momenta of the reconstructed $h_1$ candidates, and the $\chi^{2,(6)}$ and $\chi^{2,(4)}$ variables defined in Eqs.~\eqref{eq:chi26} and~\eqref{eq:chi24}.}}
\end{figure*}

{We investigate the impact of non-resonant components and interference effects on our analysis, since these effects are model- and phase-space-dependent and should not be assumed to be small in general. For the TRSM points that reach a 95\% C.L. exclusion at the high-luminosity LHC in our analysis (green crosses in Fig.~\ref{fig:resscan}), we compare the full leading-order triple Higgs boson production process, including non-resonant and interference effects, with double-resonant Monte Carlo event samples, applying the same phenomenological analysis to both.}

\begin{figure*}[!t]
\centering
\includegraphics[width=0.45\textwidth]{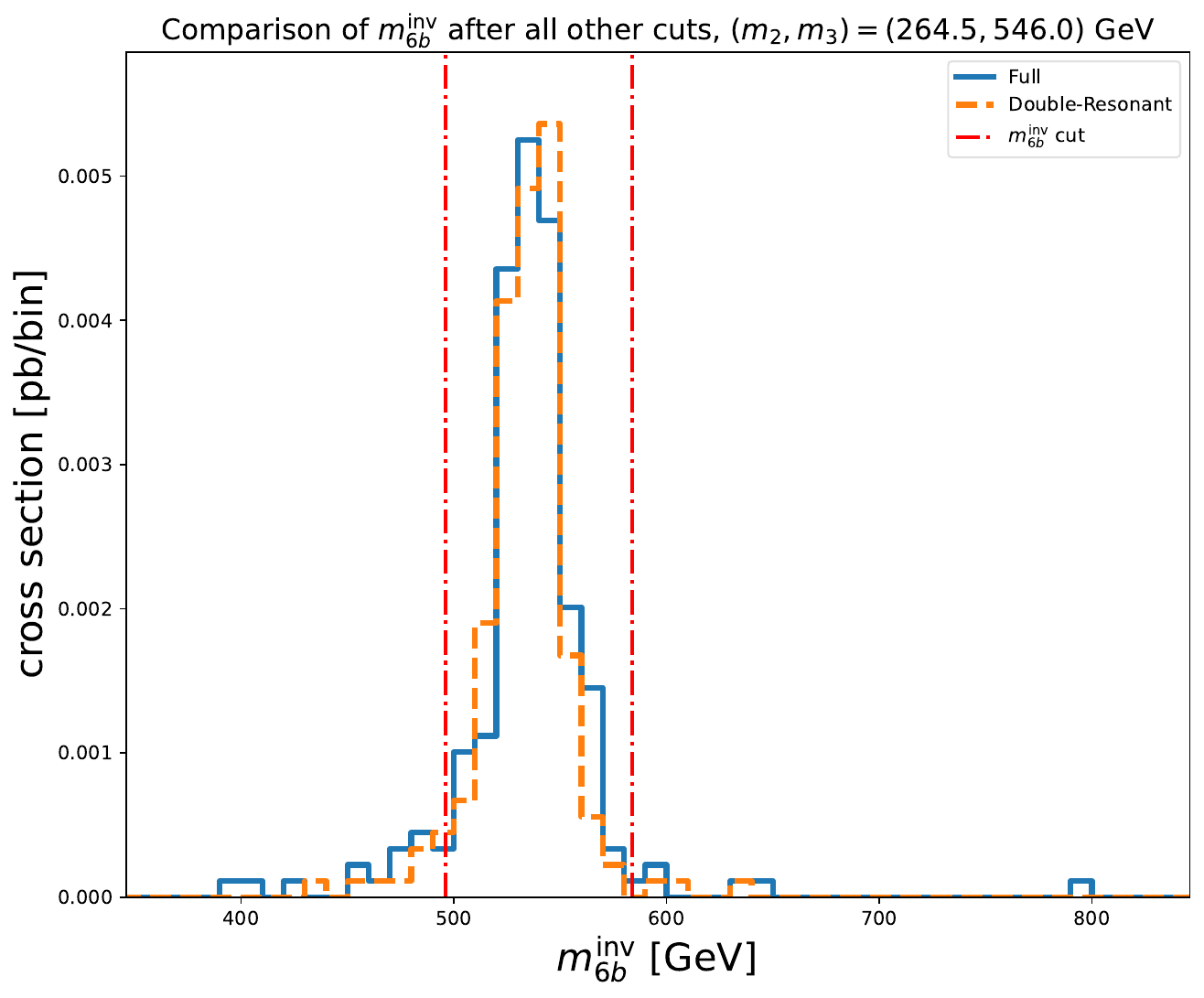}
\includegraphics[width=0.45\textwidth]{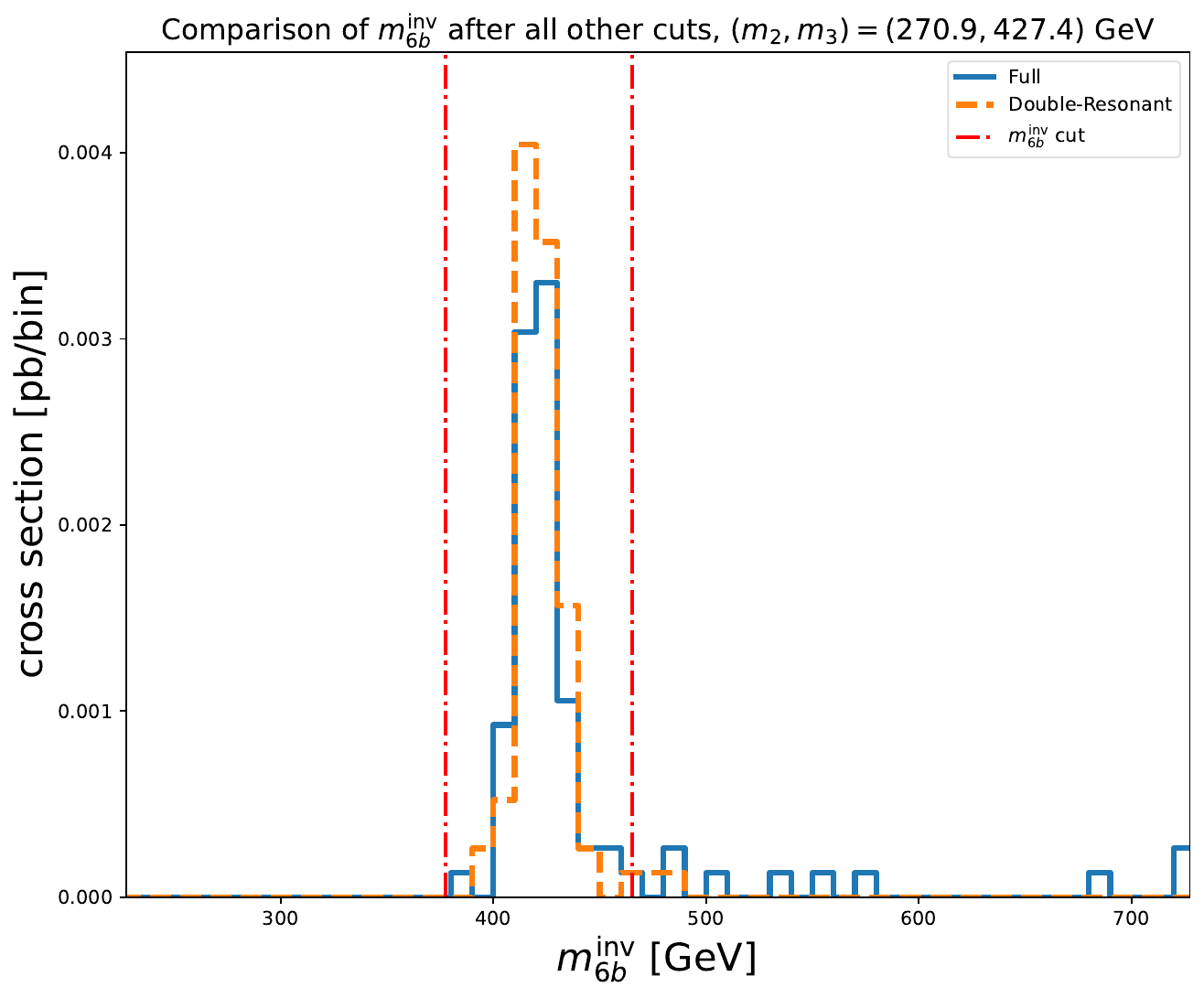}
\includegraphics[width=0.45\textwidth]{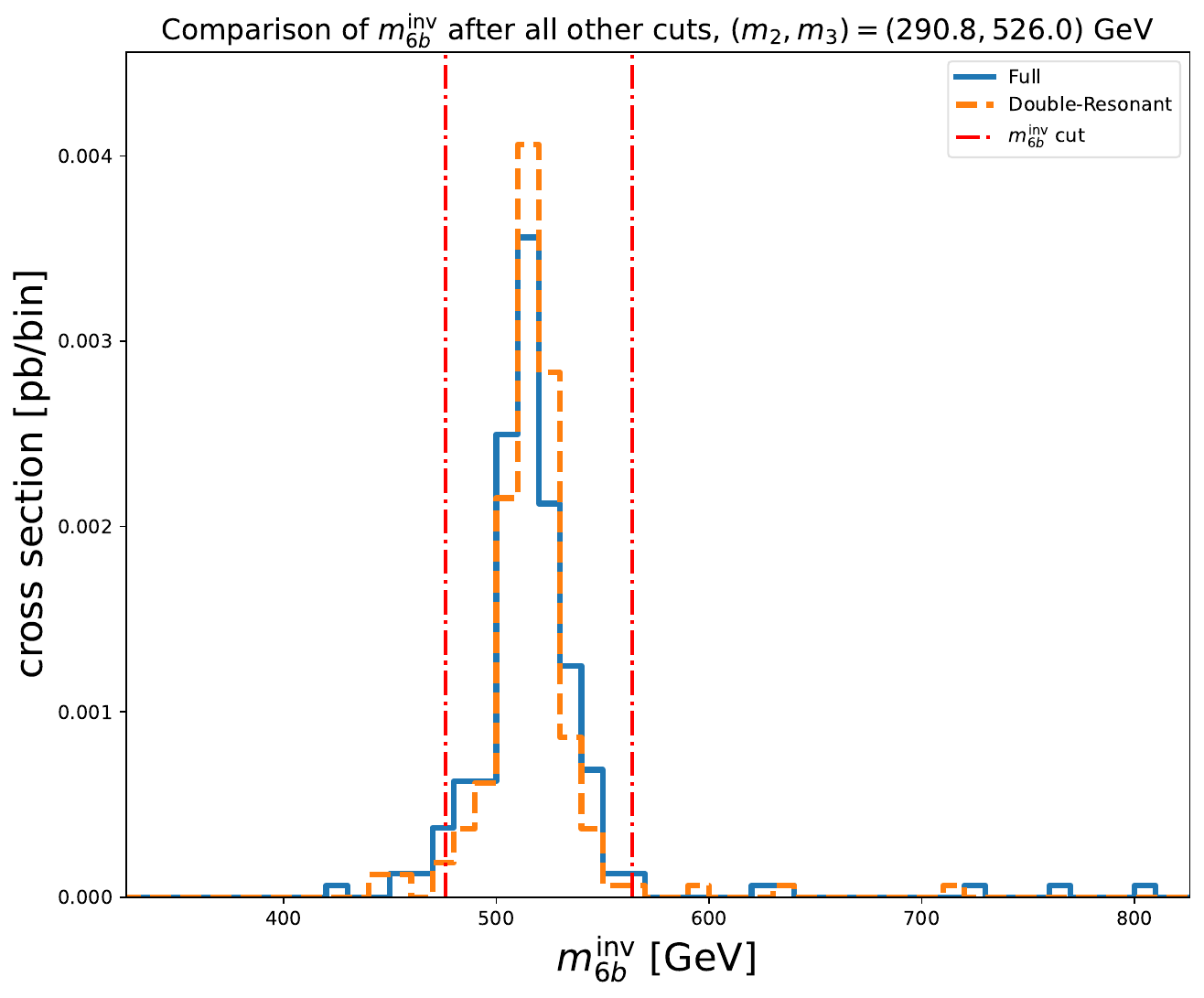}
\includegraphics[width=0.45\textwidth]{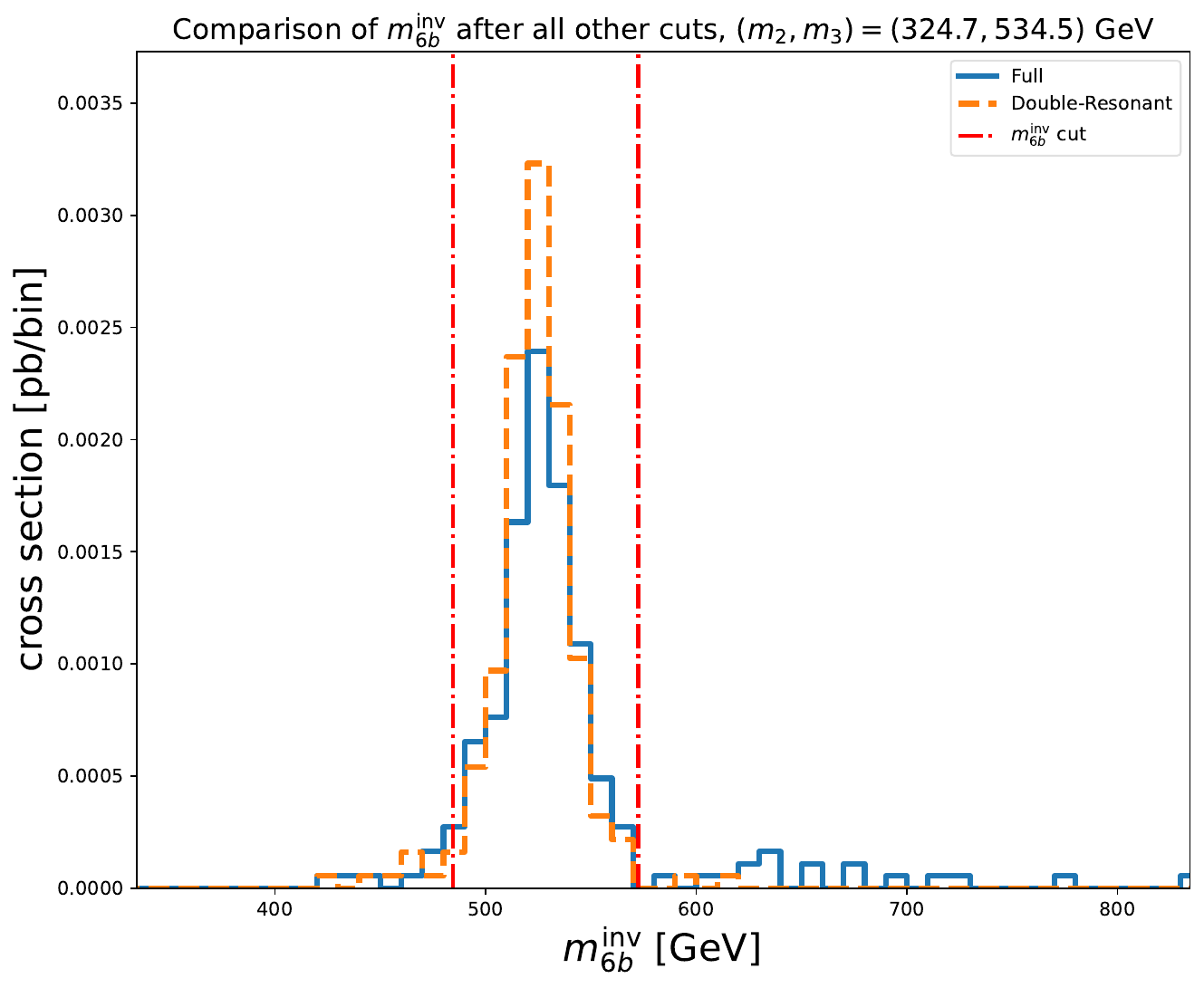}
\caption{\label{fig:m6b} A comparison of the 6 $b$-jet invariant mass distributions, $m_{6b}^\mathrm{inv}$, obtained through the analysis of the full process, including all non-resonant effects, and the double-resonant process \textit{only}. All other cuts have been applied, apart from the $m_{6b}^\mathrm{inv}$ window, shown in the red dashed-dotted lines. {For the benchmark points shown, the distributions indicate that the non-resonant part of the cross section has a limited impact on this cut-based analysis observable.}}
\end{figure*}

{The distributions in Fig.~\ref{fig:nonres_precut_324p7_534p5} provide an explicit check, for one specific TRSM point, of observables that enter the reconstruction and selection strategy, and that could also be used in a multivariate analysis, after requiring 6 $b$-jets with $p_{T,b} > 37$~GeV and $|\eta_b| < 2.5$, and before the optimized cuts are imposed. Recent work has explored this direction explicitly for resonant triple Higgs boson searches in the fully hadronic 6 $b$ channel using deep-learning techniques~\cite{Chiang:2025dlh}. For this representative benchmark, the full and double-resonant samples show broadly similar behaviour in these pre-cut variables. We therefore complement this point-level pre-cut comparison with the post-selection comparison shown below (Fig.~\ref{fig:m6b}), performed for several parameter-space points that can be excluded at the high-luminosity LHC.}

{For the post-selection comparison,} we have constructed the resulting 6 $b$-jet invariant mass distribution, \textit{after all other cuts} have been applied, for the full process that includes all interference effects and off-shell effects, and the double-resonant process \textit{only}. These are shown in Fig.~\ref{fig:m6b}, for four sample parameter-space points that can be excluded at the high-luminosity LHC via our phenomenological analysis. For the excluded TRSM points considered in this study, we find a limited impact on the cut-based significance. This does not preclude non-resonant contributions from being useful in a dedicated multivariate experimental analysis, especially before selection cuts. We have also estimated the expected significance for both cases after all cuts, and the fractional change between them. A frequency plot of the expected change in statistical significance is shown in Fig.~\ref{fig:sigchange} for the $\mathcal{O}(100)$ parameter-space points, demonstrating that the change is expected to be marginal, with most points possessing a $\mathcal{O}(10\%)$ change in the number of standard deviations. We emphasize the fact that since no smearing was applied to simulate detector effects, beyond those appearing due to the Monte Carlo simulation of hadronization, we expect any actual experimental results to exhibit even smaller differences in a more realistic analysis.

\FloatBarrier
\twocolumn[{%
\begin{center}
  \includegraphics[width=0.52\textwidth]{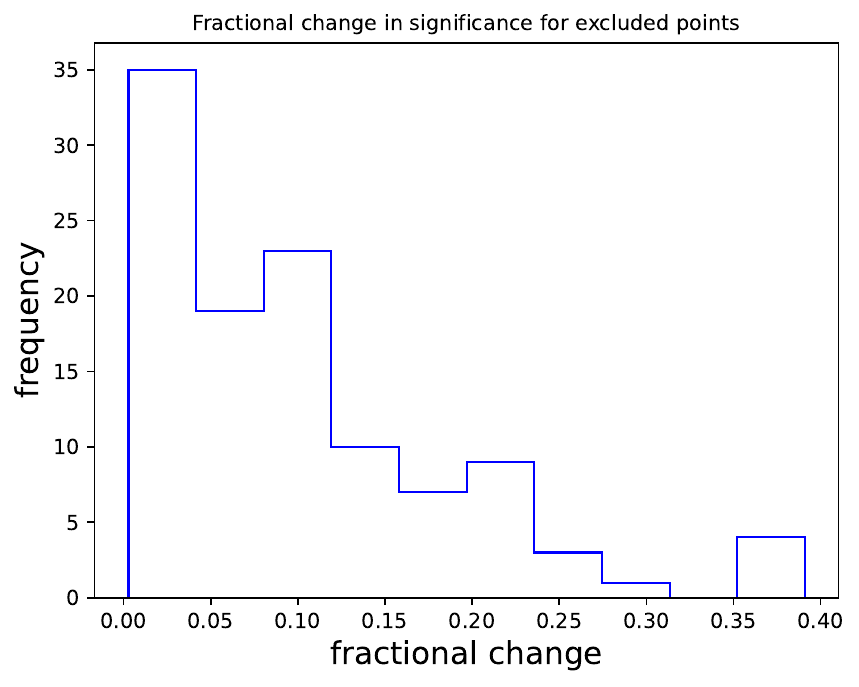}
\captionof{figure}{\label{fig:sigchange} Frequency histogram of the fractional change in statistical significance between full and double-resonant samples for the high-luminosity LHC analysis at $3000$~fb$^{-1}$, for points expected to be excluded at 95\% C.L. {The comparison is performed after applying the optimized selection cuts summarized in Table~\ref{tb:cuts}.} Most points show $\mathcal{O}(10\%)$ changes, indicating a limited impact of non-resonant effects on this cut-based sensitivity.}
\end{center}
}]

\section{Conclusions} \label{sec:conclusions}

We have developed, and applied, a simplified approach to investigating double-resonant triple Higgs boson production, $ pp \rightarrow h_3 \rightarrow h_2 h_1 \rightarrow h_1 h_1 h_1$, in models with extended scalar sectors that include at least two new, narrow, scalar resonances that mix with the SM-like Higgs boson. The two main assumptions are that:
\begin{itemize}
\item two new scalars $h_{2,3}$ are present with $m_3 > m_2 + m_1$ and $m_2 > 2m_1$, and
\item the two scalars possess narrow widths, $\Gamma_i / m_i \ll 1$.
\end{itemize}
The process is then characterized by a single rescaling parameter, which we dub $\rho^2$, that constitutes a function of all relevant parameters that determine the cross section for the process. The kinematic distributions in this simplified approach are otherwise independent of the scalar couplings $\lambda_{112}$, $\lambda_{123}$, the mixing parameter $\kappa_3$, and the widths of the new scalars, as long as the latter are small enough compared to the scalar masses (i.e.\ $\Gamma_{2,3} \ll m_{2,3}$), as demonstrated in~\ref{app:ME_NWA_gg_h3_h2_3h1}. We have used this approach to study the parameter space of the TRSM, a model with two new singlet scalar fields $S$ and $X$, with discrete symmetries $\mathbb{Z}^S_2$ and $\mathbb{Z}^X_2$. Through investigations of scalar-dominated deformations of parameter-space points, we find that the narrow-width approximation remains valid in light of the expected limits of our analysis, that effectively drive the width-to-mass ratios to $\Gamma_i/m_i \lesssim 10^{-2}$.

We also observe, empirically, within the context of our analyses, that significantly-enhanced triple Higgs boson production in the TRSM is limited to the range of masses $m_3 \lesssim 650$~GeV or $m_2 \lesssim 450$~GeV, and that it can be excluded for a subset of these points, through the $pp \rightarrow 6$ $b$-jets final state, while single heavy scalar production (i.e.\ $pp\rightarrow h_2$ or $pp\rightarrow h_3$) will exclude the majority of the investigated points. We also find that some of these points could remain viable, even at the end of the high-luminosity LHC, and that triple Higgs boson production is unlikely to play the role of a discovery channel for the TRSM. We would like to reiterate that these observations have been made considering the simple phenomenological analyses of the present article, and will need to be reevaluated in concrete and fully optimized experimental analyses. In addition, more final states of the triple Higgs boson production, particularly $4b+2\tau$ and $4b+\gamma\gamma$, will increase the reach.

Finally, we investigated the contribution of the non-resonant part of the process to our analysis, finding that, for the TRSM benchmarks and the cut-based 6 $b$-jet analysis considered here, it has a limited impact on the expected sensitivity. Our method can therefore be used as a fast approximate reinterpretation tool for models with two narrow scalar resonances realizing the double-resonant decay chain, provided that the dominance of this topology and the validity of the NWA are checked in the model under consideration. The approach could be applied in the first instance to obtain fast and approximate constraints on the rescaling parameter, $\rho^2$, over the theoretical parameter space of the model. We leave the detailed extraction of measurements in the scenario of discovery to future work.

\section*{Acknowledgments}

We would like to thank Tania Robens for useful comments and discussions during the mini-workshop on ``HHH and other extended scalar sector signatures'' at the Ru\dj er Bo\v{s}kovi\'{c} Institute, Zagreb, Croatia, in July 2024, and beyond. We would also like to thank Ian Lewis for useful discussions during the ``Extended Scalar Sectors From All Angles'' workshop at CERN, in October 2024. A.P. acknowledges support by the National Science Foundation under Grant No.\ PHY 2210161. G.T.X. received support for this project from the European Union's Horizon
2020 research and innovation programme under the Marie Sklodowska-Curie grant agreement No
945422. This research was supported by the Deutsche Forschungsgemeinschaft (DFG, German Research Foundation), under grant 396021762 - TRR 257. We acknowledge the usage of the Siegen OMNI cluster, where the parallel computing calculations corresponding to this project were performed.

\FloatBarrier
\appendix
\renewcommand{\theHequation}{\Alph{section}.\arabic{equation}}

\section{Matrix Element and Narrow-Width Factorization for \texorpdfstring{$gg \rightarrow h_3 \rightarrow (h_2 \rightarrow h_1 h_1) h_1$}{gg to h3 to (h2 to h1 h1) h1}}
\label{app:ME_NWA_gg_h3_h2_3h1}

This appendix collects the tree-level matrix element for the double-resonant process
\begin{equation}
g(p_1,a,\varepsilon_1)\;g(p_2,b,\varepsilon_2)\;\to\;h_3\;\to\;h_2\,h_1\;\to\;h_1\,h_1\,h_1\,,
\label{eq:process_app}
\end{equation}
both away from and within the narrow-width approximation (NWA) for $h_3$ and $h_2$, and shows the corresponding factorization of the partonic differential cross section as represented in the main text by the $\rho^2$ factor, defined in Eq.~\eqref{eq:rhosqdef}. In particular, in the double NWA the fully differential partonic rate is proportional to
\begin{equation}
d\hat\sigma(gg\to 3h_1)\;\propto\;\frac{\kappa_3^2 \lambda_{123}^2\,\lambda_{112}^2}{\Gamma_3\,\Gamma_2}\,,
\label{eq:scaling_statement_app}
\end{equation}
up to the $gg\to h_3$ form factor and kinematic/phase-space factors that we derive below.

\subsection{Conventions and Couplings}

We assume the scalar interactions
\begin{equation}
\mathcal{L}_{\rm int}\supset -\lambda_{123}\,h_1 h_2 h_3 \;-\;\lambda_{112}\,h_1^2 h_2\,,
\label{eq:Lint_app}
\end{equation}
The $gg\to h_3$ coupling is encoded in an $\hat s$-dependent form factor $C(\hat s)$ via the effective vertex
\begin{align}
V^{\mu\nu}_{ggh_3}(p_1,p_2) \;=\; i\, \kappa_3 C(\hat s)\,\delta^{ab}\,
\left[ (p_1\!\cdot\!p_2)\,g^{\mu\nu}-p_1^{\nu}p_2^{\mu}\right],\nonumber \\
\label{eq:ggh3vertex_app}
\end{align}
with $\hat s\equiv (p_1+p_2)^2$ as usual, and we use Breit-Wigner propagators for the scalar particles $h_2$ and $h_3$:
\begin{equation}
D_j(q^2)\;=\;\frac{i}{q^2-m_j^2+i\,m_j\Gamma_j}\,,
\qquad j\in\{2,3\}.
\label{eq:BW_app}
\end{equation}
We denote the three final-state $h_1$ momenta by $q_1,q_2,q_3$.

\subsection{Matrix Element without the NWA}

The tree-level amplitude for~\eqref{eq:process_app} is the sum over the three assignments of which final-state $h_1$ was produced directly in $h_3\to h_2h_1$ (the three $h_1$ are identical). Defining
\begin{equation}
T^{\mu\nu}\equiv (p_1\!\cdot\!p_2)\,g^{\mu\nu}-p_1^{\nu}p_2^{\mu},
\qquad s_{ij}\equiv (q_i+q_j)^2,
\label{eq:Tmunu_sij_app}
\end{equation}
the amplitude can be written compactly as
\begin{align}
\mathcal{M}
&= -i\,\kappa_3\,C(\hat s)\,\lambda_{123}\,(2\lambda_{112})\;
\delta^{ab}\;\Big(\varepsilon_{1\mu}\varepsilon_{2\nu}\,T^{\mu\nu}\Big)\;
D_3(\hat s)\nonumber\\
&\quad\times\Big[\, D_2(s_{12})+D_2(s_{13})+D_2(s_{23})\,\Big]\,.
\label{eq:M_full_app}
\end{align}
where $\varepsilon_{1\mu}$ and $\varepsilon_{2\nu}$ are the gluon polarizations.

Summing over gluon colours and physical polarizations, and using gauge invariance to take
$\sum_{\rm pol}\varepsilon_\mu \varepsilon_\nu^\ast=-g_{\mu\nu}$, one obtains
\begin{align}
\sum_{\rm col,pol}|\mathcal{M}|^2
&= 32\,|C(\hat s)|^2\,\kappa_3^2\,\lambda_{123}^2\lambda_{112}^2\;
\Big(T^{\mu\nu}T_{\mu\nu}\Big)\;|D_3(\hat s)|^2\nonumber\\
&\quad\times \Big|\,D_2(s_{12})+D_2(s_{13})+D_2(s_{23})\,\Big|^2\,.
\label{eq:Msq_sum_general_app}
\end{align}
Using
\begin{equation}
T^{\mu\nu}T_{\mu\nu}=2(p_1\!\cdot\!p_2)^2=\frac{\hat s^2}{2},
\label{eq:Tcontract_app}
\end{equation}
we arrive at
\begin{align}
\sum_{\rm col,pol}|\mathcal{M}|^2
&= 32\,|C(\hat s)|^2\,\kappa_3^2\,\lambda_{123}^2\lambda_{112}^2\;
\frac{\hat s^2}{2}\;|D_3(\hat s)|^2\nonumber\\
&\quad\times \Big|\,D_2(s_{12})+D_2(s_{13})+D_2(s_{23})\,\Big|^2\,.
\label{eq:Msq_sum_4D_app}
\end{align}
The spin/colour-averaged squared matrix element is
\begin{equation}
\overline{|\mathcal{M}|^2}\;=\;
\frac{1}{4\times 8^2}\;\sum_{\rm col,pol}|\mathcal{M}|^2
\;=\;\frac{1}{256}\;\sum_{\rm col,pol}|\mathcal{M}|^2\,,
\label{eq:avg_app}
\end{equation}
where the factor $4$ averages over the two initial gluon helicities and $8^2$ over the initial gluon colours.

\subsection{Double Narrow-width Approximation for $h_3$ and $h_2$}

In the NWA, one replaces
\begin{align}
|D_j(q^2)|^2
&=\frac{1}{(q^2-m_j^2)^2+(m_j\Gamma_j)^2}
\nonumber\\
&\xrightarrow{\rm NWA}
\frac{\pi}{m_j\Gamma_j}\,\delta(q^2-m_j^2),
\label{eq:NWA_id_app}
\end{align}
for $j\in\{2,3\}$. This can be derived by considering the integral of the Breit-Wigner factor over the virtuality of the particle, $q_i^2$, as follows (see, e.g.~\cite{lundlecturenotes}):
\begin{eqnarray}
\int_{-\infty}^{+\infty} \frac{ \mathrm{d} q^2 } { (q^2 - m_j^2)^2 + (m_j \Gamma_j)^2} = \frac{ \pi }{m_j \Gamma_j}\;.
\end{eqnarray}

For the $h_2$ contribution, the NWA further implies that the overlap of different resonance assignments is negligible in the phase-space integral, so the interference terms between different $s_{ij}$ channels can be dropped at leading NWA accuracy. One then has
\begin{align}
\left|\sum_{(ij)}D_2(s_{ij})\right|^2
&\;\xrightarrow{\rm NWA}\;
\sum_{(ij)} |D_2(s_{ij})|^2 \nonumber \\
&\;\xrightarrow{\rm NWA}\;
\frac{\pi}{m_2\Gamma_2}\sum_{(ij)}\delta(s_{ij}-m_2^2),
\label{eq:S_NWA_app}
\end{align}
with $(ij)\in\{(12),(13),(23)\}$. Therefore,
\begin{align}
\sum_{\rm col,pol}|\mathcal{M}|^2_{\rm NWA}
&=\;32\,|C(\hat s)|^2\,\kappa_3^2
\lambda_{123}^2\lambda_{112}^2\;\nonumber \\
&\quad\times\frac{\hat s^2}{2}\;
\left(\frac{\pi}{m_3\Gamma_3}\right)\delta(\hat s-m_3^2)\; \nonumber \\
&\quad\times \left(\frac{\pi}{m_2\Gamma_2}\right)
\sum_{(ij)}\delta(s_{ij}-m_2^2)\,.
\label{eq:Msq_NWA_app}
\end{align}
Eq.~\eqref{eq:Msq_NWA_app} makes the parametric dependence on the widths explicit: at fixed kinematics the squared matrix element is proportional to $\rho^2 = \kappa_3^2 \lambda_{123}^2\lambda_{112}^2/(\Gamma_3\Gamma_2)$ (up to the overall $|C(\hat s)|^2$ and the $\delta$-functions enforcing on-shell kinematics).

\subsection{Partonic Differential Cross Section and Explicit $\rho^2$ Scaling}

The partonic differential cross section for a $2\to 3$ process is
\begin{equation}
d\hat\sigma(gg\to 3h_1)\;=\;\frac{1}{2\hat s}\;\overline{|\mathcal{M}|^2}\;d\Phi_3(p_1+p_2;q_1,q_2,q_3)\,,
\label{eq:dsigma_general_app}
\end{equation}
where $d\Phi_3$ is the three-body Lorentz-invariant phase space. Substituting~\eqref{eq:Msq_NWA_app} into~\eqref{eq:dsigma_general_app} yields the fully differential double-NWA form
\begin{align}
\mathrm{d}\hat\sigma_{\rm NWA}(gg\to 3h_1)
&=\frac{1}{512\hat s}\;
\sum_{\rm col,pol}|\mathcal{M}|^2_{\rm NWA}\;
\mathrm{d}\Phi_3
\nonumber\\[2mm]
&=
\frac{\kappa_3^2 \lambda_{123}^2\lambda_{112}^2}{\Gamma_3\,\Gamma_2}\;
\left[
\frac{|C(\hat s)|^2}{2\hat s}\;
\frac{\hat s^2}{2}\;
\frac{\pi^2}{m_3 m_2}\;\right. \nonumber \\
&\left. \times \delta(\hat s-m_3^2)\;\sum_{(ij)}\delta(s_{ij}-m_2^2)
\right]\;
\frac{\mathrm{d}\Phi_3}{8}\,,
\label{eq:dsigma_NWA_explicit_scaling_app}
\end{align}
where we have used $32/256=1/8$. At fixed $\hat{s}$, the factors appearing in the square brackets only depend on the masses $m_2$ and $m_3$ as required by Eq.~\eqref{eq:rhosqdef}. Therefore, the proportionality
\begin{equation}
\mathrm{d}\hat\sigma_{\rm NWA}(gg\to 3h_1)\;\propto\;\frac{\kappa_3^2 \lambda_{123}^2\lambda_{112}^2}{\Gamma_3\,\Gamma_2}
\label{eq:dsigma_scaling_explicit_app}
\end{equation}
is now explicit at the differential level.

\section{Understanding the Shape of \texorpdfstring{$m_{h_1 h_1}$}{m(h1 h1)}}\label{app:Broad_Peak_Limits}

When considering the invariant mass $m_{h_1 h_1}$ obtained from pairing \textit{any} two Higgs bosons in the double-resonant process, a characteristic double-peak structure arises: a sharp peak at $m_{h_1 h_1}\simeq m_2$ when the two Higgs bosons originate from the $h_2\to h_1 h_1$ decay, and a broader structure when the $h_1$ produced directly in $h_3 \to h_2 h_1$ is combined with one of the $h_1$ bosons from the subsequent $h_2 \to h_1 h_1$ decay. This pattern was demonstrated at the stable-Higgs boson level in Fig.~\ref{fig:restest}.

\begin{figure}[htp]
\begin{center}
  \includegraphics[width=0.50\columnwidth]{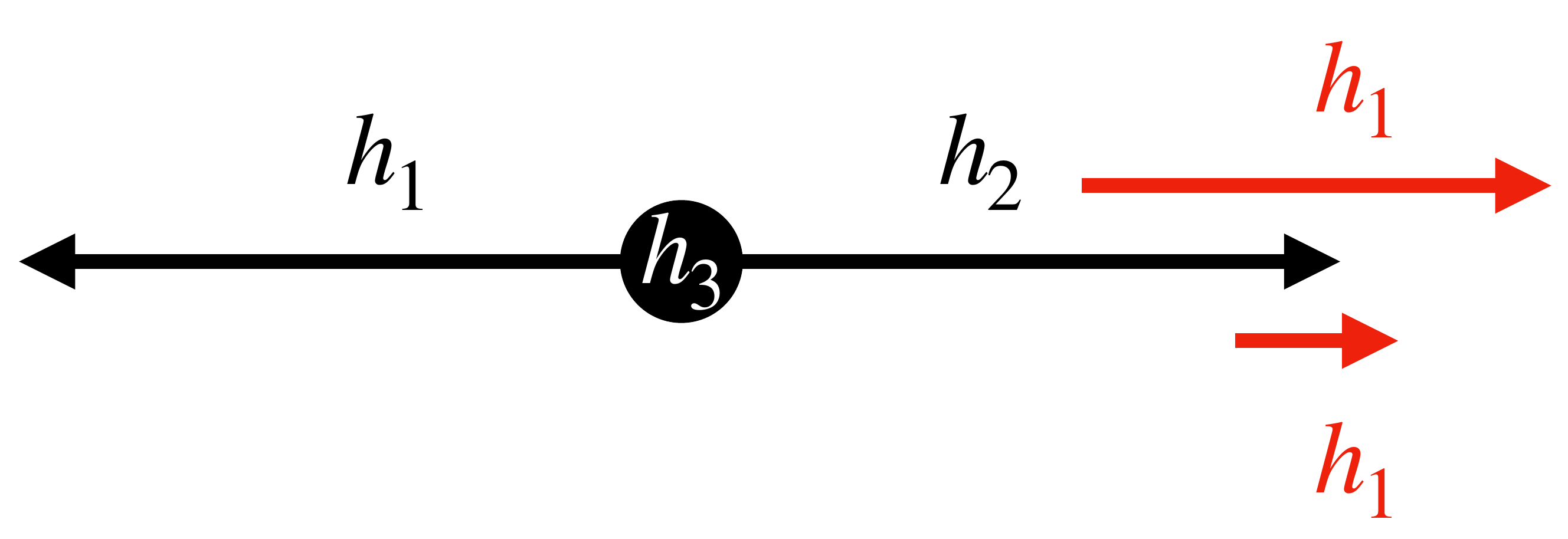}
\caption{\label{fig:peaklimitsdiag} The decay $h_3 \rightarrow h_2 h_1 \rightarrow (h_1 h_1)\, h_1$ in the centre-of-mass frame of $h_3$. The decay $h_2 \rightarrow h_1 h_1$ yields two Higgs bosons with unequal momenta in the $h_3$ frame due to the boost of $h_2$.}
\end{center}
\end{figure}

The upper and lower edges of the broader ``wrong''-combination structure can be derived by considering the process $h_3 \rightarrow h_2 h_1 \rightarrow (h_2 \rightarrow h_1 h_1)\, h_1$ in the rest frame of $h_3$, as shown in Fig.~\ref{fig:peaklimitsdiag}. In this frame, the magnitude of the back-to-back three-momentum of the ``direct'' $h_1$ and the $h_2$ is
\begin{equation}
p \equiv |\vec{p}|  = \frac{1}{2m_3}\,\sqrt{\left(m_1^2+m_2^2-m_3^2\right)^2-4m_1^2m_2^2}\;.
\end{equation}
If the four-momenta of the ``direct'' $h_{1,2}$ in the $h_3$ frame are denoted as
$q_{1,2} = (E_{1,2}, 0,0, \mp p)$, respectively, and the four-momenta of the Higgs bosons coming from $h_2$ as $k_+ = (\varepsilon_+, 0, 0, k_{z,+})$ and $k_- = (\varepsilon_-, 0, 0, k_{z,-})$, then the endpoints of the broad structure follow from the minimum and maximum of the invariant
\begin{equation}
(q_1 + k_\pm)^2.
\end{equation}
The extrema are obtained for collinear configurations of the three-momenta of $q_1$ and $k_\pm$. To determine the possible values of $k_\pm$, we work in the $h_2$ rest frame, where the four-momenta of the two Higgs bosons are
\begin{equation}
k^*_{\pm} = \left(\frac{m_2}{2}, 0, 0, \pm k^*\right)\,,
\end{equation}
with $k^* = \frac{1}{2} \sqrt{m_2^2 - 4m_1^2}$. Boosting back to the $h_3$ frame using the velocity of $h_2$, $\beta_2 = p/E_2$, yields
\begin{equation}
k_{\pm} = \left(\frac{E_2}{2} \pm \frac{p}{m_2} k^*,\, 0,0,\, \frac{p}{2} \pm \frac{E_2}{m_2}k^*\right)\,,
\end{equation}
i.e.\ $\varepsilon_{\pm} = \frac{E_2}{2} \pm \frac{p}{m_2} k^*$ and $k_{z,\pm} = \frac{p}{2} \pm \frac{E_2}{m_2}k^*$.

Then, the two possible values of $m_{h_1 h_1, \pm} \equiv \sqrt{(q_1 + k_\pm)^2}$ are
\begin{equation}
    m_{h_1 h_1, \pm} = \sqrt{2 \left(m_1^2 + E_1 \varepsilon_{\pm} +  p\, k_{z,\pm} \right)}\;.
\end{equation}
We have verified explicitly that these expressions reproduce the edges of the broad structures of Fig.~\ref{fig:restest} as follows:
\begin{itemize}
    \item \textbf{BM0}: $m_{h_1 h_1, -}=304.9$~GeV, $m_{h_1 h_1, +}=363.1$~GeV,
    \item \textbf{BM7}: $m_{h_1 h_1, -}=309.3$~GeV, $m_{h_1 h_1, +}=448.7$~GeV.
\end{itemize}
The code snippet to obtain the wrong combination edges for any of the allowed $(m_2,m_3)$ combinations is available on the associated gitlab repository~\cite{gitlabrepoTwoSingletScan}.

\clearpage
\onecolumn
\section{Selected Benchmark Points}\label{app:benchmarks}
\begin{table}[H]
\centering
\small
\setlength{\tabcolsep}{5pt}
\renewcommand{\arraystretch}{1.15}
\caption{Selected benchmark points (model parameters). A sample of selected benchmark points obtained during the scan of Ref.~\cite{Karkout:2024ojx} for the TRSM. A complete list is provided in the ancillary files of Ref.~\cite{Karkout:2024ojx}.}
\label{tab:benchmarks_model}
\begin{tabular}{@{}l r r r r r r r@{}}
\toprule
Name & $m_2$ & $m_3$ & $v_S$ & $v_X$ & $\theta_{12}$ & $\theta_{13}$ & $\theta_{23}$\\
\midrule
BM0 & 259.0 & 495.0 & 215.8 & 180.8 & 6.191 & 0.1629 & 5.691 \\
BM1 & 270.6 & 444.7 & 122.4 & 847.2 & 0.2679 & 0.03007 & 0.5219 \\
BM2 & 268.6 & 452.7 & 137.8 & 784.8 & 0.2632 & 0.02293 & 0.6451 \\
BM3 & 272.6 & 480.7 & 928.3 & 143.7 & 3.098 & 2.900 & 2.375 \\
BM4 & 269.0 & 409.8 & 138.0 & 599.4 & 0.2436 & 0.004146 & 0.7730 \\
BM5 & 269.1 & 486.9 & 227.5 & 307.9 & 0.07443 & 6.149 & 2.631 \\
BM6 & 259.2 & 577.0 & 289.0 & 275.6 & 0.1368 & 6.148 & 2.324 \\
BM7 & 283.7 & 575.0 & 259.4 & 330.4 & 0.1373 & 6.152 & 2.299 \\
BM8 & 264.3 & 469.3 & 207.3 & 359.5 & 0.2847 & 6.277 & 0.6918 \\
BM9 & 266.5 & 461.9 & 653.1 & 229.0 & 2.889 & 3.046 & 1.015 \\
BM10 & 259.2 & 399.7 & 444.5 & 217.0 & 2.917 & 3.046 & 1.047 \\
\bottomrule
\end{tabular}
\end{table}

\begin{table}[H]
\centering
\small
\setlength{\tabcolsep}{4pt}
\renewcommand{\arraystretch}{1.15}
\caption{Selected benchmark points corresponding to Table~\ref{tab:benchmarks_model}, focusing on the corresponding phenomenological parameters. The columns $\rho^2$ and $\hat{\sigma}_u$ are quoted in $[10^6\,\mathrm{GeV}^2]$ and $[10^{-9}\,\mathrm{pb/GeV}^2]$, respectively.}
\label{tab:benchmarks_pheno}
\begin{tabular}{@{}l r r r r r r r r r@{}}
\toprule
Name & $\sigma/\sigma_{\rm SM}$ & R.F. & $\Gamma_2$ & $\Gamma_3$ & $\kappa_3$ & $\lambda_{123}$ & $\lambda_{112}$ & $\rho^2$ & $\hat{\sigma}_u$\\
\midrule
BM0 & 306.0 & 0.9552 & 0.003514 & 3.927 & 0.1854 & -191.8 & 8.167 & 6.110 & 2.018 \\
BM1 & 302.4 & 0.9290 & 0.5078 & 2.586 & 0.1571 & -204.3 & 67.52 & 3.574 & 3.408 \\
BM2 & 275.6 & 0.9536 & 0.3805 & 3.142 & 0.1741 & -203.6 & 57.78 & 3.509 & 3.165 \\
BM3 & 267.2 & 0.9476 & 0.2009 & 4.758 & 0.2024 & -224.6 & 41.39 & 3.703 & 2.908 \\
BM4 & 266.4 & 0.9761 & 0.2836 & 1.995 & 0.1713 & -180.3 & 48.89 & 4.031 & 2.663 \\
BM5 & 157.6 & 0.9557 & 0.0003346 & 2.017 & 0.1527 & 103.3 & -2.477 & 2.264 & 2.805 \\
BM6 & 145.5 & 0.7810 & 0.0006274 & 5.790 & 0.1908 & 196.3 & -3.701 & 5.289 & 1.108 \\
BM7 & 122.5 & 0.7791 & 0.001056 & 5.587 & 0.1884 & 193.5 & -3.578 & 2.885 & 1.711 \\
BM8 & 119.1 & 0.9988 & 0.3916 & 2.941 & 0.1746 & -144.3 & 55.88 & 1.721 & 2.789 \\
BM9 & 112.8 & 0.8632 & 0.3092 & 2.042 & 0.1635 & 142.8 & 39.98 & 1.381 & 3.290 \\
BM10 & 103.7 & 0.9730 & 0.2188 & 0.9312 & 0.1463 & 121.2 & 35.41 & 1.936 & 2.159 \\
\bottomrule
\end{tabular}
\end{table}

\clearpage
\twocolumn

\FloatBarrier
\bibliographystyle{JHEP}
\bibliography{biblio.bib}

\end{document}